\documentclass[%
reprint,amssymb,
amsmath,nobibnotes,siunitx,aps,prb,
floatfixaps,citeautoscript,showkeys,
floatfix,balance,longbibliography,
superscriptaddress,
bibnotes,
]{revtex4-2}
\usepackage[table,dvipsnames]{xcolor}
\usepackage[]{inputenc}
\usepackage[T1]{fontenc}
\usepackage{graphicx}
\graphicspath{{./figure/}}
\usepackage{grffile}
\usepackage{wrapfig}
\usepackage{rotating}
\usepackage[normalem]{ulem}
\usepackage{amsmath}
\usepackage{textcomp}
\usepackage{amssymb}
\usepackage{capt-of}
\usepackage{fancyhdr}
\usepackage[explicit]{titlesec}
\usepackage{hyperref}
\usepackage{caption}
\usepackage{setspace}
\usepackage[]{xifthen}
\usepackage[]{pgf,pgfplots}
\usepackage{helvet}
\usepackage[eulergreek]{sansmath}
\pgfplotsset{
  tick label style = {font=\small\sansmath\sffamily},
  every axis label = {font=\small\sansmath\sffamily},
  legend style = {font=\small\sansmath\sffamily},
  label style = {font=\small\sansmath\sffamily}
}

\usepackage[paperwidth=210mm,paperheight=297mm,centering,hmargin=1.14cm,vmargin=1.4cm]{geometry}
\usepackage{titlesec}

\titlespacing\section{0pt}{4pt plus 4pt minus 2pt}{4pt plus 2pt minus 2pt}
\titlespacing\subsection{0pt}{4pt plus 4pt minus 2pt}{1pt plus 2pt minus 2pt}
\titlespacing\subsubsection{0pt}{4pt plus 4pt minus 2pt}{4pt plus 2pt minus 2pt}
\titlespacing\paragraph{0pt}{1pt plus 4pt minus 2pt}{2pt plus 2pt minus 2pt}

\setcitestyle{super}

\usepackage{graphicx}
\usepackage{natbib}
\usepackage{dcolumn}
\usepackage{amsmath, amsbsy, amssymb}
\usepackage{bm}
\usepackage{float,array}
\usepackage{textcomp}
\usepackage[version=3]{mhchem}
\usepackage{times}
\usepackage{helvet}
\usepackage{multirow}
\usepackage{booktabs}
\usepackage{mathtools}
\usepackage[]{hyperref}
\hypersetup{colorlinks,pdfborder={0,0,0},linkbordercolor={white},citecolor=MidnightBlue,filecolor=MidnightBlue,linkcolor=MidnightBlue,urlcolor=MidnightBlue}
\usepackage{pbox}
\usepackage{diagbox}
\usepackage{makecell}
\usepackage{tikz}
\usepackage{array}
\usepackage{booktabs,adjustbox}
\usepackage[UKenglish,USenglish,english,american]{babel}
\usepackage[]{upgreek}

\newcolumntype{L}[1]{>{\raggedright\let\newline\\\arraybackslash\hspace{0pt}}m{#1}}
\newcolumntype{C}[1]{>{\centering\let\newline\\\arraybackslash\hspace{0pt}}m{#1}}
\newcolumntype{R}[1]{>{\raggedleft\let\newline\\\arraybackslash\hspace{0pt}}m{#1}}

\usepackage{forest}
\usetikzlibrary{arrows,arrows.meta,decorations.markings,shapes.arrows,shapes.geometric,calc,shadows}

\usepackage{verbatim}
\makeatletter
\newcommand{\nextverbatimspread}[2]{%
  \def\verbatim@font{%
    \linespread{#1}\normalfont\ttfamily#2
    \gdef\verbatim@font{\normalfont\ttfamily#2}}
}

\renewcommand{\p@subsection}{}
\renewcommand{\p@subsubsection}{}
\makeatother

\usefont{T1}{cmr}{m}{n}

\usepackage{nameref}

\title{}
\hypersetup{
 pdfauthor={zxwush28},
 pdftitle={The Taming of the Screw: Dislocation Cores in BCC Metals and Alloys},
 pdfkeywords={},
 pdfsubject={},
 pdfcreator={Emacs 27.2 (Org mode 9.4.5)},
 pdflang={English}}

\makeatletter
\newcommand*{\balancecolsandclearpage}{%
  \close@column@grid
  \cleardoublepage
}
\renewcommand\frontmatter@abstractwidth{\dimexpr\textwidth-0.2in\relax}
\makeatother

\usepackage{datatool}
\DTLloaddb[noheader=false]{hs_gp_v_vs_solute_conc}{hs_gp_v_data.csv}

\usepackage{filecontents}

\usepackage{datatool}
\DTLloaddb[noheader=false]{hs_gp_vi_vs_solute_conc}{hs_gp_vi_data.csv}

\usepackage{filecontents}

\begin{document}
\begin{filecontents*}{hs_gp_v_data.csv}
solv,sol,conc,sorg,ssol,xer,yer,theta,temp,solgp,bibkey,bksx,bksy
%Nb,Mo,0,  19.230769230769234, 19.230769230769234,\telipa,\telipb,0,77,6,foxall_1970_am,2,20
Nb,Mo,2,   19.230769230769234, 28.065934065934066,\telipa,\telipb,0,77,6,foxall_1970_am,3,10
Nb,Mo,5,   19.230769230769234, 30.307692307692307,\telipa,\telipb,0,77,6,foxall_1970_am,2,20
Nb,Mo,8.5, 19.230769230769234, 40.72527472527473, \telipa,\telipb,0,77,6,foxall_1970_am,2,20
Nb,Mo,16,  19.230769230769234, 53.38461538461539, \telipa,\telipb,0,77,6,foxall_1970_am,2,20
Nb,Re,9,   19.230769230769234, 64.98901098901099, \telipa,\telipb,0,77,7,foxall_1970_am,2,20
%----------------------------------------------------------------------------------------------------
%Nb,Ta,0,                  34.72568578553616, 34.72568578553616,\telipa,\telipb,0,77,5, peters_1970_mt,2,20
Nb,Ta, 14.406779661016948, 34.72568578553616, 42.50623441396509,\telipa,\telipb,0,77,5, peters_1970_mt,2,20
Nb,Ta,30.790960451977405,  34.72568578553616, 41.309226932668324,\telipa,\telipb,0,77,5,peters_1970_mt,2,20
Nb,Ta,46.61016949152542,   34.72568578553616, 42.80548628428927,\telipa,\telipb,0,77,5, peters_1970_mt,2,20
Nb,Ta,61.29943502824859,   34.72568578553616, 41.408977556109726,\telipa,\telipb,0,77,5,peters_1970_mt,2,20
Nb,Ta,100,                 34.72568578553616, 40.11221945137157,\telipa,\telipb,0,77,5, peters_1970_mt,2,20

%Nb,Mo,0,                  34.72568578553616, 34.72568578553616,\telipa,\telipb,0,77,6, peters_1970_mt,2,20
Nb,Mo,0.1315789473684208,  34.72568578553616, 36.53266331658291,\telipa,\telipb,0,77,6, peters_1970_mt,1,30
Nb,Mo,2.5,                 34.72568578553616, 36.834170854271356,\telipa,\telipb,0,77,6,peters_1970_mt,2,20
Nb,Mo,5.06578947368421,    34.72568578553616, 36.934673366834176,\telipa,\telipb,0,77,6,peters_1970_mt,2,20
Nb,Mo,7.9605263157894735,  34.72568578553616, 46.28140703517588,\telipa,\telipb,0,77,6, peters_1970_mt,2,20
Nb,Mo,11.118421052631579,  34.72568578553616, 49.69849246231156,\telipa,\telipb,0,77,6, peters_1970_mt,2,20

% ----------------------------------------------------------------------------------------------------
%Ta,Re,0,                   22.42857142857143, 22.42857142857143,\telipa,\telipb,0,77,7,ulitchny_1971_iaea,2,20
Ta,Re,1.2771084337349399,  22.42857142857143, 31.095238095238095,\telipa,\telipb,0,77,7,ulitchny_1971_iaea,2,20
Ta,Re,2.216867469879518,  22.42857142857143, 34.904761904761905,\telipa,\telipb,0,77,7, ulitchny_1971_iaea,2,20
Ta,Re,3.903614457831326,  22.42857142857143, 44.523809523809526,\telipa,\telipb,0,77,7, ulitchny_1971_iaea,2,20

% ----------------------------------------------------------------------------------------------------
%Nb,Mo,0,                   22.62942779291553, 22.62942779291553,\telipa,\telipb,0,77,6,statham_1971_sm,2,20
Nb,Mo,4.4524495677233435,  22.62942779291553, 29.931880108991827,\telipa,\telipb,0,77,6,statham_1971_sm,2,20
%Nb,Re,0,                   18.81856540084388, 18.81856540084388,\telipa,\telipb,0,77,7,statham_1971_sm,2,20
Nb,Re,5.034013605442177,   18.81856540084388, 49.87341772151898,\telipa,\telipb,0,77,7, statham_1971_sm,2,20
Nb,Re,9.047619047619047,   18.81856540084388, 63.71308016877637,\telipa,\telipb,0,77,7, statham_1971_sm,2,20

% ----------------------------------------------------------------------------------------------------
%Nb,Zr,0,                  171, 171,\telipa,\telipb,0,77,4,stephens_1975_jlcm,2,20
Nb,Zr,4.3,                 171, 229,\telipa,\telipb,0,77,4,stephens_1975_jlcm,2,35
Nb,Zr,10.1,                171, 273,\telipa,\telipb,0,77,4,stephens_1975_jlcm,2,20
Nb,Zr,20.5,                171, 339,\telipa,\telipb,0,77,4,stephens_1975_jlcm,2,20
Nb,Zr,30.6,                171, 342,\telipa,\telipb,0,77,4,stephens_1975_jlcm,2,20
%Nb,Hf,0,                  171, 171,\telipa,\telipb,0,77,4,stephens_1975_jlcm,2,20
Nb,Hf,0.99,                171, 173,\telipa,\telipb,0,77,4,stephens_1975_jlcm,4,20
Nb,Hf,4.2,                 171, 181,\telipa,\telipb,0,77,4,stephens_1975_jlcm,2,30
Nb,Hf,11.3,                171, 248,\telipa,\telipb,0,77,4,stephens_1975_jlcm,2,20
Nb,Hf,17.9,                171, 248,\telipa,\telipb,0,77,4,stephens_1975_jlcm,2,20
Nb,Hf,25.8,                171, 303,\telipa,\telipb,0,77,4,stephens_1975_jlcm,2,20
Nb,Hf,33.1,                171, 303,\telipa,\telipb,0,77,4,stephens_1975_jlcm,2,20
%Nb,Mo,0.82,               171, 175,\telipa,\telipb,0,77,6,stephens_1975_jlcm,2,20
Nb,Mo,1.0,                 171, 185,\telipa,\telipb,0,77,6,stephens_1975_jlcm,3,20
Nb,Mo,4.2,                 171, 217,\telipa,\telipb,0,77,6,stephens_1975_jlcm,3,10
Nb,Mo,7.9,                 171, 254,\telipa,\telipb,0,77,6,stephens_1975_jlcm,2,20
Nb,Mo,9.2,                 171, 276,\telipa,\telipb,0,77,6,stephens_1975_jlcm,2,20
Nb,Mo,15.5,                171, 272,\telipa,\telipb,0,77,6,stephens_1975_jlcm,2,20
Nb,Mo,21.0,                171, 335,\telipa,\telipb,0,77,6,stephens_1975_jlcm,2,20
Nb,Mo,27.0,                171, 368,\telipa,\telipb,0,77,6,stephens_1975_jlcm,2,20
Nb,Mo,44.7,                171, 435,\telipa,\telipb,0,77,6,stephens_1975_jlcm,2,20
%Nb,W,0,                   171, 171,\telipa,\telipb,0,77,6,stephens_1975_jlcm,2,20
Nb,W,1.5,                  171, 216,\telipa,\telipb,0,77,6,stephens_1975_jlcm,2,20
Nb,W,9.6,                  171, 307,\telipa,\telipb,0,77,6,stephens_1975_jlcm,-2,20
Nb,W,21.0,                 171, 383,\telipa,\telipb,0,77,6,stephens_1975_jlcm,2,20
Nb,W,30.7,                 171, 448,\telipa,\telipb,0,77,6,stephens_1975_jlcm,2,20
%Nb,Re,0,                  171, 171,\telipa,\telipb,0,77,7,stephens_1975_jlcm,2,20
Nb,Re,1.0,                 171, 188,\telipa,\telipb,0,77,7,stephens_1975_jlcm,3,30
Nb,Re,2.2,                 171, 211,\telipa,\telipb,0,77,7,stephens_1975_jlcm,2,40
Nb,Re,4.0,                 171, 247,\telipa,\telipb,0,77,7,stephens_1975_jlcm,2,20
Nb,Re,10.4,                171, 337,\telipa,\telipb,0,77,7,stephens_1975_jlcm,2,20
Nb,Re,19.5,                171, 482,\telipa,\telipb,0,77,7,stephens_1975_jlcm,2,20
%Nb,Ru,0,                  171, 171,\telipa,\telipb,0,77,8,stephens_1975_jlcm,2,20
Nb,Ru,0.63,                171, 268,\telipa,\telipb,0,77,8,stephens_1975_jlcm,2,20
Nb,Ru,2.8,                 171, 324,\telipa,\telipb,0,77,8,stephens_1975_jlcm,2,20
Nb,Ru,7.5,                 171, 453,\telipa,\telipb,0,77,8,stephens_1975_jlcm,2,20
%Nb,Os,0,                  171, 171,\telipa,\telipb,0,77,8,stephens_1975_jlcm,2,20
Nb,Os,0.56,                171, 202,\telipa,\telipb,0,77,8,stephens_1975_jlcm,2,40
Nb,Os,2.3,                 171, 259,\telipa,\telipb,0,77,8,stephens_1975_jlcm,2,20
Nb,Os,6.5,                 171, 377,\telipa,\telipb,0,77,8,stephens_1975_jlcm,2,20
Nb,Os,9.9,                 171, 445,\telipa,\telipb,0,77,8,stephens_1975_jlcm,2,20
%Nb,Rh,0,                  171, 171,\telipa,\telipb,0,77,9,stephens_1975_jlcm,2,20
Nb,Rh,0.53,                171, 221,\telipa,\telipb,0,77,9,stephens_1975_jlcm,2,20
Nb,Rh,2.0,                 171, 297,\telipa,\telipb,0,77,9,stephens_1975_jlcm,2,20
Nb,Rh,5.0,                 171, 334,\telipa,\telipb,0,77,9,stephens_1975_jlcm,2,20
Nb,Rh,7.2,                 171, 395,\telipa,\telipb,0,77,9,stephens_1975_jlcm,2,20
%Nb,Ir,0,                  171, 171,\telipa,\telipb,0,77,9,stephens_1975_jlcm,2,20
Nb,Ir,0.49,                171, 254,\telipa,\telipb,0,77,9,stephens_1975_jlcm,2,20
Nb,Ir,2.0,                 171, 309,\telipa,\telipb,0,77,9,stephens_1975_jlcm,2,20
Nb,Ir,5.0,                 171, 325,\telipa,\telipb,0,77,9,stephens_1975_jlcm,2,15
Nb,Ir,7.6,                 171, 381,\telipa,\telipb,0,77,9,stephens_1975_jlcm,2,20

% ----------------------------------------------------------------------------------------------------
%Nb,Zr,0,                  185.32423208191125, 185.32423208191125,\telipa,\telipb,0,77,4,botta_1988_pma,2,20
Nb,Zr,0.51,                185.32423208191125, 164.07849829351534,\telipa,\telipb,0,77,4,botta_1988_pma,2,-20
Nb,Zr,1.03,                185.32423208191125, 193.00341296928323,\telipa,\telipb,0,77,4,botta_1988_pma,2,20
Nb,Zr,0.33,                185.32423208191125, 237.03071672354946,\telipa,\telipb,0,77,4,botta_1988_pma,2,40

% ----------------------------------------------------------------------------------------------------
%Ta,Nb,0,                   23.06903991370011, 22.36785329018339,\telipa,\telipb,0,77,5, arsenault_1966_am,2,-20
Ta,Nb,9,                   23.06903991370011, 22.36785329018339,\telipa,\telipb,0,77,5,  arsenault_1966_am,2,20
%Ta,W,0,                    23.791079812206576,23.791079812206576,\telipa,\telipb,0,77,6,arsenault_1966_am,2,20
Ta,W,9,                    23.791079812206576,32.30046948356808,\telipa,\telipb,0,77,6,  arsenault_1966_am,2,20
Ta,Re,11,                  23.791079812206576,55.305164319248824,\telipa,\telipb,0,77,7, arsenault_1966_am,2,20
% ----------------------------------------------------------------------------------------------------
%Ta,Nb,0,                   28.944146901300687, 28.944146901300687,\telipa,\telipb,0,77,5,mordike_1970_msj,2,20
Ta,Nb,4.0,                 28.944146901300687, 32.5248661055853,\telipa,\telipb,0,77,5,   mordike_1970_msj,3,20
Ta,Mo,3.0,                 28.944146901300687, 34.42234123947972,\telipa,\telipb,0,77,6,  mordike_1970_msj,2,20
Ta,W,3.75,                 28.944146901300687, 36.68706962509563,\telipa,\telipb,0,77,6,  mordike_1970_msj,2,20
Ta,Re,4.0,                 28.944146901300687, 46.72532517214996,\telipa,\telipb,0,77,7,  mordike_1970_msj,2,20
% ----------------------------------------------------------------------------------------------------
%Ta,Nb,0,                   642.8346456692911, 642.8346456692911,\telipa,\telipb,0,77,5,gypen_1982_jlcm,2,20
Ta,Nb,3.9,                 642.8346456692911, 666.1417322834644,\telipa,\telipb,0,77,5, gypen_1982_jlcm,-2,20
Ta,Nb,7.9,                 642.8346456692911, 668.031496062992,\telipa,\telipb,0,77,5,  gypen_1982_jlcm,2,20
% ----------------------------------------------------------------------------------------------------
%Ta,W,0,                   644.7224394057857, 644.7224394057857,\telipa,\telipb,0,77,6,gypen_1982_jlcm,2,20
Ta,W,1.6,                 644.7224394057857, 742.9241594996091,\telipa,\telipb,0,77,6, gypen_1982_jlcm,2,20
Ta,W,3.1,                 644.7224394057857, 726.6614542611416,\telipa,\telipb,0,77,6, gypen_1982_jlcm,2,20
Ta,W,6.3,                 644.7224394057857, 788.5848318999218,\telipa,\telipb,0,77,6, gypen_1982_jlcm,2,20
Ta,W,10.3,                644.7224394057857, 946.2079749804535,\telipa,\telipb,0,77,6, gypen_1982_jlcm,2,20
%Ta,Mo,0,                  646.1305007587254, 646.1305007587254,\telipa,\telipb,0,77,6,gypen_1982_jlcm,2,20
Ta,Mo,0.6,                646.1305007587254, 723.8239757207891,\telipa,\telipb,0,77,6, gypen_1982_jlcm,2,40
Ta,Mo,2.5,                646.1305007587254, 794.8406676783004,\telipa,\telipb,0,77,6, gypen_1982_jlcm,2,20
Ta,Mo,4.9,                646.1305007587254, 901.6691957511381,\telipa,\telipb,0,77,6, gypen_1982_jlcm,2,20
%Ta,Hf,0,                  645.3416149068321, 645.3416149068321,\telipa,\telipb,0,77,4,gypen_1982_jlcm,2,20
Ta,Hf,1.2,                645.3416149068321, 632.2981366459626,\telipa,\telipb,0,77,4, gypen_1982_jlcm,4,20
Ta,Hf,1.6,                645.3416149068321, 625.4658385093167,\telipa,\telipb,0,77,4, gypen_1982_jlcm,4,20
Ta,Hf,3.0,                645.3416149068321, 590.062111801242,\telipa,\telipb,0,77,4,  gypen_1982_jlcm,2,20

% ----------------------------------------------------------------------------------------------------
%Ta,Re,0,                  641.9117647058824,641.9117647058824,\telipa,\telipb,0,77,7,gypen_1982_jlcm,2,20
Ta,Re,1.0,                641.9117647058824,742.6470588235295,\telipa,\telipb,0,77,7, gypen_1982_jlcm,2,20
Ta,Re,2.1,                641.9117647058824,778.6764705882354,\telipa,\telipb,0,77,7, gypen_1982_jlcm,2,-20
Ta,Re,3.9,                641.9117647058824,1002.9411764705883,\telipa,\telipb,0,77,7,gypen_1982_jlcm,4,20
% ----------------------------------------------------------------------------------------------------
V,Ti,0.75,                24.128843338213763,24.128843338213763,\telipa,\telipb,0,77,4,pink_1972_msj,-2,-20
V,Ti,1.6,                 24.128843338213763,22.26207906295754,\telipa,\telipb,0,77,4, pink_1972_msj,3,20
V,Ti,4.0,                 24.128843338213763,25.117130307467058,\telipa,\telipb,0,77,4,pink_1972_msj,2,-20
V,Ti,4.0,                 24.128843338213763,27.386530014641288,\telipa,\telipb,0,77,4,pink_1972_msj,-2,20
V,Ti,20.0,                24.128843338213763,52.86237188872621,\telipa,\telipb,0,77,4, pink_1972_msj,2,20
% ----------------------------------------------------------------------------------------------------
%V,Cr,0,                   643.4042553191489,643.4042553191489,\telipa,\telipb,0,77,6,owen_1987_mta,2,20
V,Cr,2.5,                 643.4042553191489,617.0212765957447,\telipa,\telipb,0,77,6, owen_1987_mta,2,20
V,Cr,5,                   643.4042553191489,620.4255319148937,\telipa,\telipb,0,77,6, owen_1987_mta,3,20
V,Cr,10,                  643.4042553191489,744.6808510638298,\telipa,\telipb,0,77,6, owen_1987_mta,-2,20
V,Cr,20,                  643.4042553191489,925.1063829787234,\telipa,\telipb,0,77,6, owen_1987_mta,2,20
% ----------------------------------------------------------------------------------------------------
%V,Ti,0,                   494.5273631840795,494.5273631840795,\telipa,\telipb,0,77,4,owen_1985_mmta,2,20
V,Ti,1,                   494.5273631840795,494.5273631840795,\telipa,\telipb,0,77,4, owen_1985_mmta,2,-25
V,Ti,5,                   494.5273631840795,445.7711442786069,\telipa,\telipb,0,77,4, owen_1985_mmta,3,20
V,Ti,10,                  494.5273631840795,641.7910447761194,\telipa,\telipb,0,77,4, owen_1985_mmta,-2,20
V,Ti,20,                  494.5273631840795,771.1442786069651,\telipa,\telipb,0,77,4, owen_1985_mmta,2,20
V,Ti,30,                  494.5273631840795,914.4278606965174,\telipa,\telipb,0,77,4, owen_1985_mmta,2,20
% ----------------------------------------------------------------------------------------------------
%V,Ti,0,                   126.14321608040203,126.14321608040203,\telipa,\telipb,0,77,4,fraser_1962_cmq,2,20
V,Ti,0.1,                 126.14321608040203,146.0175879396985,\telipa,\telipb,0,77,4,  fraser_1962_cmq,2,35
V,Ti,0.3,                 126.14321608040203,127.55025125628143,\telipa,\telipb,0,77,4, fraser_1962_cmq,2,30
V,Ti,0.5,                 126.14321608040203,125.79145728643218,\telipa,\telipb,0,77,4, fraser_1962_cmq,2,20
V,Ti,0.7,                 126.14321608040203,113.30402010050253,\telipa,\telipb,0,77,4, fraser_1962_cmq,2,10
V,Ti,1.0,                 126.14321608040203,108.81909547738695,\telipa,\telipb,0,77,4, fraser_1962_cmq,2,-20
V,Ti,2.0,                 126.14321608040203,85.51507537688443,\telipa,\telipb,0,77,4,  fraser_1962_cmq,2,20
V,Ti,5.0,                 126.14321608040203,83.84422110552765,\telipa,\telipb,0,77,4,  fraser_1962_cmq,2,20

% ----------------------------------------------------------------------------------------------------
%Ta,Re,0,                  22.38050253988678,22.38050253988678,\telipa,\telipb,0,77,7,smialek_1970_sm,2,20
Ta,Re,1.179559878245802,  22.38050253988678,31.134213393087023,\telipa,\telipb,0,77,7,smialek_1970_sm,2,35
Ta,Re,2.0944490835547263, 22.38050253988678,34.81457792901347,\telipa,\telipb,0,77,7, smialek_1970_sm,2,40
Ta,Re,3.785560764027009,  22.38050253988678,44.538595065810284,\telipa,\telipb,0,77,7,smialek_1970_sm,2,35

\end{filecontents*}

\begin{filecontents*}{hs_gp_vi_data.csv}

solv,sol,conc,sorg,ssol,xer,yer,theta,temp,solgp,bibkey,bksx,bksy
%W,Re,0,                    640.5797101449276,  640.5797101449276,\telipa,\telipb,0,77,7,stephens_1971_jlcm,2,20
W,Re,0.8544303797468356,   640.5797101449276,  626.0869565217392,\telipa,\telipb,0,77,7,stephens_1971_jlcm,5,0,
W,Re,1.7658227848101262,   640.5797101449276,  627.536231884058, \telipa,\telipb,0,77,7,stephens_1971_jlcm,2,20
W,Re,2.791139240506329,    640.5797101449276,  605.7971014492754,\telipa,\telipb,0,77,7,stephens_1971_jlcm,2,20
W,Re,3.2468354430379733,   640.5797101449276,  589.8550724637681,\telipa,\telipb,0,77,7,stephens_1971_jlcm,2,-20
W,Re,5.012658227848101,    640.5797101449276,  571.0144927536232,\telipa,\telipb,0,77,7,stephens_1971_jlcm,-2,20
W,Re,7.632911392405061,    640.5797101449276,  514.4927536231885,\telipa,\telipb,0,77,7,stephens_1971_jlcm,2,-20
W,Re,8.943037974683543,    640.5797101449276,  523.1884057971015,\telipa,\telipb,0,77,7,stephens_1971_jlcm,-2,25
W,Re,11.563291139240505,   640.5797101449276,  530.4347826086956,\telipa,\telipb,0,77,7,stephens_1971_jlcm,2,20
W,Re,15.379746835443036,   640.5797101449276,  549.2753623188406,\telipa,\telipb,0,77,7,stephens_1971_jlcm,2,20
W,Re,20.22151898734177,    640.5797101449276,  562.3188405797102,\telipa,\telipb,0,77,7,stephens_1971_jlcm,2,20
W,Re,22.044303797468356,   640.5797101449276,  568.1159420289855,\telipa,\telipb,0,77,7,stephens_1971_jlcm,2,20
W,Re,23.183544303797465,   640.5797101449276,  579.7101449275362,\telipa,\telipb,0,77,7,stephens_1971_jlcm,2,20
W,Re,25.23417721518987,    640.5797101449276,  582.608695652174, \telipa,\telipb,0,77,7,stephens_1971_jlcm,2,20
W,Re,28.70886075949367,    640.5797101449276,  576.8115942028985,\telipa,\telipb,0,77,7,stephens_1971_jlcm,2,20
W,Re,30.36075949367088,    640.5797101449276,  588.4057971014493,\telipa,\telipb,0,77,7,stephens_1971_jlcm,2,20
W,Re,33.49367088607595,    640.5797101449276,  644.9275362318841,\telipa,\telipb,0,77,7,stephens_1971_jlcm,2,20

%-----------------------------------------------------------------------------------------------------------------------
%Cr,Re,0,                   450,450,\telipa,\telipb,0,77,7,stephens_1971_jlcm,2,20
Cr,Re,0.42168674698795217, 450,434.05063291139237,\telipa,\telipb,0,77,7,stephens_1971_jlcm,2,-20
Cr,Re,0.903614457831325,   450,423.4177215189873,\telipa,\telipb,0,77,7, stephens_1971_jlcm,3,-20
Cr,Re,1.927710843373493,   450,391.5189873417721,\telipa,\telipb,0,77,7, stephens_1971_jlcm,3,20
Cr,Re,2.831325301204819,   450,388.4810126582278,\telipa,\telipb,0,77,7, stephens_1971_jlcm,2,20
Cr,Re,4.638554216867468,   450,352.7848101265823,\telipa,\telipb,0,77,7, stephens_1971_jlcm,2,20
Cr,Re,6.807228915662649,   450,339.873417721519,\telipa,\telipb,0,77,7,  stephens_1971_jlcm,2,-25
Cr,Re,9.277108433734938,   450,357.34177215189874,\telipa,\telipb,0,77,7,stephens_1971_jlcm,2,-20
Cr,Re,12.590361445783131,  450,366.4556962025316,\telipa,\telipb,0,77,7, stephens_1971_jlcm,2,20
Cr,Re,15.722891566265059,  450,405.9493670886076,\telipa,\telipb,0,77,7, stephens_1971_jlcm,2,20
Cr,Re,21.927710843373493,  450,411.2658227848101,\telipa,\telipb,0,77,7, stephens_1971_jlcm,2,20
Cr,Re,29.096385542168672,  450,421.1392405063291,\telipa,\telipb,0,77,7, stephens_1971_jlcm,2,20
Cr,Re,33.97590361445783,   450,435.56962025316454,\telipa,\telipb,0,77,7,stephens_1971_jlcm,2,20
Cr,Re,37.5301204819277,    450,437.8481012658228,\telipa,\telipb,0,77,7, stephens_1971_jlcm,2,20

% -----------------------------------------------------------------------------------------------------------------------
Mo,Re,0.06791171477079816, 412.28501228501233, 412.28501228501233,\telipa,\telipb,0,77,7,stephens_1971_jlcm,2,25
Mo,Re,0.6112054329371817,  412.28501228501233, 406.26535626535633,\telipa,\telipb,0,77,7,stephens_1971_jlcm,2,25
Mo,Re,1.9694397283531413,  412.28501228501233, 395.08599508599514,\telipa,\telipb,0,77,7,stephens_1971_jlcm,2,20
Mo,Re,2.9881154499151106,  412.28501228501233, 379.60687960687966,\telipa,\telipb,0,77,7,stephens_1971_jlcm,2,20
Mo,Re,4.210526315789474,   412.28501228501233, 358.96805896805904,\telipa,\telipb,0,77,7,stephens_1971_jlcm,2,20
Mo,Re,5.161290322580646,   412.28501228501233, 364.9877149877151,\telipa,\telipb,0,77,7, stephens_1971_jlcm,2,20
Mo,Re,5.9083191850594226,  412.28501228501233, 347.78869778869785,\telipa,\telipb,0,77,7,stephens_1971_jlcm,2,20
Mo,Re,7.062818336162987,   412.28501228501233, 315.1105651105652,\telipa,\telipb,0,77,7, stephens_1971_jlcm,2,-20
Mo,Re,8.01358234295416,    412.28501228501233, 315.1105651105652,\telipa,\telipb,0,77,7, stephens_1971_jlcm,2,-20
Mo,Re,9.711375212224109,   412.28501228501233, 335.74938574938585,\telipa,\telipb,0,77,7,stephens_1971_jlcm,-2,20
Mo,Re,13.921901528013585,  412.28501228501233, 352.0884520884522,\telipa,\telipb,0,77,7, stephens_1971_jlcm,2,20
Mo,Re,19.558573853989817,  412.28501228501233, 395.08599508599514,\telipa,\telipb,0,77,7,stephens_1971_jlcm,2,20
Mo,Re,23.76910016977929,   412.28501228501233, 428.62407862407866,\telipa,\telipb,0,77,7,stephens_1971_jlcm,2,20
Mo,Re,32.190152801358245,  412.28501228501233, 448.4029484029485,\telipa,\telipb,0,77,7, stephens_1971_jlcm,2,20

% -----------------------------------------------------------------------------------------------------------------------
%W,Re,0.0,               195.35175879396982,195.35175879396982,\telipa,\telipb,0,77,7,raffo_1969_jlcm,2,20
W,Re,0.984251968503937, 195.35175879396982,188.19095477386932,\telipa,\telipb,0,77,7, raffo_1969_jlcm,4,-20
W,Re,1.8700787401574805,195.35175879396982,174.62311557788942,\telipa,\telipb,0,77,7, raffo_1969_jlcm,2,20
W,Re,6.840551181102363, 195.35175879396982,150.50251256281405,\telipa,\telipb,0,77,7, raffo_1969_jlcm,2,20
W,Re,24.753937007874015,195.35175879396982,142.2110552763819,\telipa,\telipb,0,77,7,  raffo_1969_jlcm,2,20

%-----------------------------------------------------------------------------------------------------------
%Mo,Re,0.0, 391,391, \telipa,\telipb,0,77,7,stephens_1972_jlcm,2,20
Mo,Re,2.0, 391,375, \telipa,\telipb,0,77,7,stephens_1972_jlcm,2,-20
Mo,Re,5.3, 391,332, \telipa,\telipb,0,77,7,stephens_1972_jlcm,2,20
Mo,Re,8.2, 391,315, \telipa,\telipb,0,77,7,stephens_1972_jlcm,2,20
Mo,Re,10.2,391,336, \telipa,\telipb,0,77,7,stephens_1972_jlcm,2,20
Mo,Re,20.3,391,383, \telipa,\telipb,0,77,7,stephens_1972_jlcm,2,20
Mo,Re,32.6,391,438, \telipa,\telipb,0,77,7,stephens_1972_jlcm,2,20

% -----------------------------------------------------------------------------------------------------------
%Mo,Pt,0.0, 391,391,  \telipa,\telipb,0,77,10,stephens_1972_jlcm,2,20
Mo,Pt,0.2, 391,368,  \telipa,\telipb,0,77,10,stephens_1972_jlcm,2,-35
Mo,Pt,0.71, 391,401, \telipa,\telipb,0,77,10,stephens_1972_jlcm,2,25
Mo,Pt,1.2, 391,475,  \telipa,\telipb,0,77,10,stephens_1972_jlcm,2,20
Mo,Pt,2.0, 391,513,  \telipa,\telipb,0,77,10,stephens_1972_jlcm,2,20
Mo,Pt,2.5, 391,592,  \telipa,\telipb,0,77,10,stephens_1972_jlcm,2,20
Mo,Pt,4.8, 391,830,  \telipa,\telipb,0,77,10,stephens_1972_jlcm,2,20
Mo,Pt,7.6, 391,976,  \telipa,\telipb,0,77,10,stephens_1972_jlcm,2,20

% -----------------------------------------------------------------------------------------------------------
%Mo,Ir,0.0,   391,391,  \telipa,\telipb,0,77,9,stephens_1972_jlcm,2,20
Mo,Ir,0.63,  391,350,  \telipa,\telipb,0,77,9,stephens_1972_jlcm,2,-20
Mo,Ir,1.12,  391,405,  \telipa,\telipb,0,77,9,stephens_1972_jlcm,2,30
Mo,Ir,1.71,  391,453,  \telipa,\telipb,0,77,9,stephens_1972_jlcm,2,20
Mo,Ir,2.49,  391,520,  \telipa,\telipb,0,77,9,stephens_1972_jlcm,2,20
Mo,Ir,3.31,  391,567,  \telipa,\telipb,0,77,9,stephens_1972_jlcm,2,20
Mo,Ir,6.67,  391,730,  \telipa,\telipb,0,77,9,stephens_1972_jlcm,2,20
Mo,Ir,10.46, 391,805,  \telipa,\telipb,0,77,9,stephens_1972_jlcm,2,20

% -----------------------------------------------------------------------------------------------------------
%Mo,Os,0.0,   391,391,  \telipa,\telipb,0,77,8,stephens_1972_jlcm,2,20
Mo,Os,1.11,  391,349,  \telipa,\telipb,0,77,8,stephens_1972_jlcm,2,-20
Mo,Os,1.72,  391,335,  \telipa,\telipb,0,77,8,stephens_1972_jlcm,2,-20
Mo,Os,2.61,  391,375,  \telipa,\telipb,0,77,8,stephens_1972_jlcm,2,20
Mo,Os,3.64,  391,414,  \telipa,\telipb,0,77,8,stephens_1972_jlcm,2,20
Mo,Os,5.23,  391,452,  \telipa,\telipb,0,77,8,stephens_1972_jlcm,2,20
%Mo,Os,9.85, 391,600,  \telipa,\telipb,0,77,8,stephens_1972_jlcm,2,20
Mo,Os,15.76, 391,716,  \telipa,\telipb,0,77,8,stephens_1972_jlcm,2,20

% -----------------------------------------------------------------------------------------------------------
%Mo,W,0.0,   391,391, \telipa,\telipb,0,77,6,stephens_1972_jlcm,2,20
Mo,W,1.1,   391,401,  \telipa,\telipb,0,77,6,stephens_1972_jlcm,3,20
Mo,W,3.4,   391,405,  \telipa,\telipb,0,77,6,stephens_1972_jlcm,2,-20
Mo,W,5.5,   391,423,  \telipa,\telipb,0,77,6,stephens_1972_jlcm,2,20
Mo,W,12.1,  391,429,  \telipa,\telipb,0,77,6,stephens_1972_jlcm,2,20
Mo,W,18.8,  391,462,  \telipa,\telipb,0,77,6,stephens_1972_jlcm,2,20
Mo,W,37.2,  391,508,  \telipa,\telipb,0,77,6,stephens_1972_jlcm,2,20
Mo,W,58.3,  391,550,  \telipa,\telipb,0,77,6,stephens_1972_jlcm,2,20
Mo,W,82.4,  391,627,  \telipa,\telipb,0,77,6,stephens_1972_jlcm,2,20

% -----------------------------------------------------------------------------------------------------------
%Mo,Ta,0.0,   391,391, \telipa,\telipb,0,77,5,stephens_1972_jlcm,2,20
Mo,Ta,0.23,  391,414,  \telipa,\telipb,0,77,5,stephens_1972_jlcm,2,20
Mo,Ta,2.2,   391,418,  \telipa,\telipb,0,77,5,stephens_1972_jlcm,2,20
Mo,Ta,4.9,   391,443,  \telipa,\telipb,0,77,5,stephens_1972_jlcm,2,-20
Mo,Ta,7.9,   391,472,  \telipa,\telipb,0,77,5,stephens_1972_jlcm,2,20
Mo,Ta,17.1,  391,524,  \telipa,\telipb,0,77,5,stephens_1972_jlcm,2,20
Mo,Ta,25.3,  391,557,  \telipa,\telipb,0,77,5,stephens_1972_jlcm,2,20
Mo,Ta,38.2,  391,531,  \telipa,\telipb,0,77,5,stephens_1972_jlcm,2,20
Mo,Ta,59.3,  391,513,  \telipa,\telipb,0,77,5,stephens_1972_jlcm,2,20
Mo,Ta,81.5,  391,436,  \telipa,\telipb,0,77,5,stephens_1972_jlcm,2,20

% -----------------------------------------------------------------------------------------------------------
%Mo,Hf,0.0,  391,391,  \telipa,\telipb,0,77,4,stephens_1972_jlcm,2,20
Mo,Hf,0.92,  391,421,  \telipa,\telipb,0,77,4,stephens_1972_jlcm,2,15
Mo,Hf,2.7,   391,437,  \telipa,\telipb,0,77,4,stephens_1972_jlcm,2,20
Mo,Hf,3.9,   391,455,  \telipa,\telipb,0,77,4,stephens_1972_jlcm,2,20
Mo,Hf,4.8,   391,490,  \telipa,\telipb,0,77,4,stephens_1972_jlcm,2,20
%Mo,Hf,9.9,  391,601,  \telipa,\telipb,0,77,4,stephens_1972_jlcm,2,20
Mo,Hf,15.9,  391,689,  \telipa,\telipb,0,77,4,stephens_1972_jlcm,2,20

% -----------------------------------------------------------------------------------------------------------
%Mo,W,0.0,               38.60092260636298,38.60092260636298, \telipa,\telipb,0,77,7, davidson_1970_am,2,20
Mo,W,7.89329856845589,   38.60092260636298,31.97432747511721, \telipa,\telipb,0,77,7, davidson_1970_am,2,20
Mo,W,16.620452779100958, 38.60092260636298,37.392132775089635, \telipa,\telipb,0,77,7,davidson_1970_am,2,20

% -----------------------------------------------------------------------------------------------------------
%Mo,Hf,0.0,               389.1472868217054, 389.1472868217054, \telipa,\telipb,0,77,4,pohl_2013_jac,2,20
Mo,Hf,0.15202702702702697,389.1472868217054, 366.8217054263566, \telipa,\telipb,0,77,4,pohl_2013_jac,1,-80
Mo,Hf,0.27871621621621623,389.1472868217054, 370.95607235142114,\telipa,\telipb,0,77,4,pohl_2013_jac,3,-29
Mo,Hf,0.6233108108108109, 389.1472868217054, 382.1188630490956, \telipa,\telipb,0,77,4,pohl_2013_jac,-3,5
Mo,Hf,0.947635135135135,  389.1472868217054, 387.906976744186,  \telipa,\telipb,0,77,4,pohl_2013_jac,4,15
Mo,Hf,1.9206081081081083, 389.1472868217054, 401.13695090439273,\telipa,\telipb,0,77,4,pohl_2013_jac,2,20
Mo,Hf,2.8783783783783785, 389.1472868217054, 420.9819121447028, \telipa,\telipb,0,77,4,pohl_2013_jac,2,20

% -----------------------------------------------------------------------------------------------------------------------
%Cr,Re,0,                  526.7376125947395,526.7376125947395,\telipa,\telipb,0,77,7, klopp_1968_nasa,2,20
Cr,Re,5.0523772776855775,  526.7376125947395,501.8328293853235,\telipa,\telipb,0,77,7, klopp_1968_nasa,2,20
Cr,Re,7.650737219674003,   526.7376125947395,514.9604998843972,\telipa,\telipb,0,77,7, klopp_1968_nasa,2,20
%Cr,Re,10.044250767503735, 526.7376125947395,481.40695649861294,\telipa,\telipb,0,77,7,klopp_1968_nasa,2,20
Cr,Re,12.370054120970684,  526.7376125947395,431.7951373056361,\telipa,\telipb,0,77,7, klopp_1968_nasa,2,20
Cr,Re,15.074708558765883,  526.7376125947395,452.3934492049146,\telipa,\telipb,0,77,7, klopp_1968_nasa,2,20
Cr,Re,19.986024718832184,  526.7376125947395,402.0927318672014,\telipa,\telipb,0,77,7, klopp_1968_nasa,2,20
Cr,Re,24.983903547396487,  526.7376125947395,376.4392116019713,\telipa,\telipb,0,77,7, klopp_1968_nasa,2,20
Cr,Re,29.92301584680208,   526.7376125947395,401.94195496939636,\telipa,\telipb,0,77,7,klopp_1968_nasa,2,20
Cr,Re,34.94229941846872,   526.7376125947395,407.65560583358604,\telipa,\telipb,0,77,7,klopp_1968_nasa,2,20
Cr,Re,39.94830690197871,   526.7376125947395,399.92609143814565,\telipa,\telipb,0,77,7,klopp_1968_nasa,2,20

\end{filecontents*}

\setlength{\abovedisplayskip}{2pt}
\setlength{\belowdisplayskip}{2pt}

\setlength{\tabcolsep}{0.5pt}
\definecolor{contiYellow}{RGB}{255,165,0}
\renewcommand{\arraystretch}{1.2}
\newcolumntype{x}[1]{>{\centering\arraybackslash\hspace{0pt}}p{#1}}

\captionsetup[figure]{justification=raggedright,labelfont={bf},labelformat={default},labelsep=space,name={Fig.}}

\pagenumbering{gobble}

\title{\Large The Taming of the Screw: Dislocation Cores in BCC Metals and Alloys}
\author{Rui Wang}
\author{Lingyu Zhu}
\affiliation{Department of Materials Science and Engineering, City University of Hong Kong, Hong Kong, China}
\author{Subrahmanyam Pattamatta}
\author{David J. Srolovitz}
\affiliation{Department of Mechanical Engineering, The University of Hong Kong, Hong Kong, China}
\author{Zhaoxuan Wu}
\email[Corresponding author: ]{zhaoxuwu@cityu.edu.hk}
\affiliation{Department of Materials Science and Engineering, City University of Hong Kong, Hong Kong, China}
\affiliation{Hong Kong Institute for Advanced Study, City University of Hong Kong, Hong Kong, China}

\begin{abstract}
Body-centred cubic (BCC) transition metals tend to be brittle at low temperatures, posing significant challenges in processing and major concerns for damage tolerance.  The brittleness is largely dictated by the screw dislocation core; the nature and control of which remain a puzzle after nearly a century.  Here, we introduce a physics-based material index \(\chi\), the energy difference between BCC and face-centred-cubic structures, that guides engineering of core properties.  The lattice friction and nucleation barrier have near-linear scaling with \(\chi\) and the core transforms from non-degenerate to degenerate when \(\chi\) drops below a threshold in BCC alloys. \(\chi\) is related to solute valence electron concentrations and can be quantitatively predicted by first-principles calculations for any alloy composition, providing a robust path for screening and design of ductile and tough BCC alloys.
\end{abstract}

\maketitle
\clearpage
\pagenumbering{roman}

\clearpage
\pagenumbering{arabic}

\vspace{-0.5cm}

\begin{figure*}[!htbp]
  \centering
  \includegraphics[width=0.91\textwidth]{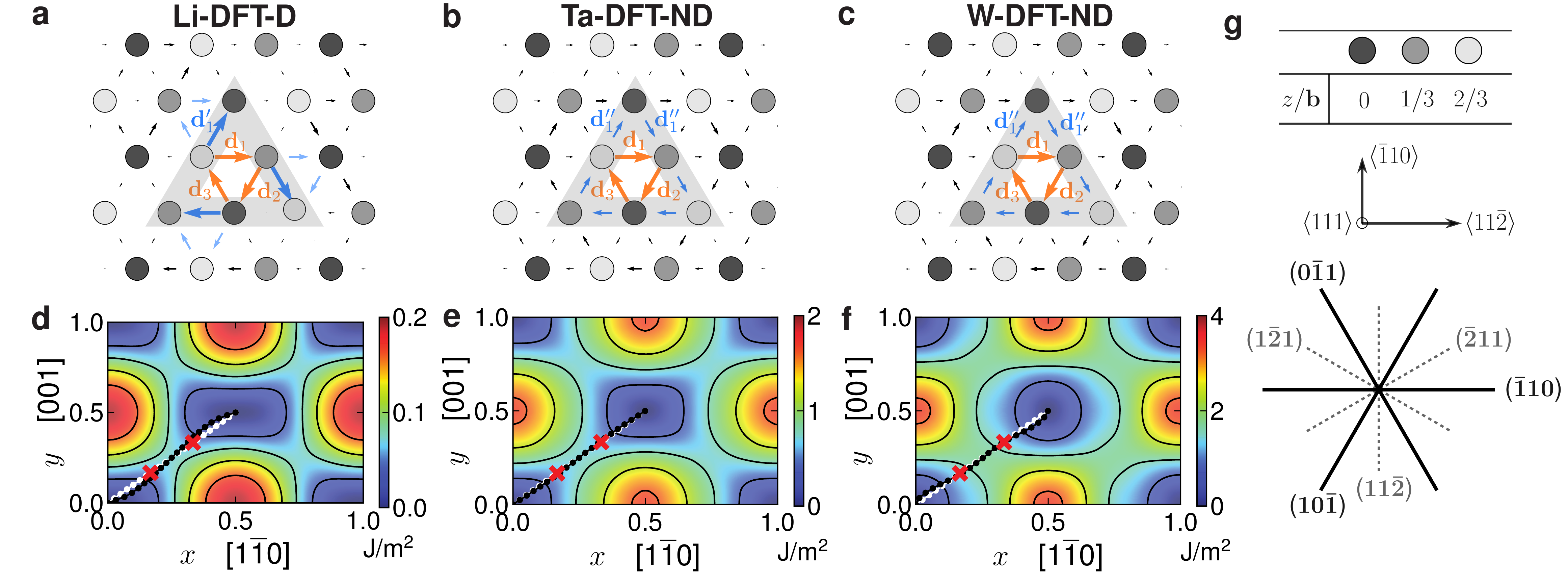}
  \caption{\label{fig:gamma_disl_Li_Ta_W} \textbf{Dislocation core structures and \(\gamma\)-surfaces of BCC Li, Ta and W}.
 \textbf{a} The D-core of Li. \textbf{b-c} The ND-cores of Ta and W. \textbf{d-f} \(\gamma\)-surfaces on \(\{110\}\) planes of Li, Ta and W.
 \textbf{g} Atom colour scheme, crystallographic orientations and traces of atomic planes in the BCC \(\langle 111 \rangle\)-zone. In (\textbf{a}-\textbf{c}), atoms are shaded based on their positions  in the \(\langle 111 \rangle\) (viewing) direction (see \textbf{g}).
 The core is enclosed by the grey triangle.  Arrows are  DDs in the Burgers vector direction.
The D-core has three DD vectors \(\mathbf{d}_1^\prime \approx \mathbf{b}/3\) (blue arrows) with a \(\langle 111 \rangle\) threefold screw axis.
 The ND-core has six DD vectors \(\mathbf{d}_1^{\prime\prime} \approx \mathbf{b}/6\) (blue arrows) and possesses both a \(\langle 111 \rangle\) threefold screw axis  and a \(\langle 110 \rangle\) diad axis.
In (\textbf{d}-\textbf{f}), the black lines are MEPs between two perfect lattice positions along the Burgers vector (white lines).
The MEP deviates slightly from the Burgers vector direction.
The red $\times$s are at \(\mathbf{b}/3\) and \(2\mathbf{b}/3\).}
\end{figure*}

BCC transition metals (TMs) play pivotal roles in a wide-range of structural and functional applications, including in current and future transportation, electronic, chemical and energy generation technologies.
They are often employed to withstand extreme mechanical, thermal, chemical and radiative loadings.
Their mechanical behaviour (e.g., yielding, hardening and fracture) is application-critical and, hence, has been under intensive research for a century~\cite{vitek_2008_dis,schade_2010_ijrmhm}.
Extensive experimental investigations show that BCC TMs exhibit a drastic decrease in ductility/toughness below some critical temperatures (i.e., ductile-to-brittle transition, DBT~\cite{gumbsch_2003_jnm,joseph_2007_jnm}).
This temperature-dependent behaviour is governed by the core of the \(1/2 \langle 111 \rangle\) screw dislocation~\cite{gumbsch_1998_science,butler_2018_ijrmhm} which carry most of the plastic deformation.
For example, screw dislocations in pure W adopt a non-degenerate (ND) core structure and preferentially glide on \(\{110\}\) planes (the core structure nomenclature~\cite{duesbery_1998_am} is described in Fig.~\ref{fig:gamma_disl_Li_Ta_W}).
This core exhibits gigapascal-scale lattice friction~\cite{brunner_2000_ml,cereceda_2013_jpcm}, making stress-relieving dislocation plasticity difficult. Below the DBT temperature, toughness is also limited by dislocation nucleation at crack-tips~\cite{gumbsch_1998_science}.
The ductility of W can be dramatically increased by alloying with Re, which reduces the lattice friction~\cite{trinkle_2005_science,caillard_2020_am,zhao_2018_msmse} and the barrier for dislocation nucleation, changes the ND core to a degenerate (D) core~\cite{romaner_2010_prl} at high Re-concentrations and promotes slips on alternating \(\{110\}\) planes~\cite{stephens_1970_mmtb,li_2012_am}.
Controlling the screw core structures and associated behaviour is thus central to enhancing the mechanical response of BCC TMs. However, the physical origin governing the screw dislocation core structure and property remains a mystery, despite decades of intense research~\cite{ren_2018_ijrmhm,vitek_2008_dis,clouet_2009_prl,weinberger_2013_imr,dezerald_2014_prb}.

In this work, we provide a universal approach to predicting and controlling all three critical properties of  screw dislocations in all BCC metals and alloys, i.e., (i) core structure, (ii) lattice friction, and (iii) dislocation nucleation barrier.
These properties are dictated by atomic slip processes on BCC \(\{110\}\) planes in which the moving core transforms through a quasi-close-packed face-centred cubic (FCC-like) structure.
This insight leads to a new material index \(\chi\) that captures the energetics of this core transformation through  crystal geometry and atomic bonding~\cite{kroupa_1964_cjpb}; i.e., the difference in energy between FCC and BCC structures \(\Delta E=E_\text{FCC}-E_\text{BCC}\).
For BCC TM alloys, \(\chi\) can be normalized as \(\chi = \Delta E^\text{A} / \Delta E^\text{P} \), where superscripts A/P represent alloyed/pure (or ideal) BCC TMs.
We show that all three properties are governed by \(\chi\), which in turn can be manipulated via alloying and predicted \textit{a priori} by rapid density-functional theory (DFT) calculations.
The parameter \(\chi\) is physics-based and universally applicable to \textit{all} BCC metals and alloys, as we demonstrate through  extensive DFT calculations and comparisons with extant experiments.
It provides a fresh engineering paradigm for computational composition design of tough, ductile  BCC TM alloys.

\section*{Results}
\subsection*{Degenerate vs Non-Degenerate cores}
A dislocation is characterised by its Burgers vector \(\mathbf{b}\) and line direction \(\boldsymbol{\xi}\)~\cite{anderson_2017_disl}.
On a continuum level, \(\mathbf{b}\) is defined as \(\oint_C \left(\partial \mathbf{u}/\partial l \right) \text{d} l\), where \(\mathbf{u}\) is the displacement vector and the integral \(\text{d}l\) is along an arbitrary closed path \(C\) enclosing the dislocation core.
On a discrete level, the contour integral is replaced by a summation of discrete segments \(k\) along a closed path,
\begin{equation}
\label{eqn:dd_dfn}
\mathbf{b} = \sum_{k=1}^N  \left( \mathbf{u}^{(k)}_{ij} - \mathbf{U}^{(k)}_{ij} \right)=\sum_{k=1}^N \mathbf{d}_k,
\end{equation}
where \(\mathbf{u}_{ij}\) and \(\mathbf{U}_{ij}\) are  displacement vectors between atom \(i\) and \(j\) before and after the introduction of the dislocation and \(\mathbf{d}_k\) is the differential displacement (DD) between atoms \(i\) and \(j\).
Each \(\mathbf{d}_k\) contributes a fraction of the total Burgers vector \(\mathbf{b}\)~\cite{anderson_2017_disl}.

In the BCC structure, the  screw dislocation Burgers vector \(\mathbf{b}\) can split into three \(\mathbf{d}_k\) enclosing the core.
Figure~\ref{fig:gamma_disl_Li_Ta_W} shows the core structures and \(\gamma\)-surfaces (energy  associated with translating one part of a crystal with respect to another along a crystal plane) of BCC Li, Ta and W calculated using DFT~\cite{supp_mat}.
Lithium exhibits a D-core while TMs Ta and W show ND-cores.
The core structures are dictated by a core energy \(E_\text{c}\), which is primarily associated with the shear displacements \(\mathbf{d}_1\),  \(\mathbf{d}_2\), \(\mathbf{d}_3\) in the \(\langle 111\rangle\) direction between neighbouring atoms enclosing the core centre (orange arrows in Fig.~\ref{fig:gamma_disl_Li_Ta_W}a-c).
For Li, Ta and W, the current DFT calculations show equal splitting of \(\mathbf{d}_1 \cong \mathbf{d}_2 \cong \mathbf{d}_3 \cong \mathbf{b}/3\) on three equivalent \(\{110\}\) planes at the core centre, i.e., all core centres are identical (Fig.~\ref{fig:gamma_disl_Li_Ta_W}).
This is consistent with nearly all previous BCC TMs core structure simulation results,  irrespective of the description of atomic interactions~\cite{rodney_2017_am} (DFT or interatomic potentials).
This core-splitting implies that the core centre (screw components) fulfils the \(\langle 110 \rangle\)-dyad symmetry of the BCC lattice and is thus ND.
Nevertheless, examination of the \(\gamma\)-surface shows that between the two ground states, the minimum energy path (MEP, black line in Fig.~\ref{fig:gamma_disl_Li_Ta_W}d-f) deviates from \(\langle 111 \rangle\) (white line).
The core energy can thus be further reduced by core splitting along the MEP with some edge components in \(\mathbf{d}_k\).

The principal differences between the D-/ND-cores lie in the details of the \(\mathbf{d}_k\) outside the core centre.
Applying the topological requirement of Eq.~\eqref{eqn:dd_dfn}, each \(\mathbf{d}_k\) must be compensated by two additional DD vectors in the triangular loop formed by one orange and two blue arrows (i.e., \(\sum \mathbf{d}_k=0\)).
The D-core (Fig.~\ref{fig:gamma_disl_Li_Ta_W}a) resolves each \(\mathbf{d}_k\) by a DD vector \(\mathbf{d}^\prime_k\) (blue arrow) of \textasciitilde{}\(\mathbf{b}/3\) along a \(\{110\}\) plane and a small DD (not visible in Fig.~\ref{fig:gamma_disl_Li_Ta_W}a).
For the ND-cores (Fig.~\ref{fig:gamma_disl_Li_Ta_W}b,c), each \(\mathbf{d}_k\) is compensated by two identical \(\mathbf{d}^{\prime\prime}_k\) (blue arrows) of \textasciitilde{}\(\mathbf{b}/6\) on two equivalent \(\{110\}\) planes. The D-/ND-core preference thus appears to be determined by the relative energy costs of the structures outside the core centre: one with a single \(\mathbf{d}^\prime_k\approx\mathbf{b}/3\) and the other with a pair of \(\mathbf{d}^{\prime\prime}_k\approx\mathbf{b}/6\).  These energy costs can be approximated by the stacking fault energy \(\gamma\) at \(\mathbf{b}/3\) and \(\mathbf{b}/6\); the ND core structure is preferred when \(2\gamma(\mathbf{b}/6) <\gamma(\mathbf{b}/3)\), as a well-known \(\gamma\)-line criterion~\cite{duesbery_1998_am}.

\begin{figure*}[!htbp]
  \centering
  \includegraphics[width=0.7\textwidth]{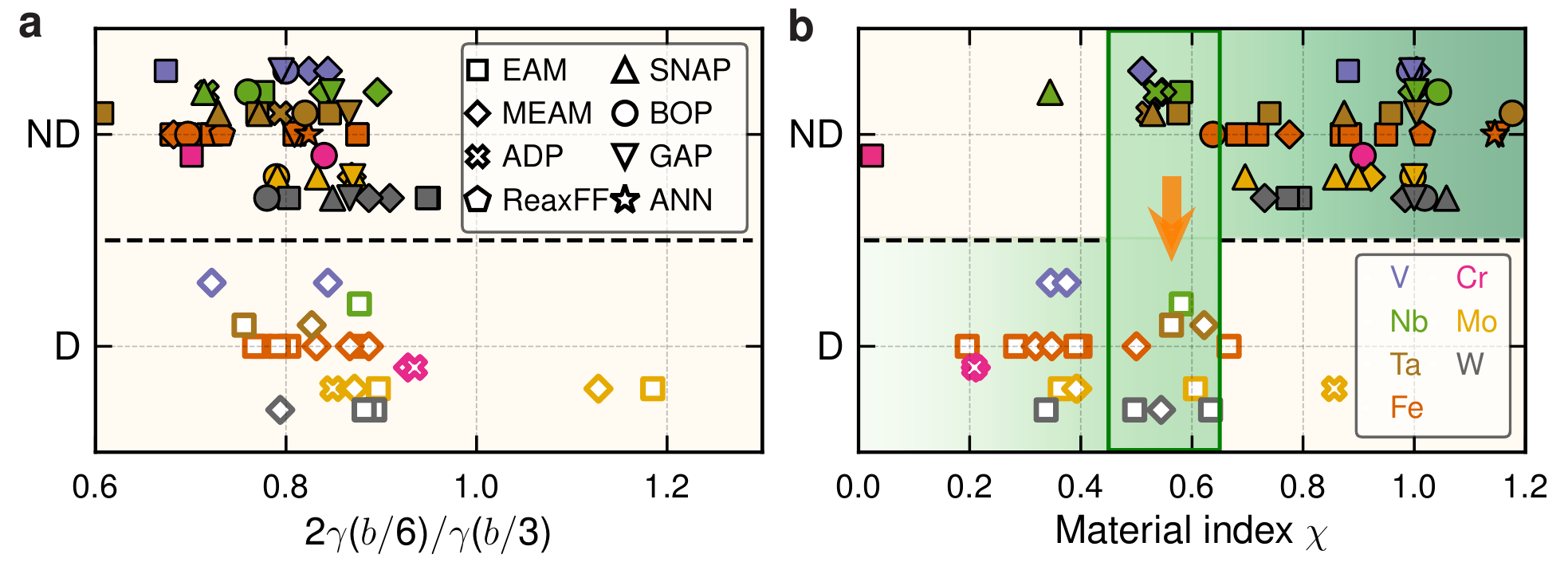}
  \caption{\label{fig:prop_core_type} \textbf{Dislocation core structure versus proposed governing criteria}. \textbf{a} \(\gamma\)-line criterion. \textbf{b} Material index \(\chi\)-criterion.
 Dislocation cores transform from non-degenerate (filled symbols) to degenerate (empty symbols) as \(\chi\) drops below \textasciitilde{}0.65. }
\end{figure*}

The above \(\gamma\)-line based criterion provides some rationalisation in pure BCC TMs~\cite{duesbery_1998_am,rodney_2017_am}.  However, the DFT-based result of Li violates this criterion (Supplementary Fig.~\ref{fig:gamma_line_li_ta_w}), so do other cases calculated by interatomic potentials~\cite{fellinger_2010_prb} or DFT~\cite{romaner_2010_prl}.  To determine the true mechanism governing the cores, we calculate the core structures of 7 BCC TMs (V, Nb, Ta, Cr, Mo, W and Fe) based upon 72 different interatomic potentials (Supplementary Table~\ref{tab:potential}).  While the accuracies of the potentials vary, we may view them as 72 different types of atoms/average alloys with properties perturbed from respective pure BCC TM elements.  These potentials yield both D- and ND-cores, thus providing a broad examination over the proposed \(\gamma\)-line and \(\Delta E\) criteria.  To consolidate data across all elements, we use the material index \(\chi\) by normalising \(\Delta E\) of interatomic potentials with the intrinsic \(\Delta E^\text{P}\) of the pure element calculated by DFT.

Figure~\ref{fig:prop_core_type}a shows that the $\gamma$-line criterion does not distinguish different core structures.  No correlation is seen between the core structure and the ratio \(2\gamma(\mathbf{b}/6)/\gamma(\mathbf{b}/3)\), or element type, or interatomic potential formalism.  Rather, Fig.~\ref{fig:prop_core_type}b shows that the material index \(\chi\) provides a prediction of core structure.
More specifically, D-cores are seen when \(\chi < 0.45\) and ND-cores are observed for \(\chi>0.65\). When \(\chi\) approaches 1, the intrinsic value of pure TMs, all potentials yield the ND core in agreement  with DFT calculations~\cite{rodney_2017_am}.
This simple \(\chi\)-criterion is highly predictive, failing in only 3 of 72 cases examined~\cite{supp_mat}.
Near the threshold \( 0.45 < \chi < 0.65\), both D- and ND-cores are seen, indicating that other subtle factors (e.g., elastic constants or the shapes of the \(\gamma\)-surfaces) may play a role.
The \(\chi\)-criterion thus appears to be broadly applicable (DFT validations are provided below).
While empirical, it is rooted in core structure geometry~\cite{kroupa_1964_cjpb}, as illustrated below.

\begin{figure*}[!htbp]
  \centering
  \includegraphics[width=0.8\textwidth]{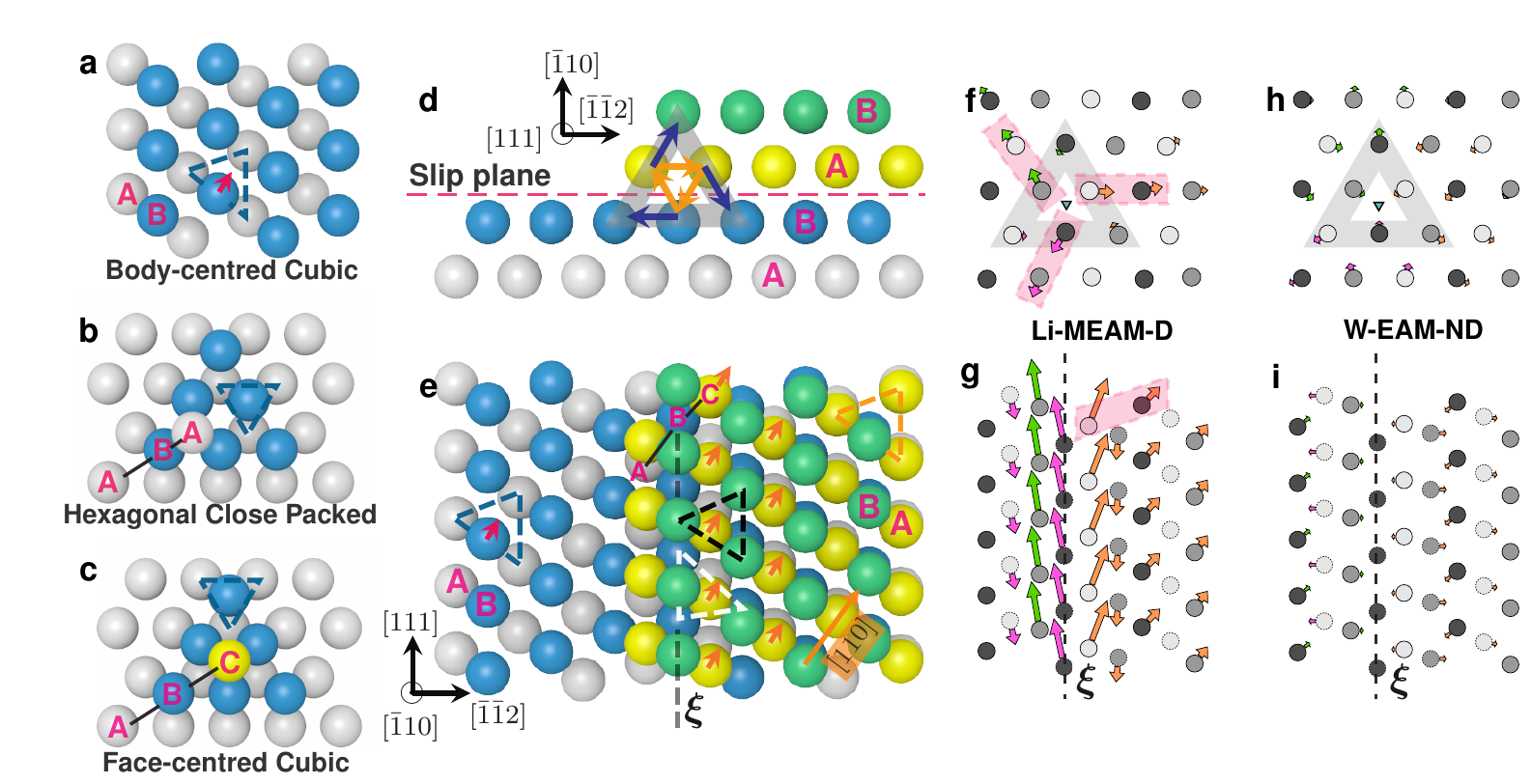}
  \caption{\label{fig:bcc_fcc} \textbf{Atomic plane stacking sequences in BCC, HCP and FCC structures and EADs of screw dislocations.}
\textbf{a-c} Atomic stacking of BCC \(\{110\}\), HCP \(\{0001\}\) and FCC \(\{111\}\) planes.
\textbf{d-e} Atomic structure of the screw dislocation core viewed from two directions.
Displacements of atoms on the second A-layer in the \([110]\) direction on the \((1\bar{1}0)\) plane (orange arrows) creates a quasi-ABC stacking, similar to that for  \(\{111\}\) planes in  FCC.
\textbf{f-i} EADs (arrows) in cores of Li-MEAM and W-EAM.  Atoms are coloured based on their positions in the \(\langle  111 \rangle\) direction.}
\end{figure*}

\subsection*{Screw Dislocation Core Geometry}
On BCC \(\{110\}\) planes,  the atomic stacking follows an ABAB stacking sequence, similar to the ABAB close-packing of  \(\{0001\}\) planes in hexagonal close-packed (HCP) structures (Fig.~\ref{fig:bcc_fcc}a-c).
This stacking can be converted to quasi close-packing if atoms in the B-layer are displaced to reside above the centroids of the triangles formed by the A-layer atom triplets (blue dashed triangle in Fig.~\ref{fig:bcc_fcc}a).
The required displacements are along the \(\langle 110 \rangle\)-direction on \(\{110\}\) planes (orange arrows in Fig.~\ref{fig:bcc_fcc}e) and can be provided by non-screw components of the dislocation displacement field.
Each fractional Burgers vector \(\mathbf{d}_k\) has an excess displacement relative to the elastic field~\cite{supp_mat}; the details vary between the D- and ND-cores.
We illustrate these excessive atomic displacements (EADs) (Fig.~\ref{fig:bcc_fcc}, f-i) in the D-core of Li computed by an MEAM potential~\cite{supp_mat} and in the ND-core of W described by an EAM potential~\cite{setyawan_2018_jap}.
In the D-core of Li (Fig.~\ref{fig:bcc_fcc}f), the EADs mainly shift 6 atoms (3 at the core centre and 3 neighbouring them) in \(\langle 110 \rangle\) directions on three \(\{110\}\) planes.
Taking any pair of atoms in the pink shaded box as an example (Fig.~\ref{fig:bcc_fcc}f, others have similar EADs by symmetry), both atoms have EADs along \(\langle 110 \rangle\). These displacements  (arrows in Fig.~\ref{fig:bcc_fcc}f,g) move the two atoms on the A-layer (yellow atoms in Fig.~\ref{fig:bcc_fcc}e) towards the centroids of triangles formed by the B-layer atom triplets above them (black triangle), thus forming a local close-packed structure similar to that in FCC/HCP (\emph{cf.} Fig.~\ref{fig:bcc_fcc}b,c,e).
The displacements also put these two atoms in the C-layer position with respect to the A and B layers beneath them, forming a local quasi-ABC stacking sequence.

Kinematically, the EADs lead to similar differential displacements (\(\mathbf{d}_k \approx \mathbf{d}^{\prime}_k \approx \mathbf{b}/3\)) for the two atoms on the same \(\{110\}\) plane and hence the D-core structure.
Energetically, the resulting close packed, quasi-ABC stacking is favourable at the core of Li since \(\Delta E = - 0.8\) meV/atom.
In contrast, the EADs of W-EAM-ND (Fig.~\ref{fig:bcc_fcc}h,i) are relatively small and not aligned on \(\{110\}\) planes.
In this case, displacements in the \(\langle 110 \rangle \) directions on \(\{110\}\) planes incur a large energy penalty since the ABC close-packed structure is highly unfavourable (\(\Delta E = 384\) meV/atom).
Therefore, the D-/ND-core competition is determined by the energy cost associated with displacements along \(\langle  110 \rangle\) on \(\{110\}\) planes that lead to locally close packed structures.
Li and W represent extreme scenarios of the D- and ND-cores; similar examination of the Fe-GAP-ND~\cite{dragoni_2018_prm} core (\(\Delta E= 159\) meV/atom) and Fe-MEAM-D~\cite{asadi_2015_prb} core (\(\Delta E= 44\) meV/atom) shows that their EADs fall between these two cases (Supplementary Fig.~\ref{fig:fe_core_eecd}).

The structure formed at the D-core is not precisely FCC, but only similar in close-packing sequence; the geometric description here should be viewed as suggestive.
Nevertheless, the model explains the  trend in Fig.~\ref{fig:prop_core_type}, where the transition from ND- to D-core starts as \(\chi\) drops below a critical value \textasciitilde{}0.65 (towards smaller \(\Delta E\)).
Smaller \(\Delta E\) implies that FCC structure becomes less unfavourable (relative to BCC) and forming the quasi-ABC structure is increasingly favourable.
Our extensive study of core structures using different interatomic potentials indicates that relating the energy of the quasi-ABC structure to that of the perfect FCC structure is of sufficient accuracy to predict D- and ND-cores.
The current model, based on \(\Delta E\), is also consistent with DFT calculations for all BCC TMs performed to date~\cite{rodney_2017_am}.
For all BCC TMs,  \(\Delta E\) is large (from 138 to 483 meV/atom, Supplementary Fig.~\ref{fig:deltaE}) and  FCC structures are highly unfavourable, suggesting limited EADs in \(\langle 110 \rangle\) directions on \(\{110\}\) planes.
Hence all screw dislocations prefer the ND-core structure in pure BCC TMs (Fig.~\ref{fig:gamma_disl_Li_Ta_W}b,c and other DFT calculations~\cite{dezerald_2014_prb}).
On the other hand, alkaline metals have nearly zero \(\Delta E\) and show D-cores (Fig.~\ref{fig:gamma_disl_Li_Ta_W}a for Li-DFT and Supplementary Fig.~\ref{fig:alkali_core} for Li, Na and K-interatomic potentials).

\begin{figure*}[!htbp]
  \centering
  \includegraphics[width=0.75\textwidth]{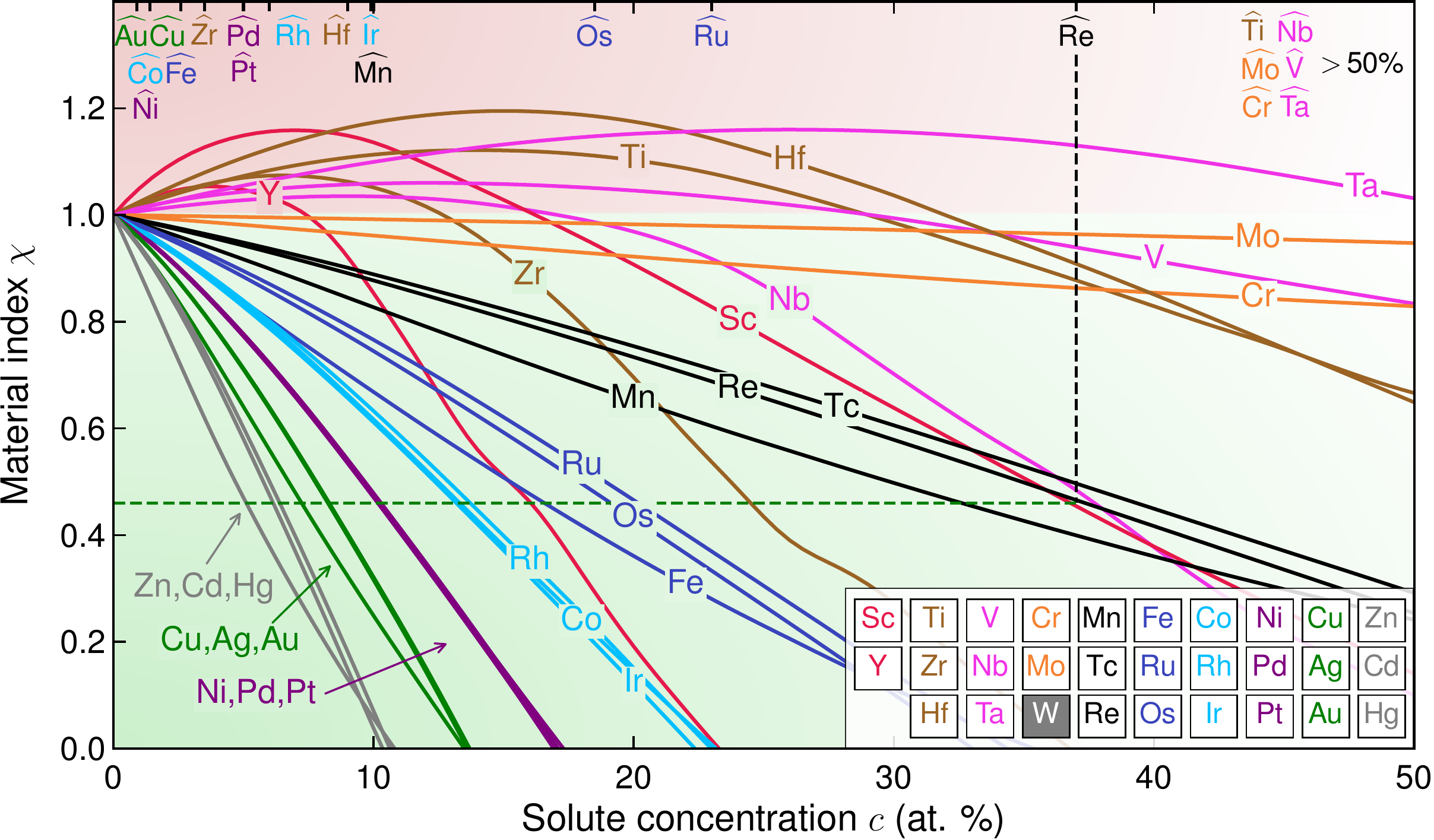}
  \caption{\label{fig:vca} \textbf{Prediction of \(\chi\) as a function of solute concentrations for binary W-TM alloys from VCA DFT calculations.}
  The solubility of each element (El) in W is denoted by \(\widehat{\text{El}}\) at the top. The elements are colour-coded by their group numbers in the periodic table.
  }
\end{figure*}

\begin{figure*}[!htbp]
  \centering
  \includegraphics[width=0.90\textwidth]{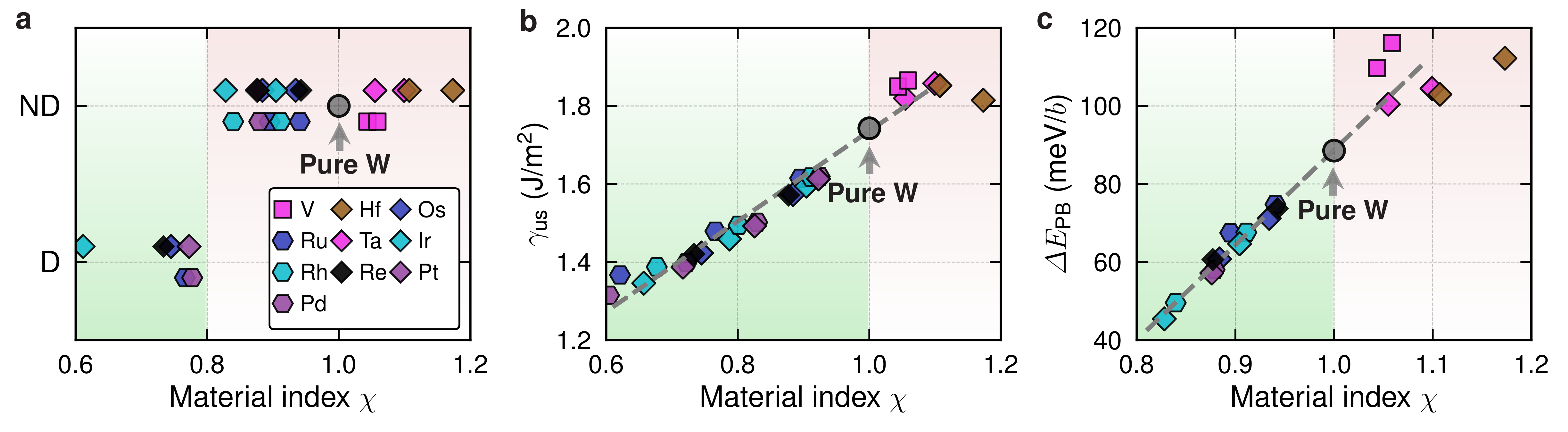}
  \caption{\label{fig:chi_prop} \textbf{Prediction of key material properties as a function of \(\chi\) for binary W-TM alloys. } \textbf{a} ND and D cores. \textbf{b} Unstable stacking fault energy \(\gamma_{\text{us}}\). \textbf{c} Peierls barrier \(\Delta E_{\text{PB}}\) of the screw dislocation. All properties are calculated by VCA DFT. Lower \(\chi\) favours the degenerate core, reduces \(\gamma_{\text{us}}\) and \(\Delta E_{\text{PB}}\), all independent of solute types.
  }
\end{figure*}

\subsection*{Validation and Prediction in Transition Metal Alloys}
The above analysis provides a geometric link between the material index \(\chi\) and the screw dislocation core structure;  \(\chi\) can be thought of as a surrogate for local bonding in the core.
It serves as an indicator of core structure and provides a physical/thermodynamic basis for understanding core structure in all BCC metals.  Since slip occurs along the Burgers vector direction on \(\{110\}\) planes, the slip process can be related to a transformation between the BCC structure and the quasi-FCC structure (Fig.~\ref{fig:bcc_fcc}).  Therefore, the energy barriers associated with dislocation nucleation \(\gamma_\text{us}\) and glide (Peierls barrier) \(\Delta E_\text{PB}\) should also be related to the transformation energy barrier or \(\Delta E\)/\(\chi\). Solid solution alloying can effectively modulate \(\Delta E\) and thus be used as a means of controlling the screw dislocation core structure and slip behaviour.  The material index \(\chi\) thus has practical, engineering application for BCC TM alloy design. Since \(\chi\) depends on the energy difference of two elementary structures, DFT calculations can be used to predict \(\chi\) as a function of solute concentration \(c\) and identify favourable solute types/concentrations. We demonstrate this strategy in binary BCC TM alloys, and extensively in W-TMs, a well-studied system in experiments~\cite{stephens_1970_mmtb,li_2012_am} and simulations~\cite{romaner_2010_prl}.

We employ the DFT-based virtual crystal approximation (VCA~\cite{nordheim_1931_ap}) to predict \(\chi\) and core properties for alloys.
VCA replaces all atoms in an alloy by virtual atoms with a composition-weighted pseudopotential and valence electron number based upon the alloy constituent elements~\cite{bellaiche_2000_prb}.
While VCA accuracy is limited to average, homogeneous alloy properties, it retains a general first-principles DFT framework~\cite{supp_mat}.
Figure~\ref{fig:vca} shows the predicted \(\chi\) calculated for 28 binary W-TM alloys (Supplementary Fig.~\ref{fig:vca_mo_vasp} for Mo-TM alloys).
At low solute concentrations ($<$10 \%),  \(\chi\) increases for elements to the left of the solvent TM  in the periodic table and decreases sharply (favourably) for elements to its right.
For W/Mo, increasing solute valence electron concentration (VEC~\cite{stephens_1972_jlcm,klopp_1975_jlcm,guo_2011_jap}) decreases \(\chi\) and its slope.

Figure~\ref{fig:chi_prop} shows the core structure, \(\gamma_\text{us}\) and \(\Delta E_\text{PB}\) as a function of \(\chi\)  for binary W-TM alloys.
For all alloys, ND- and D-cores are seen when \(\chi\) is above and below $0.8$, similar to predictions by interatomic potentials (Fig.~\ref{fig:prop_core_type}b).
Both \(\gamma_\text{us}\) and \(\Delta E_\text{PB}\) scale  linearly with \(\chi\); i.e., 
\begin{align}
  \label{eqn:chi_scaling}
  \gamma_\text{us} (\chi) &= \gamma_\text{us}^\text{P}\left[1+ k_\text{us} (\chi - 1)\right]  \\
  \Delta E_\text{PB} (\chi) &= \Delta E_\text{PB}^\text{P} \left[1 +  k_\text{PB} (\chi - 1)\right] .
\end{align}
where \(k_\text{us} = 0.66\) and \(k_\text{PB} = 2.75\) as fit from the data.
The above expressions can be extrapolated to Group VI-based TM alloys.
Solutes to the left/right of the solvent increase/decrease \(\gamma_\text{us}\) and \(\Delta E_\text{PB}\), which is broadly consistent with solute hardening/softening effects measured in Group VI binary BCC TM alloys (Supplementary Fig.~\ref{fig:hs_gp_vi_exp}) and DFT-based calculations in Mo/W-TMs~\cite{trinkle_2005_science,hu_2017_am}.

The VCA calculations for W-Re show that \(\chi\) decreases with increasing Re for all concentrations.
At 20\% Re (\(\chi=0.73\)), the ND-core transforms to the D-core (Supplementary Fig.~\ref{fig:wre_core}), in agreement with the transformation seen at 25\% Re in explicit solute-based DFT calculations~\cite{romaner_2010_prl} and the observation of a change to alternating \(\{110\}\) slips
in experiments~\cite{li_2012_am}.
The core transformation is also seen in Mo-20\% Re (Supplementary Fig.~\ref{fig:more_core}).
In addition, reducing \(\chi\) can favourably decrease the barrier for dislocation nucleation (\(\gamma_\text{us}\)) at crack tips and lattice friction (\(\Delta E_\text{PB}\)) of the screw dislocation, the two limiting factors governing DBT in BCC TMs~\cite{gumbsch_1998_science}.  At sufficiently low \(\chi\) (e.g., 0.66, or 25 at.\% Re), the gap between the stress intensity factor \(K_\text{Ie}\) for dislocation emission and \(K_\text{Ic}\) for cleavage may be reduced to enable ductile behaviour in W at higher temperatures (Supplementary Fig.~\ref{fig:kie_kic_wre_calc})~\cite{supp_mat}.  In fact, increasing the Re concentration to  $\geq25$\% is the basis of many commercial W-Re alloys which exhibit high ductility and a DBT below room temperature.
For W-Ta alloys, \(\chi\) is always greater than 1 and no core transformation is predicted, in agreement with DFT calculations~\cite{li_2012_am} and experiments where a switch in slip behaviour is not observed~\cite{li_2012_am}.
Ta also increases \(\gamma_\text{us}\) and \(\Delta E_\text{PB}\) (Fig.~\ref{fig:chi_prop}b,c), and thus has deleterious effects on W alloy ductility, again consistent with experiments~\cite{wurster_2011_jnm}.  While Fig.~\ref{fig:vca} suggests many solutes (Tc, Fe, Os, Ru, Rh, Co, Ir etc.) possessing the ductilizing/softening effects of Re~\cite{klopp_1975_jlcm,caillard_2020_am}, their practical application is limited by their low solubility~\cite{thaddeus_1986_bapd} and/or cost.
Re is thus one of the most practical solutes for ductilizing W.
The present predictions provide explicit recommendations of solutes and concentrations that may be explored through non-equilibrium techniques (e.g., mechanical alloying or plasma sintering) to overcome solubility limits.
Overall, the \(\chi\)-based approach identifies solutes capable of ductilizing/softening Group VI TMs that are broadly consistent with nearly all experiments performed to date~\cite{ren_2018_ijrmhm} (Supplementary Fig.~\ref{fig:hs_gp_vi_exp}).

The applicability of the \(\chi\)-approach is further examined in binary TM alloys with solvents from Group V.
For V-TM alloys, solutes to the left of V in the periodic table (decreasing VEC) reduce \(\chi\) (Supplementary Fig.~\ref{fig:vca_v_vasp}) and are expected to have softening effects, while solutes to its right increase \(\chi\) and lead to hardening.
The V-results are opposite to that in Group VI but fully consistent with hardening/softening measurements of Group V TMs (Supplementary Fig.~\ref{fig:hs_gp_v_exp}), as well as DFT calculations based upon inter-string potential~\cite{medvedeva_2007_prb} and a double-kink nucleation~\cite{zhang_2015_sm} models.
As another example, Ti was observed to ductilize V at 77 K~\cite{fraser_1962_cmq}, in agreement with  \(\chi\)-model predictions.
While solute/VEC effects vary among elements, the proposed material index \(\chi\) succinctly and quantitatively captures such effects of quantum-mechanics origin via inexpensive DFT calculations.

\section*{Discussion}
The importance of \(\chi\) for \(\gamma_\text{us}\) and \(\Delta E_\text{PB}\) and thus overall material plasticity and toughness is indisputable~\cite{gumbsch_1998_science}.
Effects of \(\chi\) on the screw dislocation core structures are more subtle but significant.
A change from the ND-core to D-core can have multiple effects on dislocation behaviour; the most important of which is the activation of additional slip systems.
The D-core slips on alternating \(\{110\}\) planes with net slip on multiple \(\{112\}\) planes, while the ND-core is highly restricted to a single \(\{110\}\) plane~\cite{vitek_2008_dis}.
A complete switching of slip plane only occurs at high solute concentrations (corresponding to \(\chi\)<0.8) and may be influenced by additional factors such as stress state and temperature~\cite{weinberger_2013_imr,caillard_2020_am}.
At moderate solute concentrations, both D- and ND-core dislocations can co-exist, depending on local solute concentrations.
Their co-existence activates 24 slip systems; i.e., twice the number of ND core-favoured \{110\} slip systems.
The importance of an increase in the number of active slip systems should not be underestimated.  The ductility of W depends on texture and grain size~\cite{zhang_2009_msea}; single crystals are generally ductile and polycrystals brittle.
Activation of additional slip systems can be critical; providing the flexibility required for strain compatibility at grain boundaries/junctions and reducing the texture-sensitivity of ductility in structural alloys.  

At low temperature, pure BCC TMs tend to be brittle as slip is inhibited by high lattice friction and limited to \(\{110\}\) systems.
With increasing temperatures, the ductility of these pure TMs increases dramatically as \(\Delta E_\text{PB}\)/\(\gamma_\text{us}\) is reduced and additional slip planes (\(\{112\}\) and \(\{123\}\)) become thermally activated.
Such effects of entropy can be equivalently achieved by solute-addition changes to the core structure, \(\gamma_\text{us}\) and \(\Delta E_\text{PB}\), as captured by \(\chi\).
In other words, plastic deformation of pure TMs observed at high temperature can be achieved by alloying with ductilizing solutes (e.g., Re in W and Mo) at low temperature.
This implies that appropriate choice of solutes can increase ductility and lower the DBT temperature, as seen in W/Mo-Re alloys.

The \(\chi\)-based predictions are also fully consistent with general ductile/brittle behaviour in emerging compositionally complex alloys. Refractory high entropy alloy NbTaMoW can be considered as adding group V BCC TM to group VI TM, which raises \(\chi\). The ND core is expected to be prevalent, consistent with recent DFT calculations~\cite{yin_2020_npjcm}. NbTaMoW is also generally brittle at low temperatures.  In contrast, HfTiVNb can be considered as adding group IV elements to the group V TM, which favorably reduces \(\chi\).  Preliminary DFT calculations show that HfTiVNb adopts both the ND and D core, depending on the local solute environment. HfTiVNb possesses tensile ductility above 25\% at room temperature in experiments~\cite{an_2021_mh}.

Solutes also have local effects, which influence \(\chi\) and all core-related properties locally; e.g., dislocation lattice friction by modifying double-kink nucleation/propagation, dislocation cross-slip and interactions~\cite{trinkle_2005_science}.
These local effects vary between BCC TMs and should not be overlooked.  The \(\chi\)-approach does not replace established temperature-dependent solid-solution softening/hardening models~\cite{trinkle_2005_science,zhao_2018_msmse,ghafarollahi_2021_am}.
Instead, \(\chi\) offers a fundamental understanding of the physical basis for controlling dislocation behaviour and quantitatively predicts solute effects at low temperatures where softening/toughening is most needed.
Since \(\chi\) is built upon crystal geometry and bonding characteristics (captured through energy differences of two simple crystal structures), and is validated in extensive DFT calculations and across nearly all interatomic potentials available, it is expected to be general and can be used to rapidly identify favourable solutes in the entire family of BCC TM alloys (e.g., Nb/Fe-based alloys~\cite{romaner_2014_pml}).

In summary, we revealed the mechanistic origin of the critical screw dislocation properties and presented a general model with a fundamental material index \(\chi\) for predicting and controlling screw dislocation average properties in all BCC materials.  The \(\chi\)-model rationalises core structures in pure alkaline metals, BCC TMs and their alloys, and is quantitatively related to lattice friction and barrier for dislocation nucleation.  More importantly, \(\chi\) can be computed rapidly for any alloy composition using DFT, providing a practical computational approach for ductile and tough BCC TM alloy design.  This approach was tested in several binary BCC TM systems and, for example, correctly identifies Re as the most practical solute for ductilizing W and the appropriate Re concentration range.  The material index \(\chi\) does not predict ductility or toughness directly, but does, nevertheless, provide a fundamental constraint on the nature of plasticity in these materials. The proposed \(\chi\) approach is first-principles-based; hence it is applicable to a wide-range of complex alloys where ductility is controlled/limited by lattice friction, dislocation nucleation barrier or nature's choice of slip systems.

\newpage
\section*{Methods}
An extensive range of calculations are performed using various methods and models in the current work.  We also used two density functional theory (DFT) packages to overcome some limitations in determining some specific properties and to cross-validate the generality of the \(\chi\)-model.  We therefore first describe the general methods, followed by details of individual models.  The parameters in the general methods apply to all the calculations unless mentioned otherwise in the respective models.

\subsection*{Density-functional theory calculations using VASP}
\label{sec:dft_method}
First-principles calculations are performed within the DFT framework using the Vienna Ab initio Simulation Package (VASP~\cite{kresse_1996_prb,kresse_1999_prb}) and Quantum ESPRESSO (QE~\cite{giannozzi_2009_jpcm}).

In VASP, the exchange-correlation functional is described by the generalized gradient approximation (GGA) with the Perdew-Burke-Ernzerhof (PBE~\cite{perdew_1996_prl}) parameterization.  In the standard DFT calculations, the core electrons are replaced by the projector augmented wave (PAW~\cite{blochl_1994_prb}) pseudopotentials with the valence states shown in Supplementary Table~\ref{tab:dft_details}. A first-order Methfessel-Paxton method~\cite{methfessel_1989_prb} is used to smooth eigenstate occupancy.  The plane-wave cutoff energy, \(\Gamma\)-centered Monkhorst-Pack k-point sampling mesh~\cite{monkhorst_1976_prb} and the smearing parameters sigma are established through convergence tests of 2-atom BCC unit cells with a threshold energy variation \(\Delta E_\text{c} < 1\) meV/atom. The final converged parameters are shown in Supplementary Table~\ref{tab:dft_details}.

In the virtual crystal approximation (VCA~\cite{bellaiche_2000_prb}) DFT calculations, virtual atoms are created with effective cores and valence electrons constructed based on the alloy constituents. In particular, the virtual atoms have their core electrons replaced by pseudopotentials \(\phi_\text{virtual}\), which are obtained via mixing that of the solvent and solute atoms and weighted by their atomic fraction \(X\).  For example, for a binary \(\text{A}_{X} \text{B}_{1-X}\) alloy, a pseudopotential is created for the core electrons of the virtual atom representing the average alloy property, i.e.,
\begin{equation}\label{eqn:vca_weighted_core_ve}
  \phi_\text{virtual} =  X \phi_\text{A} + (1-X) \phi_\text{B}
\end{equation}
where \(\phi_{i}\) are the pseudopotentials of the respective atoms.   In addition, the valence electron numbers of virtual atoms are calculated according to Eqn.~\ref{eqn:vca_weighted_core_ve}. The plane-wave kinetic cutoff energy is set at 450 eV and a fine \(\Gamma\)-centered Monkhorst-Pack k-point  mesh~\cite{monkhorst_1976_prb} (k-spacing \(2\pi \times 0.02\)\AA\(^{-1}\)) is used for Brillouin zone integration.  A first order Methfessel-Paxton smearing with a width of 0.2 eV is used to smooth the partial electron occupancies. The valence states used for different elements are listed in Supplementary Table~\ref{tab:vs_state_vca_vasp_qe}.

\subsection*{Molecular dynamics/statics simulations using LAMMPS}
\label{sec:md_method}
Molecular dynamics and static relaxation simulations are performed using the Large-scale Atomic/Molecular Massively Parallel Simulator (LAMMPS~\cite{plimpton_1995_jcp}). Structure optimization is performed using the conjugate gradient method.  Convergence is assumed when forces on all atoms drop below \(10^{-12}\) eV/\AA{} in calculations of \(\Delta E\) and \(\gamma\) surfaces,  and \(10^{-10}\) eV/\AA{} in calculations of dislocation core structures.  For the GAP potential for Fe~\cite{dragoni_2018_prm,bartok_2010_prl,bartok_2013_prb,bartok_2015_ijqc,gap_www}, convergence tolerance is relaxed to \(10^{-12}\) eV for energy variation and \(10^{-6}\) eV/\AA{} for atomic forces, due to its slow convergence and high computational cost.

\subsection*{Interatomic potential models for BCC metals}

The screw dislocation has been extensively studied in elementary BCC transition metals using DFT calculations. A rich set of interatomic potentials have also been developed using various formalisms over the past several decades.  In this work, we examined nearly all interatomic potentials (publically accessible) for BCC metals developed up to date.  These potentials are available from the National Institute of Standards and Technology (NIST) potential repository~\cite{nist_2013}, the Open Knowledgebase of Interatomic Models (OpenKIM) project~\cite{openkim_2011} or literature.  In total, we examined 72 potentials, as shown in Supplementary Table~\ref{tab:potential}.  These potentials use different functional forms in their descriptions of interatomic interactions. The choice made here is not biased toward any particular form, element or fitting procedure.

In addition, we developed two new MEAM potentials for BCC V (V1 and V2). The potential parameters are fitted with target properties shown in Supplementary Fig.~\ref{fig:meam_prop_v1_v2} using the particle swarm optimization algorithm.  The two fit-for-purpose potentials have nearly identical properties, except for \(\Delta E\), which is intentionally controlled at 91 meV/atom and 124 meV/atom, respectively.  With \(\Delta E\) from DFT at 243 meV/atom, V1 and V2 have \(\chi = 0.37 \) and 0.51, respectively. V1 adopts a D-core and V2 adopts a ND-core for the screw dislocation, consistent with the proposed \(\chi\)-criterion.  A new MEAM interatomic potential for Li is also developed with \(\Delta E\) = -0.8 meV/atom close to that from DFT.  All potential parameters for Li are shown in Supplementary Table~\ref{tab:meam_pot_param} and can be used in LAMMPS directly.  The MEAM potential for Li exhibits a D-core, in agreement with DFT and consistent with the geometric model where D-core is favored if the FCC structure is not hugely energetically unfavourable.

\subsection*{Calculation of energy difference \(\Delta E\) between FCC and BCC structures}
\label{sec:calc_chi_pure_ele}
The energy differences \(\Delta E\) per atom between the FCC and BCC structures are calculated using two-atom BCC and 4-atom FCC unit cells by DFT in VASP.   Convergence is assumed when energy variation drops below \(10^{-5}\) eV per electronic step and \(10^{-4}\) eV per ionic step.  The final \(\Delta E\) results are shown in Supplementary Table~\ref{tab:delta_E_bcc_fcc}.  Similarly, the energy difference \(\Delta E\) is also calculated using all the interatomic potentials in Supplementary Table~\ref{tab:potential} and convergence criterion described earlier. In all cases, the values of \(\Delta E\) do not depend on the employed supercell size within convergence tolerances.

\subsection*{Generalized stacking fault energy surface}
\label{sec:gamma_surf_calc_vasp}

The generalized stacking fault energy surfaces (\(\gamma\)-surface) of Li, Ta and W are calculated by DFT using VASP. A slab-supercell model is used, with its crystal orientation given in Supplementary Table~\ref{tab:supercell_dft}. The slab contains 12 \(\{110\}\)-plane atom layers and a 20 \AA{} vacuum layer.  \(\gamma\)-surfaces are calculated using the classical method by Vitek. For each specific stacking fault position, a homogeneous slip displacement \(\mathbf{s}\) is applied to atoms in the upper half block of the supercell to create the stacking fault between the sheared and un-sheared atom block.  Stacking fault energies are calculated with all atoms allowed to move in the direction perpendicular to the slip, \(\{110\}\) plane. The same convergence criteria as that for \(\Delta E\) are used for the \(\gamma\)-surface calculations.

In the calculations of \(\gamma\)-surface using interatomic potentials in LAMMPS, tilted supercells of dimensions \textasciitilde{}\(10 \times 10 \times 40\) (\AA{}\textsuperscript{3}) are used.  The stacking fault energies are calculated with atoms constrained to move in the direction perpendicular to the slip plane only (same as in DFT, see Ref.\cite{yin_2017_am} for details).

\subsection*{Dislocation core structure}
\label{sec:disl_core}

Dislocation core structures are calculated using both a dipole configuration (dipole method) and a single dislocation with fixed boundary conditions (cluster method). For the dipole method, a fully periodic prism supercell is chosen with the dislocations arranged in quadrupolar positions, as shown in Supplementary Fig.~\ref{fig:dis_sche}b.  The supercell vectors are
\begin{align}
  \mathbf{c}_1 &= \mathbf{e}_{1}, \\
  \mathbf{c}_2 &= 5\mathbf{e}_{2}, \\
  \mathbf{c}_3 &= \dfrac{1}{2}\mathbf{e}_{1} + \dfrac{5}{2}\mathbf{e}_{2}+9\mathbf{e}_{3},
\end{align}
where (\(\mathbf{e}_{1},\mathbf{e}_{2},\mathbf{e}_{3}) = (a/2[111],a[11\bar{2}],a[\bar{1}10])\) and  \(a\) is the lattice parameter of the BCC structure determined in unit cell calculations. In total, the supercell contains 135 atoms.

A dislocation dipole, i.e., a pair of dislocations with opposite Burgers vectors \(\pm \mathbf{b}\), is introduced at easy-core/hard-core positions (Supplementary Fig.~\ref{fig:dis_sche}c) using the Babel package developed by E. Clouet~\cite{clouet_2021_babel}.  In particular, the displacement field of arrays of dipoles (repeated 16 times in the \(\mathbf{c}_2\) and \(\mathbf{c}_3\) directions) is firstly applied to atoms in the supercell, followed by applying a homogeneous strain to accommodate the plastic strain introduced by the dipole (see Ref.~\cite{rodney_2017_am} for details).  During structure optimization, the cell vectors are fixed and all atoms are allowed to move until the ionic force is below \(5\times 10^{-3}\) eV/\AA. DFT calculations show that the easy-core is the minimum energy configuration and the hard-core is a local maximum. The DFT-computed core structures (ND vs D) are shown in Fig.~\ref{fig:gamma_disl_Li_Ta_W} for BCC Li, Ta and W.

The dipole approach allows modelling of dislocation cores with a fully periodic supercell of relatively small sizes.  It avoids the complexity of free surfaces, but introduces dislocation interactions and a non-negligible homogeneous strain scaling with supercell sizes.  Therefore, we also used a hexagonal prism supercell (Supplementary Fig.~\ref{fig:dis_sche}a) with a single dislocation and performed dislocation core structure calculations with interatomic potentials.  In this approach, supercells of radius \(R\) \textasciitilde{}92 \AA{} and length of one Burgers vector are first created with perfect BCC structures. The crystallographic \([11\bar{2}]\) direction is aligned with the \(x\)-axis, \([\bar{1}10]\) aligned with the \(y\)-axis and the screw dislocation line direction \([111]\) aligned with the \(z\)-axis, i.e., the hexagonal prism axis (Supplementary Fig.~\ref{fig:dis_sche}d).  The \(z\)-direction has periodic boundary conditions while the \(x\) and \(y\) directions are treated as surfaces.  This geometry keeps the \(\langle  111 \rangle\)-axis threefold symmetry of the BCC lattice.  Each supercell contains \textasciitilde{}3600 atoms (depending on element types).  A single dislocation is then introduced by displacing atoms according to the anisotropic elastic displacement field of the corresponding Volterra dislocation at the supercell center (easy-core position). Atoms within \textasciitilde{}12 \AA{} from the outer surface (grey area in Supplementary Fig \ref{fig:dis_sche}a) are always fixed to their elastic displacements.  The initial dislocation structure is first optimized using conjugate gradient algorithm at 0 K, then equilibrated for 100,000 time steps (100,000 femtoseconds) at finite temperatures (\textasciitilde{}50 K) and optimized again at 0 K.  This procedure is not always needed, but helps to obtain well-equilibrated core structures in some interatomic potentials. Only 0 K structure optimization is performed for the GAP potential for Fe~\cite{dragoni_2018_prm}.  The obtained core structures (ND vs D) are summarized in Fig.~\ref{fig:prop_core_type}.

\subsection*{Differential displacement plot and atomistic visualisation}
Dislocation core structures are visualised using the differential displacement (DD) map between neighboring atoms~\cite{vitek_1970_pma}.  For screw components (Fig.~\ref{fig:gamma_disl_Li_Ta_W}, Supplementary Figs.~\ref{fig:prop_core_type_v} and~\ref{fig:alkali_core}), the arrows are scaled so that the largest components (i.e., \textasciitilde{}\(\mathbf{b}/3\)) touch the neighboring atoms.

Excessive atomic displacements (EADs) (Fig.~\ref{fig:bcc_fcc}f-i and Supplementary Fig.~\ref{fig:fe_core_eecd}) are obtained by using the corresponding anisotropic Volterra dislocation field as references. The EADs of ND-/D-cores possess different symmetry; in the D-core, they have a \(\langle{111} \rangle\)-threefold screw axis symmetry while that in the ND-core have both the \(\langle{111} \rangle\)-threefold screw axis symmetry and the \(\langle {110} \rangle\)-diad axis symmetry. The EADs illustrate the key, subtle differences among different materials/models. They are generally very small, and are thus magnified by 6 times.

Atomic configurations (Fig.~\ref{fig:bcc_fcc}a-e) are visualised using the Open Visualisation Tool (OVITO~\cite{stukowski_2009_msmse}).

\subsection*{Calculation of \(\chi\) in binary alloys with DFT VCA in VASP }
\label{sec:chi_vca_vasp}

The values of \(\chi\) are calculated at a 2\%-increment for binary W-TM, Mo-TM and V-TM alloys using the VCA method in VASP.  The supercells and convergence criteria are the same as that used for calculating \(\Delta E\) in pure elements. The DFT settings are described in the Methods: Density-functional theory calculations using VASP, except that the plane-wave kinetic cutoff energy is increased to 550 eV for all considered elements.  We used the same cutoff energy for 84 alloys here to demonstrate the method of \(\chi\) to identify favourable solutes, as commonly employed in high throughput DFT calculations~\cite{lejaeghere_2016_science}.  Refined calculations can be performed once favourable solutes are identified. The final results are shown in Fig.~\ref{fig:vca}, Supplementary Figs.~\ref{fig:vca_mo_vasp} and ~\ref{fig:vca_v_vasp}.

\subsection*{Peierls barrier of binary W-TM alloys with DFT VCA in VASP}

Peierls barrier can be defined as the energy barrier per unit length for a straight dislocation to move from one energy valley (equilibrium position) to the next one at zero stress and zero temperature.  For the \(1/2\langle 111 \rangle\) screw dislocation in BCC structures, the two equilibrium positions correspond to the easy-core configuration in adjacent lattice positions (Supplementary Fig.~\ref{fig:dis_sche}c).  In the current work, we use the nudged elastic band (NEB~\cite{henkelman_2000_11_jcp}) method to calculate the dislocation migration path between these two adjacent easy-core configurations and thus obtain the height of the energy barrier, i.e., the Peierls barrier.  The transition path is calculated for pure W by DFT in VASP and for W-5\%Re, W-10\%Re and W-10\%Ta by DFT VCA in VASP.  The dipole supercell with 7 linearly interpolated replicas are used for all calculations (see Refs.~\cite{samolyuk_2013_jpcm,dezerald_2014_prb} for details).

Supplementary Fig.~\ref{fig:vca_wtms_pe} shows the energy variation along the transition path.  For all cases, the Peierls potential profiles have a near-sinusoidal profile with a peak in the middle replica corresponding to the saddle-core configuration.  Re and Ta respectively decreases and increases the Peierls barrier relative to that of pure W, in agreement with previous DFT calculations~\cite{samolyuk_2013_jpcm,dezerald_2014_prb}.

The NEB method requires simultaneous structure optimizations in all replicas and is computational expensive.  For calculations in a wide range of W-TM alloys, we employ the reaction coordinate method~\cite{xu_1998_cms} (i.e., the drag method) to compute their Peierls barriers as a function of solute concentrations and \(\chi\).  In the drag method, the 5 intermediate replicas are linearly interpolated between the initial and final configurations (easy-1 and easy-2 in Supplementary Fig.~\ref{fig:dis_sche}c). In the middle replica, a single atom near the core (the light grey atom in Supplementary Fig.~\ref{fig:dis_sche}c) is fixed along the dislocation line direction while all other degrees of freedom are fully optimized.  Under this constraint, the middle replica will evolve to the saddle-core configuration (peak energy in the transition path) found in the NEB method. Supplementary Table~\ref{tab:drag_vs_neb} shows the Peierls barriers computed by the NEB and the drag methods, respectively.  Their agreement shows that the drag method, as an alternative to NEB, can be used to compute the Peierls barrier in a reliable and computationally efficient manner.  The calculations for alloys are performed with the VCA method in VASP and convergence is assumed when the ionic forces fall below \(0.02\) eV/\AA .  The obtained Peierls barriers of W-TM alloys are shown in Fig.~\ref{fig:chi_prop}c.

\subsection*{Unstable stacking fault energy of binary W-TM alloys with DFT VCA in VASP}
\label{sec:usf_wre_vca_vasp}
The unstable stacking fault energy \(\gamma_\text{us}\) dictates the energy barrier to nucleate a dislocation.  In the current work, we calculate \(\gamma_\text{us}\) in the \(\langle 111 \rangle\) direction on the \(\{110\}\) plane as a function of solute concentration in binary BCC W-TM alloys by DFT VCA in VASP. In particular, \(\gamma_\text{us}\) of \{110\} and \{112\} planes are calculated for W-Re alloys and are used as material property inputs for analysis of crack tip behaviour using linear elastic fracture mechanics (LEFM). The DFT settings, supercell and convergence criterion are described in the Methods section: density-functional theory calculations using VASP, Generalized stacking fault energy surface and Supplementary Table~\ref{tab:supercell_dft}. The \(\gamma_{\text{us}}\) of \{112\} plane converges when the slab has 16 atomic layers or more. The converged \(\gamma_\text{us}\) values are shown in Fig.~\ref{fig:chi_prop}b and Supplementary Fig.~\ref{fig:usf_sf_wre_calc}a.

\subsection*{Surface energy of W-Re alloys with DFT VCA in VASP and QE}
\label{sec:wre_surf_e_vca_vasp}
The surface energies of binary W-Re alloys are calculated using VCA as implemented in both VASP and QE (see Supplementary Table~\ref{tab:supercell_dft} for supercells employed).  The pseudopotentials and parameters are described in the Methods section: Density-functional theory calculations using VASP and Cross-validation of \(\chi\) in binary W-TM alloys with DFT VCA in QE.  In addition, the kinetic cutoff energies for wavefunctions and charge densities are chosen as 952 eV (70 Ry) and 11424 eV (840 Ry) respectively in QE. For the ionic minimization, the convergence thresholds for total energy and forces are 1.36 $\times 10^{-3}$ eV (1 $\times 10^{-4}$ Ry) and 2.57 $\times 10^{-2}$ eV/\AA\ (1 $\times 10^{-3}$ Ry/bohr). The convergence criterion for electronic self-consistency is 1.36 $\times 10^{-5}$ eV (1 $\times 10^{-6}$ Ry).

For each surface, a fully periodic supercell is created with supercell vectors \((\mathbf{c}_1, \mathbf{c}_2, \mathbf{c}_3)\), where \(\mathbf{c}_1\) and \(\mathbf{c}_2\) are the in-plane supercell repeating vectors and \(\mathbf{c}_3\) is the out-of-plane vector.  We first obtain the energy per atom \(E_\text{c}\) in the bulk by optimizing the supercell and ionic positions.  The slab-vacuum supercell is then created by adding a vacuum layer in the out-of-plane direction. The slab-vacuum structure is optimized through ionic relaxation only in the out-of-plane direction while keeping all supercell vectors fixed.  In both calculations of the bulk supercell and slab-vacuum supercell, a consistent k-point sampling is used in the in-plane directions, leading to better convergence with respect to \(\mathbf{c}_3\), i.e., slab layers~\cite{sun_2013_ss}.  The surface energy \(\gamma\) is calculated as
\begin{align}
  \gamma = \dfrac{E_\text{sv} - n_\text{sv}E_\text{c}}{2A},
\end{align}
where \(E_\text{sv}\) and \(n_\text{sv}\) are the total energy and number of atoms of the slab-vacuum system and \(A\) is the area of the surface, i.e., \(A = \lvert  \mathbf{c}_1 \times \mathbf{c}_2 \rvert\).  The supercell vectors are shown in Supplementary Table~\ref{tab:supercell_dft} and a slab with 12 atomic layers gives the converged surface energies, as shown in Supplementary Fig.~\ref{fig:usf_sf_wre_calc}c,d.

For pure W in stardard DFT calculations, both VASP and QE give similar surface energies; \(\gamma_{\text{s} \{100\}}\) and \(\gamma_{\text{s}\{110\}}\) are 4.03 J/m\(^{2}\) and 3.27 J/m\(^{2}\) in VASP, and 4.15 J/m\(^{2}\) and 3.28 J/m\(^{2}\) in QE respectively.  In VCA QE, the surface energies of the \(\{100\}\) and \(\{110\}\) planes are predicted to gradually decrease with Re additions, consistent with previous calculations on W-Re alloys~\cite{yang_2018_prb,hu_2021_am}. However, in VCA VASP, the surface energies are unusual; they increase substantially with Re additions. For example, the surface energy is 11.12 J/m\(^{2}\) for the \(\{110\}\) plane of W-50\%Re, nearly 3.4 times of pure W (3.27 J/m\(^{2}\)). The VCA VASP-based results are also substantially different from that based on special quasi-random structure (SQS) supercell in standard VASP calculations~\cite{hu_2021_am}. In contrast, the results of VCA QE are close to those in SQS supercell calculations.  Therefore, we use the surface energies based on VCA QE as inputs in subsequent LEFM analysis.

\subsection*{Elastic constants of W-Re alloys with DFT VCA in VASP}
\label{sec:wre_elas_vca_vasp}
The elastic constants of binary W-Re alloys are calculated using VCA method with VASP.  The elastic contants are determined by fitting to the total energies versus strains within \(\pm\)1\%.  Supplementary Fig.~\ref{fig:usf_sf_wre_calc}b shows the elastic constants as a function of Re concentrations. Overall, Re has weak effects on the elastic constants of W.

\subsection*{Critical stress intensity factors for cleavage and dislocation emission}
\label{sec:lefm_wre}
Pure W exhibits brittle cleavage behaviour on \(\{100\}\) and \(\{110\}\) planes at low temperature~\cite{gumbsch_1998_science}.  The critical stress intensity factor \(K_\text{Ic}\) for Griffith cleavage is lower than \(K_\text{Ie}\) for dislocation emission~\cite{rice_1992_jmps} for sharp cracks on these two planes under mode-I loading. In binary W-Re alloys, \(K_\text{Ic}\) and \(K_\text{Ie}\) may change due to Re-effect. Here, we compute the respective \(K_\text{Ic}\) and \(K_\text{Ie}\) as a function of Re concentration for mode-I loaded sharp cracks.

Since W is nearly isotropic, we use isotropic LEFM theory. In particular, \(K_\text{Ic}\) for Griffith cleavage under plane strain Mode-I loading is computed as
\begin{equation}
  K_\text{Ic}=\sqrt{\frac{G}{D}},
\end{equation}
where
\begin{equation}
  G = 2 \gamma_\text{s}, \gamma_\text{s} \text{ is the surface energy}
\end{equation}
and
\begin{equation}
  D=\frac{1-{\nu}^{2}}{E}, \nu \text{ is the Poisson's ratio and } E \text{ is the Young's modulus}.
\end{equation}

The critical stress intensity for dislocation emission \(K_\text{Ie}\) is calculated according to the Rice criterion~\cite{rice_1992_jmps}
\begin{equation}
K_\text{Ie}=\frac{1}{f(\theta)} \sqrt{\frac{2\mu}{1-\nu} [1+(1-\nu)\tan^{2}\phi]\gamma_\text{us}}
\end{equation}
where \(f(\theta)=\cos^{2}(\theta/2)\sin(\theta/2)\), \(\theta\) is the angle between the crack plane and slip plane, \(\phi\) is the angle between the dislocation Burgers vector and crack front direction in the slip plane, and \(\mu\) is shear modulus, \(\nu\) is Possion's ratio, \(\gamma_\text{us}\) is the unstable stacking fault energy of the slip plane.

All material properties (\(\gamma_\text{s}, \gamma_\text{us}, E, \nu, \mu,\)) depend on Re concentrations in W and are calculated based on first-principle DFT VCA calculations. Supplementary Fig.~\ref{fig:kie_kic_wre_calc} shows the results for 4 crack systems (\(\{100\}\langle 001 \rangle\), \(\{100\}\langle 011 \rangle\), \(\{110\}\langle 001 \rangle\), \(\{110\}\langle 011 \rangle\)) susceptible to cleavage fracture in BCC W.  In pure W and for a \(\{100\}\) sharp crack (Supplementary Fig.~\ref{fig:kie_kic_wre_calc} c), \(K_\text{Ic} = 1.90 \text{ MPa}\cdot\text{m}^{1/2}\), which is \(\sim\)40\% and \(\sim\)42\% lower than \(K_\text{Ie}\) for dislocation emission in the \(\{100\}\langle 001 \rangle\) and \(\{100\}\langle 011 \rangle\) crack systems. Similar behaviours are seen in cracks on \(\{110\}\) planes (Supplementary Fig.~\ref{fig:kie_kic_wre_calc}g). In pure W, \(K_{\text{Ic}} = 1.72 \text{ Mpa}\cdot\text{m}^{1/2}\), which is \(\sim\)41\% and \(\sim\)31\% lower than \(K_\text{Ie}\) for dislocation emission in the \(\{110\}\langle 001 \rangle\) and \(\{110\}\langle 011 \rangle\) crack systems. The isotropic solution is nearly identical to that of the anisotropic solution~\cite{mak_2021_jmps} and both results suggest that \(\{100\}\) and \(\{110\}\) planes are intrinsically brittle in W.  With increasing Re concentrations, all \(K_{\text{Ie}}\) decrease sharply. However, \(K_{\text{Ic}}\) is insensitive to Re additions.  As a result, the gaps between \(K_{\text{Ie}}\) and \(K_{\text{Ic}}\) decrease sharply in all the four crack systems, which is potentially beneficial to make W ductile.  Nevertheless, no cross-over between \(K_{\text{Ie}}\) and \(K_{\text{Ic}}\) is predicted in the analysis based on material properties at 0 K.

To extend the prediction to room temperatures, we employ a finite-temperature ductility criterion \(D=K_{\text{Ie}}/K_{\text{Ic}}\) proposed recently~\cite{mak_2021_jmps} based on experimental observations. In BCC W, ductile behaviour can be achieved at room temperatures if \(D < D_\text{c}\); \(D_\text{c} \approx 1.56/1.26\) for \(\{100\}/\{110\}\) planes respectively. In W-20\%Re, \(D_{\{100\}\langle 001 \rangle} = 1.53\) and \(D_{\{110\}\langle 011 \rangle} = 1.30\), close to the respective threshold values \(D_\text{c}\) (Supplementary Fig.~\ref{fig:kie_kic_wre_calc}d,h).  With higher Re concentrations, \(D\) decreases further and satisfies the ductile criterion of both planes. At 20\% Re concentration (\(\chi=0.73\)), the ND core also transforms to the D core.  This concentration is also close to many commercial W-Re alloys exhibiting high ductility and a DBT below room temperatures.  The finite-temperature \(D_\text{c}\) is established empirically based on DBT experiments, it may include the effects of core transformation and associated change in plastic slip behaviour.  Since \(D > 1\) at 0 K, the mechanism of Re-ductilization effects is thus more likely related to core transformation.  However, we can not rule out thermal activation in dislocation nucleation at crack tips.  Nevertheless,  the above analysis is based on well-established linear elastic fracture mechanics and DFT-computed material properties. The primary assumption is linear elasticity.  Therefore, the results are robust and consistent with ductile behaviour in BCC W-Re alloys at room temperatures.  More importantly, the critical material properties (\(\gamma_\text{us}\) and ND/D core structures) and associated brittle vs ductile behaviour can be related to the material index \(\chi\).

\def\bibsection{\section*{}}
\section*{References}
\textbf{References 1-70 are for the main text.}
\vspace{-0.5cm}

\bibliography{cited_ref}

\begin{thebibliography}{145}%
\makeatletter
\providecommand \@ifxundefined [1]{%
 \@ifx{#1\undefined}
}%
\providecommand \@ifnum [1]{%
 \ifnum #1\expandafter \@firstoftwo
 \else \expandafter \@secondoftwo
 \fi
}%
\providecommand \@ifx [1]{%
 \ifx #1\expandafter \@firstoftwo
 \else \expandafter \@secondoftwo
 \fi
}%
\providecommand \natexlab [1]{#1}%
\providecommand \enquote  [1]{``#1''}%
\providecommand \bibnamefont  [1]{#1}%
\providecommand \bibfnamefont [1]{#1}%
\providecommand \citenamefont [1]{#1}%
\providecommand \href@noop [0]{\@secondoftwo}%
\providecommand \href [0]{\begingroup \@sanitize@url \@href}%
\providecommand \@href[1]{\@@startlink{#1}\@@href}%
\providecommand \@@href[1]{\endgroup#1\@@endlink}%
\providecommand \@sanitize@url [0]{\catcode `\\12\catcode `\$12\catcode
  `\&12\catcode `\#12\catcode `\^12\catcode `\_12\catcode `\%12\relax}%
\providecommand \@@startlink[1]{}%
\providecommand \@@endlink[0]{}%
\providecommand \url  [0]{\begingroup\@sanitize@url \@url }%
\providecommand \@url [1]{\endgroup\@href {#1}{\urlprefix }}%
\providecommand \urlprefix  [0]{URL }%
\providecommand \Eprint [0]{\href }%
\providecommand \doibase [0]{https://doi.org/}%
\providecommand \selectlanguage [0]{\@gobble}%
\providecommand \bibinfo  [0]{\@secondoftwo}%
\providecommand \bibfield  [0]{\@secondoftwo}%
\providecommand \translation [1]{[#1]}%
\providecommand \BibitemOpen [0]{}%
\providecommand \bibitemStop [0]{}%
\providecommand \bibitemNoStop [0]{.\EOS\space}%
\providecommand \EOS [0]{\spacefactor3000\relax}%
\providecommand \BibitemShut  [1]{\csname bibitem#1\endcsname}%
\let\auto@bib@innerbib\@empty
\bibitem [{\citenamefont {Vitek}\ and\ \citenamefont
  {Paidar}(2008)}]{vitek_2008_dis}%
  \BibitemOpen
  \bibfield  {author} {\bibinfo {author} {\bibfnamefont {V.}~\bibnamefont
  {Vitek}}\ and\ \bibinfo {author} {\bibfnamefont {V.}~\bibnamefont {Paidar}},\
  }\bibfield  {title} {\bibinfo {title} {Non-planar dislocation cores: A
  ubiquitous phenomenon affecting mechanical properties of crystalline
  materials},\ }\href {https://doi.org/10.1016/S1572-4859(07)00007-1}
  {\bibfield  {journal} {\bibinfo  {journal} {Dislocations in Solids}\ }\textbf
  {\bibinfo {volume} {14}},\ \bibinfo {pages} {439} (\bibinfo {year}
  {2008})}\BibitemShut {NoStop}%
\bibitem [{\citenamefont {Schade}(2010)}]{schade_2010_ijrmhm}%
  \BibitemOpen
  \bibfield  {author} {\bibinfo {author} {\bibfnamefont {P.}~\bibnamefont
  {Schade}},\ }\bibfield  {title} {\bibinfo {title} {100 years of doped
  tungsten wire},\ }\href {https://doi.org/10.1016/j.ijrmhm.2010.05.003}
  {\bibfield  {journal} {\bibinfo  {journal} {International Journal of
  Refractory Metals and Hard Materials}\ }\textbf {\bibinfo {volume} {28}},\
  \bibinfo {pages} {648} (\bibinfo {year} {2010})}\BibitemShut {NoStop}%
\bibitem [{\citenamefont {Gumbsch}(2003)}]{gumbsch_2003_jnm}%
  \BibitemOpen
  \bibfield  {author} {\bibinfo {author} {\bibfnamefont {P.}~\bibnamefont
  {Gumbsch}},\ }\bibfield  {title} {\bibinfo {title} {Brittle fracture and the
  brittle-to-ductile transition of tungsten},\ }\href
  {https://doi.org/10.1016/j.jnucmat.2003.08.009} {\bibfield  {journal}
  {\bibinfo  {journal} {Journal of Nuclear Materials}\ }\textbf {\bibinfo
  {volume} {323}},\ \bibinfo {pages} {304} (\bibinfo {year}
  {2003})}\BibitemShut {NoStop}%
\bibitem [{\citenamefont {Joseph}\ \emph {et~al.}(2007)\citenamefont {Joseph},
  \citenamefont {Tanaka}, \citenamefont {Wilkinson},\ and\ \citenamefont
  {Roberts}}]{joseph_2007_jnm}%
  \BibitemOpen
  \bibfield  {author} {\bibinfo {author} {\bibfnamefont {T.}~\bibnamefont
  {Joseph}}, \bibinfo {author} {\bibfnamefont {M.}~\bibnamefont {Tanaka}},
  \bibinfo {author} {\bibfnamefont {A.}~\bibnamefont {Wilkinson}},\ and\
  \bibinfo {author} {\bibfnamefont {S.}~\bibnamefont {Roberts}},\ }\bibfield
  {title} {\bibinfo {title} {Brittle{--}ductile transitions in vanadium and
  iron{--}chromium},\ }\href {https://doi.org/10.1016/j.jnucmat.2007.03.077}
  {\bibfield  {journal} {\bibinfo  {journal} {Journal of Nuclear Materials}\
  }\textbf {\bibinfo {volume} {367--370}},\ \bibinfo {pages} {637} (\bibinfo
  {year} {2007})}\BibitemShut {NoStop}%
\bibitem [{\citenamefont {Gumbsch}\ \emph {et~al.}(1998)\citenamefont
  {Gumbsch}, \citenamefont {Riedle}, \citenamefont {Hartmaier},\ and\
  \citenamefont {Fischmeister}}]{gumbsch_1998_science}%
  \BibitemOpen
  \bibfield  {author} {\bibinfo {author} {\bibfnamefont {P.}~\bibnamefont
  {Gumbsch}}, \bibinfo {author} {\bibfnamefont {J.}~\bibnamefont {Riedle}},
  \bibinfo {author} {\bibfnamefont {A.}~\bibnamefont {Hartmaier}},\ and\
  \bibinfo {author} {\bibfnamefont {H.~F.}\ \bibnamefont {Fischmeister}},\
  }\bibfield  {title} {\bibinfo {title} {Controlling factors for the
  brittle-to-ductile transition in tungsten single crystals},\ }\href
  {https://doi.org/10.1126/science.282.5392.1293} {\bibfield  {journal}
  {\bibinfo  {journal} {Science}\ }\textbf {\bibinfo {volume} {282}},\ \bibinfo
  {pages} {1293} (\bibinfo {year} {1998})}\BibitemShut {NoStop}%
\bibitem [{\citenamefont {Butler}\ \emph {et~al.}(2018)\citenamefont {Butler},
  \citenamefont {Paramore}, \citenamefont {Ligda}, \citenamefont {Ren},
  \citenamefont {Fang}, \citenamefont {Middlemas},\ and\ \citenamefont
  {Hemker}}]{butler_2018_ijrmhm}%
  \BibitemOpen
  \bibfield  {author} {\bibinfo {author} {\bibfnamefont {B.~G.}\ \bibnamefont
  {Butler}}, \bibinfo {author} {\bibfnamefont {J.~D.}\ \bibnamefont
  {Paramore}}, \bibinfo {author} {\bibfnamefont {J.~P.}\ \bibnamefont {Ligda}},
  \bibinfo {author} {\bibfnamefont {C.}~\bibnamefont {Ren}}, \bibinfo {author}
  {\bibfnamefont {Z.~Z.}\ \bibnamefont {Fang}}, \bibinfo {author}
  {\bibfnamefont {S.~C.}\ \bibnamefont {Middlemas}},\ and\ \bibinfo {author}
  {\bibfnamefont {K.~J.}\ \bibnamefont {Hemker}},\ }\bibfield  {title}
  {\bibinfo {title} {Mechanisms of deformation and ductility in tungsten {--}
  {A} review},\ }\href {https://doi.org/10.1016/j.ijrmhm.2018.04.021}
  {\bibfield  {journal} {\bibinfo  {journal} {International Journal of
  Refractory Metals and Hard Materials}\ }\textbf {\bibinfo {volume} {75}},\
  \bibinfo {pages} {248} (\bibinfo {year} {2018})}\BibitemShut {NoStop}%
\bibitem [{\citenamefont {Duesbery}\ and\ \citenamefont
  {Vitek}(1998)}]{duesbery_1998_am}%
  \BibitemOpen
  \bibfield  {author} {\bibinfo {author} {\bibfnamefont {M.}~\bibnamefont
  {Duesbery}}\ and\ \bibinfo {author} {\bibfnamefont {V.}~\bibnamefont
  {Vitek}},\ }\bibfield  {title} {\bibinfo {title} {Plastic anisotropy in
  b.c.c. transition metals},\ }\href
  {https://doi.org/10.1016/s1359-6454(97)00367-4} {\bibfield  {journal}
  {\bibinfo  {journal} {Acta Materialia}\ }\textbf {\bibinfo {volume} {46}},\
  \bibinfo {pages} {1481} (\bibinfo {year} {1998})}\BibitemShut {NoStop}%
\bibitem [{\citenamefont {Brunner}\ and\ \citenamefont
  {Glebovsky}(2000)}]{brunner_2000_ml}%
  \BibitemOpen
  \bibfield  {author} {\bibinfo {author} {\bibfnamefont {D.}~\bibnamefont
  {Brunner}}\ and\ \bibinfo {author} {\bibfnamefont {V.}~\bibnamefont
  {Glebovsky}},\ }\bibfield  {title} {\bibinfo {title} {The plastic properties
  of high-purity {W} single crystals},\ }\href
  {https://doi.org/10.1016/s0167-577x(99)00200-1} {\bibfield  {journal}
  {\bibinfo  {journal} {Materials Letters}\ }\textbf {\bibinfo {volume} {42}},\
  \bibinfo {pages} {290} (\bibinfo {year} {2000})}\BibitemShut {NoStop}%
\bibitem [{\citenamefont {Cereceda}\ \emph {et~al.}(2013)\citenamefont
  {Cereceda}, \citenamefont {Stukowski}, \citenamefont {Gilbert}, \citenamefont
  {Queyreau}, \citenamefont {Ventelon}, \citenamefont {Marinica}, \citenamefont
  {Perlado},\ and\ \citenamefont {Marian}}]{cereceda_2013_jpcm}%
  \BibitemOpen
  \bibfield  {author} {\bibinfo {author} {\bibfnamefont {D.}~\bibnamefont
  {Cereceda}}, \bibinfo {author} {\bibfnamefont {A.}~\bibnamefont {Stukowski}},
  \bibinfo {author} {\bibfnamefont {M.~R.}\ \bibnamefont {Gilbert}}, \bibinfo
  {author} {\bibfnamefont {S.}~\bibnamefont {Queyreau}}, \bibinfo {author}
  {\bibfnamefont {L.}~\bibnamefont {Ventelon}}, \bibinfo {author}
  {\bibfnamefont {M.-C.}\ \bibnamefont {Marinica}}, \bibinfo {author}
  {\bibfnamefont {J.~M.}\ \bibnamefont {Perlado}},\ and\ \bibinfo {author}
  {\bibfnamefont {J.}~\bibnamefont {Marian}},\ }\bibfield  {title} {\bibinfo
  {title} {Assessment of interatomic potentials for atomistic analysis of
  static and dynamic properties of screw dislocations in {W}},\ }\href
  {https://doi.org/10.1088/0953-8984/25/8/085702} {\bibfield  {journal}
  {\bibinfo  {journal} {Journal of Physics: Condensed Matter}\ }\textbf
  {\bibinfo {volume} {25}},\ \bibinfo {pages} {085702} (\bibinfo {year}
  {2013})}\BibitemShut {NoStop}%
\bibitem [{\citenamefont {Trinkle}\ and\ \citenamefont
  {Woodward}(2005)}]{trinkle_2005_science}%
  \BibitemOpen
  \bibfield  {author} {\bibinfo {author} {\bibfnamefont {D.~R.}\ \bibnamefont
  {Trinkle}}\ and\ \bibinfo {author} {\bibfnamefont {C.}~\bibnamefont
  {Woodward}},\ }\bibfield  {title} {\bibinfo {title} {The chemistry of
  deformation: How solutes soften pure metals},\ }\href
  {https://doi.org/10.1126/science.1118616} {\bibfield  {journal} {\bibinfo
  {journal} {Science}\ }\textbf {\bibinfo {volume} {310}},\ \bibinfo {pages}
  {1665} (\bibinfo {year} {2005})}\BibitemShut {NoStop}%
\bibitem [{\citenamefont {Caillard}(2020)}]{caillard_2020_am}%
  \BibitemOpen
  \bibfield  {author} {\bibinfo {author} {\bibfnamefont {D.}~\bibnamefont
  {Caillard}},\ }\bibfield  {title} {\bibinfo {title} {A {TEM} in situ study of
  the softening of {Tungsten by Rhenium}},\ }\href
  {https://doi.org/10.1016/j.actamat.2020.04.039} {\bibfield  {journal}
  {\bibinfo  {journal} {Acta Materialia}\ }\textbf {\bibinfo {volume} {194}},\
  \bibinfo {pages} {249} (\bibinfo {year} {2020})}\BibitemShut {NoStop}%
\bibitem [{\citenamefont {Zhao}\ and\ \citenamefont
  {Marian}(2018)}]{zhao_2018_msmse}%
  \BibitemOpen
  \bibfield  {author} {\bibinfo {author} {\bibfnamefont {Y.}~\bibnamefont
  {Zhao}}\ and\ \bibinfo {author} {\bibfnamefont {J.}~\bibnamefont {Marian}},\
  }\bibfield  {title} {\bibinfo {title} {Direct prediction of the solute
  softening-to-hardening transition in {W--Re} alloys using stochastic
  simulations of screw dislocation motion},\ }\href
  {https://doi.org/10.1088/1361-651x/aaaecf} {\bibfield  {journal} {\bibinfo
  {journal} {Modelling and Simulation in Materials Science and Engineering}\
  }\textbf {\bibinfo {volume} {26}},\ \bibinfo {pages} {045002} (\bibinfo
  {year} {2018})}\BibitemShut {NoStop}%
\bibitem [{\citenamefont {Romaner}\ \emph {et~al.}(2010)\citenamefont
  {Romaner}, \citenamefont {Ambrosch-Draxl},\ and\ \citenamefont
  {Pippan}}]{romaner_2010_prl}%
  \BibitemOpen
  \bibfield  {author} {\bibinfo {author} {\bibfnamefont {L.}~\bibnamefont
  {Romaner}}, \bibinfo {author} {\bibfnamefont {C.}~\bibnamefont
  {Ambrosch-Draxl}},\ and\ \bibinfo {author} {\bibfnamefont {R.}~\bibnamefont
  {Pippan}},\ }\bibfield  {title} {\bibinfo {title} {Effect of rhenium on the
  dislocation core structure in tungsten},\ }\href
  {https://doi.org/10.1103/physrevlett.104.195503} {\bibfield  {journal}
  {\bibinfo  {journal} {Physical Review Letters}\ }\textbf {\bibinfo {volume}
  {104}},\ \bibinfo {pages} {195503} (\bibinfo {year} {2010})}\BibitemShut
  {NoStop}%
\bibitem [{\citenamefont {Stephens}(1970)}]{stephens_1970_mmtb}%
  \BibitemOpen
  \bibfield  {author} {\bibinfo {author} {\bibfnamefont {J.~R.}\ \bibnamefont
  {Stephens}},\ }\bibfield  {title} {\bibinfo {title} {Dislocation structures
  in single-crystal tungsten and tungsten alloys},\ }\href
  {https://doi.org/10.1007/bf02900246} {\bibfield  {journal} {\bibinfo
  {journal} {Metallurgical and Materials Transactions B}\ }\textbf {\bibinfo
  {volume} {1}},\ \bibinfo {pages} {1293} (\bibinfo {year} {1970})}\BibitemShut
  {NoStop}%
\bibitem [{\citenamefont {Li}\ \emph {et~al.}(2012)\citenamefont {Li},
  \citenamefont {Wurster}, \citenamefont {Motz}, \citenamefont {Romaner},
  \citenamefont {Ambrosch-Draxl},\ and\ \citenamefont {Pippan}}]{li_2012_am}%
  \BibitemOpen
  \bibfield  {author} {\bibinfo {author} {\bibfnamefont {H.}~\bibnamefont
  {Li}}, \bibinfo {author} {\bibfnamefont {S.}~\bibnamefont {Wurster}},
  \bibinfo {author} {\bibfnamefont {C.}~\bibnamefont {Motz}}, \bibinfo {author}
  {\bibfnamefont {L.}~\bibnamefont {Romaner}}, \bibinfo {author} {\bibfnamefont
  {C.}~\bibnamefont {Ambrosch-Draxl}},\ and\ \bibinfo {author} {\bibfnamefont
  {R.}~\bibnamefont {Pippan}},\ }\bibfield  {title} {\bibinfo {title}
  {Dislocation-core symmetry and slip planes in tungsten alloys: {Ab initio}
  calculations and microcantilever bending experiments},\ }\href
  {https://doi.org/10.1016/j.actamat.2011.10.031} {\bibfield  {journal}
  {\bibinfo  {journal} {Acta Materialia}\ }\textbf {\bibinfo {volume} {60}},\
  \bibinfo {pages} {748} (\bibinfo {year} {2012})}\BibitemShut {NoStop}%
\bibitem [{\citenamefont {Ren}\ \emph {et~al.}(2018)\citenamefont {Ren},
  \citenamefont {Fang}, \citenamefont {Koopman}, \citenamefont {Butler},
  \citenamefont {Paramore},\ and\ \citenamefont {Middlemas}}]{ren_2018_ijrmhm}%
  \BibitemOpen
  \bibfield  {author} {\bibinfo {author} {\bibfnamefont {C.}~\bibnamefont
  {Ren}}, \bibinfo {author} {\bibfnamefont {Z.}~\bibnamefont {Fang}}, \bibinfo
  {author} {\bibfnamefont {M.}~\bibnamefont {Koopman}}, \bibinfo {author}
  {\bibfnamefont {B.}~\bibnamefont {Butler}}, \bibinfo {author} {\bibfnamefont
  {J.}~\bibnamefont {Paramore}},\ and\ \bibinfo {author} {\bibfnamefont
  {S.}~\bibnamefont {Middlemas}},\ }\bibfield  {title} {\bibinfo {title}
  {Methods for improving ductility of tungsten{ - A} review},\ }\href
  {https://doi.org/10.1016/j.ijrmhm.2018.04.012} {\bibfield  {journal}
  {\bibinfo  {journal} {International Journal of Refractory Metals and Hard
  Materials}\ }\textbf {\bibinfo {volume} {75}},\ \bibinfo {pages} {170}
  (\bibinfo {year} {2018})}\BibitemShut {NoStop}%
\bibitem [{\citenamefont {Clouet}\ \emph {et~al.}(2009)\citenamefont {Clouet},
  \citenamefont {Ventelon},\ and\ \citenamefont {Willaime}}]{clouet_2009_prl}%
  \BibitemOpen
  \bibfield  {author} {\bibinfo {author} {\bibfnamefont {E.}~\bibnamefont
  {Clouet}}, \bibinfo {author} {\bibfnamefont {L.}~\bibnamefont {Ventelon}},\
  and\ \bibinfo {author} {\bibfnamefont {F.}~\bibnamefont {Willaime}},\
  }\bibfield  {title} {\bibinfo {title} {Dislocation core energies and core
  fields from first principles},\ }\href
  {http://link.aps.org/doi/10.1103/PhysRevLett.102.055502} {\bibfield
  {journal} {\bibinfo  {journal} {Physical Review Letters}\ }\textbf {\bibinfo
  {volume} {102}},\ \bibinfo {pages} {055502} (\bibinfo {year}
  {2009})}\BibitemShut {NoStop}%
\bibitem [{\citenamefont {Weinberger}\ \emph {et~al.}(2013)\citenamefont
  {Weinberger}, \citenamefont {Boyce},\ and\ \citenamefont
  {Battaile}}]{weinberger_2013_imr}%
  \BibitemOpen
  \bibfield  {author} {\bibinfo {author} {\bibfnamefont {C.~R.}\ \bibnamefont
  {Weinberger}}, \bibinfo {author} {\bibfnamefont {B.~L.}\ \bibnamefont
  {Boyce}},\ and\ \bibinfo {author} {\bibfnamefont {C.~C.}\ \bibnamefont
  {Battaile}},\ }\bibfield  {title} {\bibinfo {title} {Slip planes in bcc
  transition metals},\ }\href {https://doi.org/10.1179/1743280412Y.0000000015}
  {\bibfield  {journal} {\bibinfo  {journal} {International Materials Reviews}\
  }\textbf {\bibinfo {volume} {58}},\ \bibinfo {pages} {296} (\bibinfo {year}
  {2013})}\BibitemShut {NoStop}%
\bibitem [{\citenamefont {Dezerald}\ \emph {et~al.}(2014)\citenamefont
  {Dezerald}, \citenamefont {Ventelon}, \citenamefont {Clouet}, \citenamefont
  {Denoual}, \citenamefont {Rodney},\ and\ \citenamefont
  {Willaime}}]{dezerald_2014_prb}%
  \BibitemOpen
  \bibfield  {author} {\bibinfo {author} {\bibfnamefont {L.}~\bibnamefont
  {Dezerald}}, \bibinfo {author} {\bibfnamefont {L.}~\bibnamefont {Ventelon}},
  \bibinfo {author} {\bibfnamefont {E.}~\bibnamefont {Clouet}}, \bibinfo
  {author} {\bibfnamefont {C.}~\bibnamefont {Denoual}}, \bibinfo {author}
  {\bibfnamefont {D.}~\bibnamefont {Rodney}},\ and\ \bibinfo {author}
  {\bibfnamefont {F.}~\bibnamefont {Willaime}},\ }\bibfield  {title} {\bibinfo
  {title} {\textit{Ab initio} modeling of the two-dimensional energy landscape
  of screw dislocations in bcc transition metals},\ }\href
  {https://doi.org/10.1103/physrevb.89.024104} {\bibfield  {journal} {\bibinfo
  {journal} {Physical Review B}\ }\textbf {\bibinfo {volume} {89}},\ \bibinfo
  {pages} {024104} (\bibinfo {year} {2014})}\BibitemShut {NoStop}%
\bibitem [{\citenamefont {Kroupa}\ and\ \citenamefont
  {V{\'\i}tek}(1964)}]{kroupa_1964_cjpb}%
  \BibitemOpen
  \bibfield  {author} {\bibinfo {author} {\bibfnamefont {F.}~\bibnamefont
  {Kroupa}}\ and\ \bibinfo {author} {\bibfnamefont {V.}~\bibnamefont
  {V{\'\i}tek}},\ }\bibfield  {title} {\bibinfo {title} {Splitting of
  dislocations in b.c.c. metals on $\lbrace$110$\rbrace$ planes},\ }\href
  {https://doi.org/10.1007/bf01689142} {\bibfield  {journal} {\bibinfo
  {journal} {Czechoslovak Journal of Physics}\ }\textbf {\bibinfo {volume}
  {14}},\ \bibinfo {pages} {337} (\bibinfo {year} {1964})}\BibitemShut
  {NoStop}%
\bibitem [{\citenamefont {Anderson}\ \emph {et~al.}(2017)\citenamefont
  {Anderson}, \citenamefont {Hirth},\ and\ \citenamefont
  {Lothe}}]{anderson_2017_disl}%
  \BibitemOpen
  \bibfield  {author} {\bibinfo {author} {\bibfnamefont {P.~M.}\ \bibnamefont
  {Anderson}}, \bibinfo {author} {\bibfnamefont {J.~P.}\ \bibnamefont
  {Hirth}},\ and\ \bibinfo {author} {\bibfnamefont {J.}~\bibnamefont {Lothe}},\
  }\href {https://books.google.com.sg/books?id=LK7DDQAAQBAJ} {\emph {\bibinfo
  {title} {Theory of Dislocations}}}\ (\bibinfo  {publisher} {Cambridge
  University Press},\ \bibinfo {year} {2017})\BibitemShut {NoStop}%
\bibitem [{sup(2022)}]{supp_mat}%
  \BibitemOpen
  \href@noop {} {\bibinfo {title} {\text{M}ethods and \text{C}omputational
  \text{M}odels are available as {S}upplementary {M}aterials online}} (\bibinfo
  {year} {2022})\BibitemShut {NoStop}%
\bibitem [{\citenamefont {Rodney}\ \emph {et~al.}(2017)\citenamefont {Rodney},
  \citenamefont {Ventelon}, \citenamefont {Clouet}, \citenamefont
  {Pizzagalli},\ and\ \citenamefont {Willaime}}]{rodney_2017_am}%
  \BibitemOpen
  \bibfield  {author} {\bibinfo {author} {\bibfnamefont {D.}~\bibnamefont
  {Rodney}}, \bibinfo {author} {\bibfnamefont {L.}~\bibnamefont {Ventelon}},
  \bibinfo {author} {\bibfnamefont {E.}~\bibnamefont {Clouet}}, \bibinfo
  {author} {\bibfnamefont {L.}~\bibnamefont {Pizzagalli}},\ and\ \bibinfo
  {author} {\bibfnamefont {F.}~\bibnamefont {Willaime}},\ }\bibfield  {title}
  {\bibinfo {title} {Ab initio modeling of dislocation core properties in
  metals and semiconductors},\ }\href
  {https://doi.org/10.1016/j.actamat.2016.09.049} {\bibfield  {journal}
  {\bibinfo  {journal} {Acta Materialia}\ }\textbf {\bibinfo {volume} {124}},\
  \bibinfo {pages} {633} (\bibinfo {year} {2017})}\BibitemShut {NoStop}%
\bibitem [{\citenamefont {Fellinger}\ \emph {et~al.}(2010)\citenamefont
  {Fellinger}, \citenamefont {Park},\ and\ \citenamefont
  {Wilkins}}]{fellinger_2010_prb}%
  \BibitemOpen
  \bibfield  {author} {\bibinfo {author} {\bibfnamefont {M.~R.}\ \bibnamefont
  {Fellinger}}, \bibinfo {author} {\bibfnamefont {H.}~\bibnamefont {Park}},\
  and\ \bibinfo {author} {\bibfnamefont {J.~W.}\ \bibnamefont {Wilkins}},\
  }\bibfield  {title} {\bibinfo {title} {Force-matched embedded-atom method
  potential for niobium},\ }\href
  {https://link.aps.org/doi/10.1103/PhysRevB.81.144119} {\bibfield  {journal}
  {\bibinfo  {journal} {Physical Review B}\ }\textbf {\bibinfo {volume} {81}},\
  \bibinfo {pages} {144119} (\bibinfo {year} {2010})}\BibitemShut {NoStop}%
\bibitem [{\citenamefont {Setyawan}\ \emph {et~al.}(2018)\citenamefont
  {Setyawan}, \citenamefont {Gao},\ and\ \citenamefont
  {Kurtz}}]{setyawan_2018_jap}%
  \BibitemOpen
  \bibfield  {author} {\bibinfo {author} {\bibfnamefont {W.}~\bibnamefont
  {Setyawan}}, \bibinfo {author} {\bibfnamefont {N.}~\bibnamefont {Gao}},\ and\
  \bibinfo {author} {\bibfnamefont {R.~J.}\ \bibnamefont {Kurtz}},\ }\bibfield
  {title} {\bibinfo {title} {A tungsten-rhenium interatomic potential for point
  defect studies},\ }\href {https://doi.org/10.1063/1.5030113} {\bibfield
  {journal} {\bibinfo  {journal} {Journal of Applied Physics}\ }\textbf
  {\bibinfo {volume} {123}},\ \bibinfo {pages} {205102} (\bibinfo {year}
  {2018})}\BibitemShut {NoStop}%
\bibitem [{\citenamefont {Dragoni}\ \emph {et~al.}(2018)\citenamefont
  {Dragoni}, \citenamefont {Daff}, \citenamefont {Cs{\'a}nyi},\ and\
  \citenamefont {Marzari}}]{dragoni_2018_prm}%
  \BibitemOpen
  \bibfield  {author} {\bibinfo {author} {\bibfnamefont {D.}~\bibnamefont
  {Dragoni}}, \bibinfo {author} {\bibfnamefont {T.~D.}\ \bibnamefont {Daff}},
  \bibinfo {author} {\bibfnamefont {G.}~\bibnamefont {Cs{\'a}nyi}},\ and\
  \bibinfo {author} {\bibfnamefont {N.}~\bibnamefont {Marzari}},\ }\bibfield
  {title} {\bibinfo {title} {Achieving {DFT} accuracy with a machine-learning
  interatomic potential{: T}hermomechanics and defects in bcc ferromagnetic
  iron},\ }\href {https://doi.org/10.1103/PhysRevMaterials.2.013808} {\bibfield
   {journal} {\bibinfo  {journal} {Physical Review Materials}\ }\textbf
  {\bibinfo {volume} {2}},\ \bibinfo {pages} {013808} (\bibinfo {year}
  {2018})}\BibitemShut {NoStop}%
\bibitem [{\citenamefont {Asadi}\ \emph {et~al.}(2015)\citenamefont {Asadi},
  \citenamefont {Zaeem}, \citenamefont {Nouranian},\ and\ \citenamefont
  {Baskes}}]{asadi_2015_prb}%
  \BibitemOpen
  \bibfield  {author} {\bibinfo {author} {\bibfnamefont {E.}~\bibnamefont
  {Asadi}}, \bibinfo {author} {\bibfnamefont {M.~A.}\ \bibnamefont {Zaeem}},
  \bibinfo {author} {\bibfnamefont {S.}~\bibnamefont {Nouranian}},\ and\
  \bibinfo {author} {\bibfnamefont {M.~I.}\ \bibnamefont {Baskes}},\ }\bibfield
   {title} {\bibinfo {title} {Quantitative modeling of the equilibration of
  two-phase solid-liquid {Fe} by atomistic simulations on diffusive time
  scales},\ }\href {https://doi.org/10.1103/physrevb.91.024105} {\bibfield
  {journal} {\bibinfo  {journal} {Physical Review B}\ }\textbf {\bibinfo
  {volume} {91}},\ \bibinfo {pages} {024105} (\bibinfo {year}
  {2015})}\BibitemShut {NoStop}%
\bibitem [{\citenamefont {Nordheim}(1931)}]{nordheim_1931_ap}%
  \BibitemOpen
  \bibfield  {author} {\bibinfo {author} {\bibfnamefont {L.}~\bibnamefont
  {Nordheim}},\ }\bibfield  {title} {\bibinfo {title} {Zur elektronentheorie
  der metalle{. I}},\ }\href {https://doi.org/10.1002/andp.19314010507}
  {\bibfield  {journal} {\bibinfo  {journal} {Annalen der Physik}\ }\textbf
  {\bibinfo {volume} {401}},\ \bibinfo {pages} {607} (\bibinfo {year}
  {1931})}\BibitemShut {NoStop}%
\bibitem [{\citenamefont {Bellaiche}\ and\ \citenamefont
  {Vanderbilt}(2000)}]{bellaiche_2000_prb}%
  \BibitemOpen
  \bibfield  {author} {\bibinfo {author} {\bibfnamefont {L.}~\bibnamefont
  {Bellaiche}}\ and\ \bibinfo {author} {\bibfnamefont {D.}~\bibnamefont
  {Vanderbilt}},\ }\bibfield  {title} {\bibinfo {title} {Virtual crystal
  approximation revisited{: A}pplication to dielectric and piezoelectric
  properties of perovskites},\ }\href
  {https://doi.org/10.1103/physrevb.61.7877} {\bibfield  {journal} {\bibinfo
  {journal} {Physical Review B}\ }\textbf {\bibinfo {volume} {61}},\ \bibinfo
  {pages} {7877} (\bibinfo {year} {2000})}\BibitemShut {NoStop}%
\bibitem [{\citenamefont {Stephens}\ and\ \citenamefont
  {Witzke}(1972)}]{stephens_1972_jlcm}%
  \BibitemOpen
  \bibfield  {author} {\bibinfo {author} {\bibfnamefont {J.~R.}\ \bibnamefont
  {Stephens}}\ and\ \bibinfo {author} {\bibfnamefont {W.~R.}\ \bibnamefont
  {Witzke}},\ }\bibfield  {title} {\bibinfo {title} {Alloy hardening and
  softening in binary molybdenum alloys as related to electron concentration},\
  }\href {https://doi.org/10.1016/0022-5088(72)90201-9} {\bibfield  {journal}
  {\bibinfo  {journal} {Journal of the Less Common Metals}\ }\textbf {\bibinfo
  {volume} {29}},\ \bibinfo {pages} {371} (\bibinfo {year} {1972})}\BibitemShut
  {NoStop}%
\bibitem [{\citenamefont {Klopp}(1975)}]{klopp_1975_jlcm}%
  \BibitemOpen
  \bibfield  {author} {\bibinfo {author} {\bibfnamefont {W.~D.}\ \bibnamefont
  {Klopp}},\ }\bibfield  {title} {\bibinfo {title} {A review of chromium,
  molybdenum, and tungsten alloys},\ }\href
  {https://doi.org/10.1016/0022-5088(75)90046-6} {\bibfield  {journal}
  {\bibinfo  {journal} {Journal of the Less Common Metals}\ }\textbf {\bibinfo
  {volume} {42}},\ \bibinfo {pages} {261} (\bibinfo {year} {1975})}\BibitemShut
  {NoStop}%
\bibitem [{\citenamefont {Guo}\ \emph {et~al.}(2011)\citenamefont {Guo},
  \citenamefont {Ng}, \citenamefont {Lu},\ and\ \citenamefont
  {Liu}}]{guo_2011_jap}%
  \BibitemOpen
  \bibfield  {author} {\bibinfo {author} {\bibfnamefont {S.}~\bibnamefont
  {Guo}}, \bibinfo {author} {\bibfnamefont {C.}~\bibnamefont {Ng}}, \bibinfo
  {author} {\bibfnamefont {J.}~\bibnamefont {Lu}},\ and\ \bibinfo {author}
  {\bibfnamefont {C.~T.}\ \bibnamefont {Liu}},\ }\bibfield  {title} {\bibinfo
  {title} {Effect of valence electron concentration on stability of fcc or bcc
  phase in high entropy alloys},\ }\href {https://doi.org/10.1063/1.3587228}
  {\bibfield  {journal} {\bibinfo  {journal} {Journal of Applied Physics}\
  }\textbf {\bibinfo {volume} {109}},\ \bibinfo {pages} {103505} (\bibinfo
  {year} {2011})}\BibitemShut {NoStop}%
\bibitem [{\citenamefont {Hu}\ \emph {et~al.}(2017)\citenamefont {Hu},
  \citenamefont {Fellinger}, \citenamefont {Butler}, \citenamefont {Wang},
  \citenamefont {Darling}, \citenamefont {Kecskes}, \citenamefont {Trinkle},\
  and\ \citenamefont {Liu}}]{hu_2017_am}%
  \BibitemOpen
  \bibfield  {author} {\bibinfo {author} {\bibfnamefont {Y.-J.}\ \bibnamefont
  {Hu}}, \bibinfo {author} {\bibfnamefont {M.~R.}\ \bibnamefont {Fellinger}},
  \bibinfo {author} {\bibfnamefont {B.~G.}\ \bibnamefont {Butler}}, \bibinfo
  {author} {\bibfnamefont {Y.}~\bibnamefont {Wang}}, \bibinfo {author}
  {\bibfnamefont {K.~A.}\ \bibnamefont {Darling}}, \bibinfo {author}
  {\bibfnamefont {L.~J.}\ \bibnamefont {Kecskes}}, \bibinfo {author}
  {\bibfnamefont {D.~R.}\ \bibnamefont {Trinkle}},\ and\ \bibinfo {author}
  {\bibfnamefont {Z.-K.}\ \bibnamefont {Liu}},\ }\bibfield  {title} {\bibinfo
  {title} {Solute-induced solid-solution softening and hardening in bcc
  tungsten},\ }\href {https://doi.org/10.1016/j.actamat.2017.09.019} {\bibfield
   {journal} {\bibinfo  {journal} {Acta Materialia}\ }\textbf {\bibinfo
  {volume} {141}},\ \bibinfo {pages} {304} (\bibinfo {year}
  {2017})}\BibitemShut {NoStop}%
\bibitem [{\citenamefont {Wurster}\ \emph {et~al.}(2011)\citenamefont
  {Wurster}, \citenamefont {Gludovatz}, \citenamefont {Hoffmann},\ and\
  \citenamefont {Pippan}}]{wurster_2011_jnm}%
  \BibitemOpen
  \bibfield  {author} {\bibinfo {author} {\bibfnamefont {S.}~\bibnamefont
  {Wurster}}, \bibinfo {author} {\bibfnamefont {B.}~\bibnamefont {Gludovatz}},
  \bibinfo {author} {\bibfnamefont {A.}~\bibnamefont {Hoffmann}},\ and\
  \bibinfo {author} {\bibfnamefont {R.}~\bibnamefont {Pippan}},\ }\bibfield
  {title} {\bibinfo {title} {Fracture behaviour of tungsten--vanadium and
  tungsten--tantalum alloys and composites},\ }\href
  {https://doi.org/10.1016/j.jnucmat.2011.04.025} {\bibfield  {journal}
  {\bibinfo  {journal} {Journal of Nuclear Materials}\ }\textbf {\bibinfo
  {volume} {413}},\ \bibinfo {pages} {166} (\bibinfo {year}
  {2011})}\BibitemShut {NoStop}%
\bibitem [{\citenamefont {Massalski}\ \emph {et~al.}(1990)\citenamefont
  {Massalski}, \citenamefont {Okamoto}, \citenamefont {Subramanian},\ and\
  \citenamefont {Kacprzak}}]{thaddeus_1986_bapd}%
  \BibitemOpen
  \bibfield  {author} {\bibinfo {author} {\bibfnamefont {T.~B.}\ \bibnamefont
  {Massalski}}, \bibinfo {author} {\bibfnamefont {H.}~\bibnamefont {Okamoto}},
  \bibinfo {author} {\bibfnamefont {P.}~\bibnamefont {Subramanian}},\ and\
  \bibinfo {author} {\bibfnamefont {L.}~\bibnamefont {Kacprzak}},\ }\href@noop
  {} {\emph {\bibinfo {title} {Binary Alloy Phase Diagrams}}},\ \bibinfo
  {edition} {2nd}\ ed.\ (\bibinfo  {publisher} {Materials Park, Ohio, ASM
  International},\ \bibinfo {year} {1990})\BibitemShut {NoStop}%
\bibitem [{\citenamefont {Medvedeva}\ \emph {et~al.}(2007)\citenamefont
  {Medvedeva}, \citenamefont {Gornostyrev},\ and\ \citenamefont
  {Freeman}}]{medvedeva_2007_prb}%
  \BibitemOpen
  \bibfield  {author} {\bibinfo {author} {\bibfnamefont {N.~I.}\ \bibnamefont
  {Medvedeva}}, \bibinfo {author} {\bibfnamefont {Y.~N.}\ \bibnamefont
  {Gornostyrev}},\ and\ \bibinfo {author} {\bibfnamefont {A.~J.}\ \bibnamefont
  {Freeman}},\ }\bibfield  {title} {\bibinfo {title} {Solid solution softening
  and hardening in the {group-V} and {group-VI} bcc transition metals alloys{:
  F}irst principles calculations and atomistic modeling},\ }\href
  {https://doi.org/10.1103/physrevb.76.212104} {\bibfield  {journal} {\bibinfo
  {journal} {Physical Review B}\ }\textbf {\bibinfo {volume} {76}},\ \bibinfo
  {pages} {212104} (\bibinfo {year} {2007})}\BibitemShut {NoStop}%
\bibitem [{\citenamefont {Zhang}\ \emph {et~al.}(2015)\citenamefont {Zhang},
  \citenamefont {Tang}, \citenamefont {Deng}, \citenamefont {Deng},
  \citenamefont {Xiao},\ and\ \citenamefont {Hu}}]{zhang_2015_sm}%
  \BibitemOpen
  \bibfield  {author} {\bibinfo {author} {\bibfnamefont {X.}~\bibnamefont
  {Zhang}}, \bibinfo {author} {\bibfnamefont {J.}~\bibnamefont {Tang}},
  \bibinfo {author} {\bibfnamefont {L.}~\bibnamefont {Deng}}, \bibinfo {author}
  {\bibfnamefont {H.}~\bibnamefont {Deng}}, \bibinfo {author} {\bibfnamefont
  {S.}~\bibnamefont {Xiao}},\ and\ \bibinfo {author} {\bibfnamefont
  {W.}~\bibnamefont {Hu}},\ }\bibfield  {title} {\bibinfo {title} {{Effects of
  solute size on solid-solution hardening in vanadium alloys{: A}
  first-principles calculation}},\ }\href
  {https://doi.org/10.1016/j.scriptamat.2015.01.006} {\bibfield  {journal}
  {\bibinfo  {journal} {Scripta Materialia}\ }\textbf {\bibinfo {volume}
  {100}},\ \bibinfo {pages} {106} (\bibinfo {year} {2015})}\BibitemShut
  {NoStop}%
\bibitem [{\citenamefont {Fraser}\ and\ \citenamefont
  {Lund}(1962)}]{fraser_1962_cmq}%
  \BibitemOpen
  \bibfield  {author} {\bibinfo {author} {\bibfnamefont {R.~W.}\ \bibnamefont
  {Fraser}}\ and\ \bibinfo {author} {\bibfnamefont {J.~A.}\ \bibnamefont
  {Lund}},\ }\bibfield  {title} {\bibinfo {title} {{Effect of Titanium
  Additions on the Low-Temperature Behaviour of Vanadium}},\ }\href
  {https://doi.org/10.1179/cmq.1962.1.1.1} {\bibfield  {journal} {\bibinfo
  {journal} {Canadian Metallurgical Quarterly}\ }\textbf {\bibinfo {volume}
  {1}},\ \bibinfo {pages} {1} (\bibinfo {year} {1962})}\BibitemShut {NoStop}%
\bibitem [{\citenamefont {Zhang}\ \emph {et~al.}(2009)\citenamefont {Zhang},
  \citenamefont {Ganeev}, \citenamefont {Wang}, \citenamefont {Liu},\ and\
  \citenamefont {Alexandrov}}]{zhang_2009_msea}%
  \BibitemOpen
  \bibfield  {author} {\bibinfo {author} {\bibfnamefont {Y.}~\bibnamefont
  {Zhang}}, \bibinfo {author} {\bibfnamefont {A.~V.}\ \bibnamefont {Ganeev}},
  \bibinfo {author} {\bibfnamefont {J.~T.}\ \bibnamefont {Wang}}, \bibinfo
  {author} {\bibfnamefont {J.~Q.}\ \bibnamefont {Liu}},\ and\ \bibinfo {author}
  {\bibfnamefont {I.~V.}\ \bibnamefont {Alexandrov}},\ }\bibfield  {title}
  {\bibinfo {title} {Observations on the ductile-to-brittle transition in
  ultrafine-grained tungsten of commercial purity},\ }\href
  {https://doi.org/10.1016/j.msea.2008.07.074} {\bibfield  {journal} {\bibinfo
  {journal} {Materials Science and Engineering: A}\ }\textbf {\bibinfo {volume}
  {503}},\ \bibinfo {pages} {37} (\bibinfo {year} {2009})}\BibitemShut
  {NoStop}%
\bibitem [{\citenamefont {Yin}\ \emph {et~al.}(2020)\citenamefont {Yin},
  \citenamefont {Ding}, \citenamefont {Asta},\ and\ \citenamefont
  {Ritchie}}]{yin_2020_npjcm}%
  \BibitemOpen
  \bibfield  {author} {\bibinfo {author} {\bibfnamefont {S.}~\bibnamefont
  {Yin}}, \bibinfo {author} {\bibfnamefont {J.}~\bibnamefont {Ding}}, \bibinfo
  {author} {\bibfnamefont {M.}~\bibnamefont {Asta}},\ and\ \bibinfo {author}
  {\bibfnamefont {R.~O.}\ \bibnamefont {Ritchie}},\ }\bibfield  {title}
  {\bibinfo {title} {Ab initio modeling of the energy landscape for screw
  dislocations in body-centered cubic high-entropy alloys},\ }\bibfield
  {journal} {\bibinfo  {journal} {npj Computational Materials}\ }\textbf
  {\bibinfo {volume} {6}},\ \href {https://doi.org/10.1038/s41524-020-00377-5}
  {10.1038/s41524-020-00377-5} (\bibinfo {year} {2020})\BibitemShut {NoStop}%
\bibitem [{\citenamefont {An}\ \emph {et~al.}(2021)\citenamefont {An},
  \citenamefont {Mao}, \citenamefont {Yang}, \citenamefont {Liu}, \citenamefont
  {Zhang}, \citenamefont {Ma}, \citenamefont {Zhou}, \citenamefont {Zhang},
  \citenamefont {Wang},\ and\ \citenamefont {Han}}]{an_2021_mh}%
  \BibitemOpen
  \bibfield  {author} {\bibinfo {author} {\bibfnamefont {Z.}~\bibnamefont
  {An}}, \bibinfo {author} {\bibfnamefont {S.}~\bibnamefont {Mao}}, \bibinfo
  {author} {\bibfnamefont {T.}~\bibnamefont {Yang}}, \bibinfo {author}
  {\bibfnamefont {C.~T.}\ \bibnamefont {Liu}}, \bibinfo {author} {\bibfnamefont
  {B.}~\bibnamefont {Zhang}}, \bibinfo {author} {\bibfnamefont
  {E.}~\bibnamefont {Ma}}, \bibinfo {author} {\bibfnamefont {H.}~\bibnamefont
  {Zhou}}, \bibinfo {author} {\bibfnamefont {Z.}~\bibnamefont {Zhang}},
  \bibinfo {author} {\bibfnamefont {L.}~\bibnamefont {Wang}},\ and\ \bibinfo
  {author} {\bibfnamefont {X.}~\bibnamefont {Han}},\ }\bibfield  {title}
  {\bibinfo {title} {Spinodal-modulated solid solution delivers a strong and
  ductile refractory high-entropy alloy},\ }\href
  {https://doi.org/10.1039/d0mh01341b} {\bibfield  {journal} {\bibinfo
  {journal} {Materials Horizons}\ }\textbf {\bibinfo {volume} {8}},\ \bibinfo
  {pages} {948} (\bibinfo {year} {2021})}\BibitemShut {NoStop}%
\bibitem [{\citenamefont {Ghafarollahi}\ and\ \citenamefont
  {Curtin}(2021)}]{ghafarollahi_2021_am}%
  \BibitemOpen
  \bibfield  {author} {\bibinfo {author} {\bibfnamefont {A.}~\bibnamefont
  {Ghafarollahi}}\ and\ \bibinfo {author} {\bibfnamefont {W.}~\bibnamefont
  {Curtin}},\ }\bibfield  {title} {\bibinfo {title} {Theory of kink migration
  in dilute {BCC} alloys},\ }\href
  {https://doi.org/10.1016/j.actamat.2021.117078} {\bibfield  {journal}
  {\bibinfo  {journal} {Acta Materialia}\ }\textbf {\bibinfo {volume} {215}},\
  \bibinfo {pages} {117078} (\bibinfo {year} {2021})}\BibitemShut {NoStop}%
\bibitem [{\citenamefont {Romaner}\ \emph {et~al.}(2014)\citenamefont
  {Romaner}, \citenamefont {Razumovskiy},\ and\ \citenamefont
  {Pippan}}]{romaner_2014_pml}%
  \BibitemOpen
  \bibfield  {author} {\bibinfo {author} {\bibfnamefont {L.}~\bibnamefont
  {Romaner}}, \bibinfo {author} {\bibfnamefont {V.}~\bibnamefont
  {Razumovskiy}},\ and\ \bibinfo {author} {\bibfnamefont {R.}~\bibnamefont
  {Pippan}},\ }\bibfield  {title} {\bibinfo {title} {Core polarity of screw
  dislocations in {Fe--Co} alloys},\ }\href
  {https://doi.org/10.1080/09500839.2014.904055} {\bibfield  {journal}
  {\bibinfo  {journal} {Philosophical Magazine Letters}\ }\textbf {\bibinfo
  {volume} {94}},\ \bibinfo {pages} {334} (\bibinfo {year} {2014})}\BibitemShut
  {NoStop}%
\bibitem [{\citenamefont {Kresse}\ and\ \citenamefont
  {Furthm{\"u}ller}(1996)}]{kresse_1996_prb}%
  \BibitemOpen
  \bibfield  {author} {\bibinfo {author} {\bibfnamefont {G.}~\bibnamefont
  {Kresse}}\ and\ \bibinfo {author} {\bibfnamefont {J.}~\bibnamefont
  {Furthm{\"u}ller}},\ }\bibfield  {title} {\bibinfo {title} {Efficient
  iterative schemes forab initiototal-energy calculations using a plane-wave
  basis set},\ }\href {https://doi.org/10.1103/physrevb.54.11169} {\bibfield
  {journal} {\bibinfo  {journal} {Physical Review B}\ }\textbf {\bibinfo
  {volume} {54}},\ \bibinfo {pages} {11169} (\bibinfo {year}
  {1996})}\BibitemShut {NoStop}%
\bibitem [{\citenamefont {Kresse}\ and\ \citenamefont
  {Joubert}(1999)}]{kresse_1999_prb}%
  \BibitemOpen
  \bibfield  {author} {\bibinfo {author} {\bibfnamefont {G.}~\bibnamefont
  {Kresse}}\ and\ \bibinfo {author} {\bibfnamefont {D.}~\bibnamefont
  {Joubert}},\ }\bibfield  {title} {\bibinfo {title} {From ultrasoft
  pseudopotentials to the projector augmented-wave method},\ }\href
  {https://doi.org/10.1103/PhysRevB.59.1758} {\bibfield  {journal} {\bibinfo
  {journal} {Physical Review B}\ }\textbf {\bibinfo {volume} {59}},\ \bibinfo
  {pages} {1758} (\bibinfo {year} {1999})}\BibitemShut {NoStop}%
\bibitem [{\citenamefont {Giannozzi}\ \emph {et~al.}(2009)\citenamefont
  {Giannozzi}, \citenamefont {Baroni}, \citenamefont {Bonini}, \citenamefont
  {Calandra}, \citenamefont {Car}, \citenamefont {Cavazzoni}, \citenamefont
  {Ceresoli}, \citenamefont {Chiarotti}, \citenamefont {Cococcioni},
  \citenamefont {Dabo}, \citenamefont {Corso}, \citenamefont {de~Gironcoli},
  \citenamefont {Fabris}, \citenamefont {Fratesi}, \citenamefont {Gebauer},
  \citenamefont {Gerstmann}, \citenamefont {Gougoussis}, \citenamefont
  {Kokalj}, \citenamefont {Lazzeri}, \citenamefont {Martin-Samos},
  \citenamefont {Marzari}, \citenamefont {Mauri}, \citenamefont {Mazzarello},
  \citenamefont {Paolini}, \citenamefont {Pasquarello}, \citenamefont
  {Paulatto}, \citenamefont {Sbraccia}, \citenamefont {Scandolo}, \citenamefont
  {Sclauzero}, \citenamefont {Seitsonen}, \citenamefont {Smogunov},
  \citenamefont {Umari},\ and\ \citenamefont
  {Wentzcovitch}}]{giannozzi_2009_jpcm}%
  \BibitemOpen
  \bibfield  {author} {\bibinfo {author} {\bibfnamefont {P.}~\bibnamefont
  {Giannozzi}}, \bibinfo {author} {\bibfnamefont {S.}~\bibnamefont {Baroni}},
  \bibinfo {author} {\bibfnamefont {N.}~\bibnamefont {Bonini}}, \bibinfo
  {author} {\bibfnamefont {M.}~\bibnamefont {Calandra}}, \bibinfo {author}
  {\bibfnamefont {R.}~\bibnamefont {Car}}, \bibinfo {author} {\bibfnamefont
  {C.}~\bibnamefont {Cavazzoni}}, \bibinfo {author} {\bibfnamefont
  {D.}~\bibnamefont {Ceresoli}}, \bibinfo {author} {\bibfnamefont {G.~L.}\
  \bibnamefont {Chiarotti}}, \bibinfo {author} {\bibfnamefont {M.}~\bibnamefont
  {Cococcioni}}, \bibinfo {author} {\bibfnamefont {I.}~\bibnamefont {Dabo}},
  \bibinfo {author} {\bibfnamefont {A.~D.}\ \bibnamefont {Corso}}, \bibinfo
  {author} {\bibfnamefont {S.}~\bibnamefont {de~Gironcoli}}, \bibinfo {author}
  {\bibfnamefont {S.}~\bibnamefont {Fabris}}, \bibinfo {author} {\bibfnamefont
  {G.}~\bibnamefont {Fratesi}}, \bibinfo {author} {\bibfnamefont
  {R.}~\bibnamefont {Gebauer}}, \bibinfo {author} {\bibfnamefont
  {U.}~\bibnamefont {Gerstmann}}, \bibinfo {author} {\bibfnamefont
  {C.}~\bibnamefont {Gougoussis}}, \bibinfo {author} {\bibfnamefont
  {A.}~\bibnamefont {Kokalj}}, \bibinfo {author} {\bibfnamefont
  {M.}~\bibnamefont {Lazzeri}}, \bibinfo {author} {\bibfnamefont
  {L.}~\bibnamefont {Martin-Samos}}, \bibinfo {author} {\bibfnamefont
  {N.}~\bibnamefont {Marzari}}, \bibinfo {author} {\bibfnamefont
  {F.}~\bibnamefont {Mauri}}, \bibinfo {author} {\bibfnamefont
  {R.}~\bibnamefont {Mazzarello}}, \bibinfo {author} {\bibfnamefont
  {S.}~\bibnamefont {Paolini}}, \bibinfo {author} {\bibfnamefont
  {A.}~\bibnamefont {Pasquarello}}, \bibinfo {author} {\bibfnamefont
  {L.}~\bibnamefont {Paulatto}}, \bibinfo {author} {\bibfnamefont
  {C.}~\bibnamefont {Sbraccia}}, \bibinfo {author} {\bibfnamefont
  {S.}~\bibnamefont {Scandolo}}, \bibinfo {author} {\bibfnamefont
  {G.}~\bibnamefont {Sclauzero}}, \bibinfo {author} {\bibfnamefont {A.~P.}\
  \bibnamefont {Seitsonen}}, \bibinfo {author} {\bibfnamefont {A.}~\bibnamefont
  {Smogunov}}, \bibinfo {author} {\bibfnamefont {P.}~\bibnamefont {Umari}},\
  and\ \bibinfo {author} {\bibfnamefont {R.~M.}\ \bibnamefont {Wentzcovitch}},\
  }\bibfield  {title} {\bibinfo {title} {{QUANTUM ESPRESSO}: a modular and
  open-source software project for quantum simulations of materials},\ }\href
  {https://doi.org/10.1088/0953-8984/21/39/395502} {\bibfield  {journal}
  {\bibinfo  {journal} {Journal of Physics: Condensed Matter}\ }\textbf
  {\bibinfo {volume} {21}},\ \bibinfo {pages} {395502} (\bibinfo {year}
  {2009})}\BibitemShut {NoStop}%
\bibitem [{\citenamefont {Perdew}\ \emph {et~al.}(1996)\citenamefont {Perdew},
  \citenamefont {Burke},\ and\ \citenamefont {Ernzerhof}}]{perdew_1996_prl}%
  \BibitemOpen
  \bibfield  {author} {\bibinfo {author} {\bibfnamefont {J.~P.}\ \bibnamefont
  {Perdew}}, \bibinfo {author} {\bibfnamefont {K.}~\bibnamefont {Burke}},\ and\
  \bibinfo {author} {\bibfnamefont {M.}~\bibnamefont {Ernzerhof}},\ }\bibfield
  {title} {\bibinfo {title} {Generalized gradient approximation made simple},\
  }\href {https://doi.org/10.1103/physrevlett.77.3865} {\bibfield  {journal}
  {\bibinfo  {journal} {Physical Review Letters}\ }\textbf {\bibinfo {volume}
  {77}},\ \bibinfo {pages} {3865} (\bibinfo {year} {1996})}\BibitemShut
  {NoStop}%
\bibitem [{\citenamefont {Blöchl}(1994)}]{blochl_1994_prb}%
  \BibitemOpen
  \bibfield  {author} {\bibinfo {author} {\bibfnamefont {P.~E.}\ \bibnamefont
  {Blöchl}},\ }\bibfield  {title} {\bibinfo {title} {Projector augmented-wave
  method},\ }\href {https://doi.org/10.1103/PhysRevB.50.17953} {\bibfield
  {journal} {\bibinfo  {journal} {Physical Review B}\ }\textbf {\bibinfo
  {volume} {50}},\ \bibinfo {pages} {17953} (\bibinfo {year}
  {1994})}\BibitemShut {NoStop}%
\bibitem [{\citenamefont {Methfessel}\ and\ \citenamefont
  {Paxton}(1989)}]{methfessel_1989_prb}%
  \BibitemOpen
  \bibfield  {author} {\bibinfo {author} {\bibfnamefont {M.}~\bibnamefont
  {Methfessel}}\ and\ \bibinfo {author} {\bibfnamefont {A.~T.}\ \bibnamefont
  {Paxton}},\ }\bibfield  {title} {\bibinfo {title} {High-precision sampling
  for {Brillouin-zone} integration in metals},\ }\href
  {https://doi.org/10.1103/physrevb.40.3616} {\bibfield  {journal} {\bibinfo
  {journal} {Physical Review B}\ }\textbf {\bibinfo {volume} {40}},\ \bibinfo
  {pages} {3616} (\bibinfo {year} {1989})}\BibitemShut {NoStop}%
\bibitem [{\citenamefont {Monkhorst}\ and\ \citenamefont
  {Pack}(1976)}]{monkhorst_1976_prb}%
  \BibitemOpen
  \bibfield  {author} {\bibinfo {author} {\bibfnamefont {H.~J.}\ \bibnamefont
  {Monkhorst}}\ and\ \bibinfo {author} {\bibfnamefont {J.~D.}\ \bibnamefont
  {Pack}},\ }\bibfield  {title} {\bibinfo {title} {Special points for
  brillouin-zone integrations},\ }\href
  {http://link.aps.org/doi/10.1103/PhysRevB.13.5188} {\bibfield  {journal}
  {\bibinfo  {journal} {Physical Review B}\ }\textbf {\bibinfo {volume} {13}},\
  \bibinfo {pages} {5188} (\bibinfo {year} {1976})}\BibitemShut {NoStop}%
\bibitem [{\citenamefont {Plimpton}(1995)}]{plimpton_1995_jcp}%
  \BibitemOpen
  \bibfield  {author} {\bibinfo {author} {\bibfnamefont {S.}~\bibnamefont
  {Plimpton}},\ }\bibfield  {title} {\bibinfo {title} {Fast parallel algorithms
  for short-range molecular dynamics},\ }\href
  {https://doi.org/10.1006/jcph.1995.1039} {\bibfield  {journal} {\bibinfo
  {journal} {Journal of Computational Physics}\ }\textbf {\bibinfo {volume}
  {117}},\ \bibinfo {pages} {1} (\bibinfo {year} {1995})}\BibitemShut {NoStop}%
\bibitem [{\citenamefont {Bart{\'o}k}\ \emph {et~al.}(2010)\citenamefont
  {Bart{\'o}k}, \citenamefont {Payne}, \citenamefont {Kondor},\ and\
  \citenamefont {Cs{\'a}nyi}}]{bartok_2010_prl}%
  \BibitemOpen
  \bibfield  {author} {\bibinfo {author} {\bibfnamefont {A.~P.}\ \bibnamefont
  {Bart{\'o}k}}, \bibinfo {author} {\bibfnamefont {M.~C.}\ \bibnamefont
  {Payne}}, \bibinfo {author} {\bibfnamefont {R.}~\bibnamefont {Kondor}},\ and\
  \bibinfo {author} {\bibfnamefont {G.}~\bibnamefont {Cs{\'a}nyi}},\ }\bibfield
   {title} {\bibinfo {title} {{Gaussian Approximation Potentials: T}he accuracy
  of quantum mechanics, without the electrons},\ }\href
  {https://doi.org/10.1103/physrevlett.104.136403} {\bibfield  {journal}
  {\bibinfo  {journal} {Physical Review Letters}\ }\textbf {\bibinfo {volume}
  {104}},\ \bibinfo {pages} {136403} (\bibinfo {year} {2010})}\BibitemShut
  {NoStop}%
\bibitem [{\citenamefont {Bart{\'o}k}\ \emph {et~al.}(2013)\citenamefont
  {Bart{\'o}k}, \citenamefont {Kondor},\ and\ \citenamefont
  {Cs{\'a}nyi}}]{bartok_2013_prb}%
  \BibitemOpen
  \bibfield  {author} {\bibinfo {author} {\bibfnamefont {A.~P.}\ \bibnamefont
  {Bart{\'o}k}}, \bibinfo {author} {\bibfnamefont {R.}~\bibnamefont {Kondor}},\
  and\ \bibinfo {author} {\bibfnamefont {G.}~\bibnamefont {Cs{\'a}nyi}},\
  }\bibfield  {title} {\bibinfo {title} {On representing chemical
  environments},\ }\href {https://doi.org/10.1103/physrevb.87.184115}
  {\bibfield  {journal} {\bibinfo  {journal} {Physical Review B}\ }\textbf
  {\bibinfo {volume} {87}},\ \bibinfo {pages} {184115} (\bibinfo {year}
  {2013})}\BibitemShut {NoStop}%
\bibitem [{\citenamefont {Bart{\'o}k}\ and\ \citenamefont
  {Cs{\'a}nyi}(2015)}]{bartok_2015_ijqc}%
  \BibitemOpen
  \bibfield  {author} {\bibinfo {author} {\bibfnamefont {A.~P.}\ \bibnamefont
  {Bart{\'o}k}}\ and\ \bibinfo {author} {\bibfnamefont {G.}~\bibnamefont
  {Cs{\'a}nyi}},\ }\bibfield  {title} {\bibinfo {title} {Gaussian approximation
  potentials: A brief tutorial introduction},\ }\href
  {https://doi.org/10.1002/qua.24927} {\bibfield  {journal} {\bibinfo
  {journal} {International Journal of Quantum Chemistry}\ }\textbf {\bibinfo
  {volume} {115}},\ \bibinfo {pages} {1051} (\bibinfo {year}
  {2015})}\BibitemShut {NoStop}%
\bibitem [{gap(2022)}]{gap_www}%
  \BibitemOpen
  \href@noop {} {\bibinfo {title} {This software is available for
  non-commercial use from www.libatoms.org}} (\bibinfo {year}
  {2022})\BibitemShut {NoStop}%
\bibitem [{\citenamefont {Becker}\ \emph {et~al.}(2013)\citenamefont {Becker},
  \citenamefont {Tavazza}, \citenamefont {Trautt},\ and\ \citenamefont
  {de~Macedo}}]{nist_2013}%
  \BibitemOpen
  \bibfield  {author} {\bibinfo {author} {\bibfnamefont {C.~A.}\ \bibnamefont
  {Becker}}, \bibinfo {author} {\bibfnamefont {F.}~\bibnamefont {Tavazza}},
  \bibinfo {author} {\bibfnamefont {Z.~T.}\ \bibnamefont {Trautt}},\ and\
  \bibinfo {author} {\bibfnamefont {R.~A.~B.}\ \bibnamefont {de~Macedo}},\
  }\bibfield  {title} {\bibinfo {title} {Considerations for choosing and using
  force fields and interatomic potentials in materials science and
  engineering},\ }\href {https://doi.org/10.1016/j.cossms.2013.10.001}
  {\bibfield  {journal} {\bibinfo  {journal} {Current Opinion in Solid State
  and Materials Science}\ }\textbf {\bibinfo {volume} {17}},\ \bibinfo {pages}
  {277} (\bibinfo {year} {2013})}\BibitemShut {NoStop}%
\bibitem [{\citenamefont {Tadmor}\ \emph {et~al.}(2011)\citenamefont {Tadmor},
  \citenamefont {Elliott}, \citenamefont {Sethna}, \citenamefont {Miller},\
  and\ \citenamefont {Becker}}]{openkim_2011}%
  \BibitemOpen
  \bibfield  {author} {\bibinfo {author} {\bibfnamefont {E.~B.}\ \bibnamefont
  {Tadmor}}, \bibinfo {author} {\bibfnamefont {R.~S.}\ \bibnamefont {Elliott}},
  \bibinfo {author} {\bibfnamefont {J.~P.}\ \bibnamefont {Sethna}}, \bibinfo
  {author} {\bibfnamefont {R.~E.}\ \bibnamefont {Miller}},\ and\ \bibinfo
  {author} {\bibfnamefont {C.~A.}\ \bibnamefont {Becker}},\ }\bibfield  {title}
  {\bibinfo {title} {The potential of atomistic simulations and the
  knowledgebase of interatomic models},\ }\href
  {https://doi.org/10.1007/s11837-011-0102-6} {\bibfield  {journal} {\bibinfo
  {journal} {{JOM}}\ }\textbf {\bibinfo {volume} {63}},\ \bibinfo {pages} {17}
  (\bibinfo {year} {2011})}\BibitemShut {NoStop}%
\bibitem [{\citenamefont {Yin}\ \emph {et~al.}(2017)\citenamefont {Yin},
  \citenamefont {Wu},\ and\ \citenamefont {Curtin}}]{yin_2017_am}%
  \BibitemOpen
  \bibfield  {author} {\bibinfo {author} {\bibfnamefont {B.}~\bibnamefont
  {Yin}}, \bibinfo {author} {\bibfnamefont {Z.}~\bibnamefont {Wu}},\ and\
  \bibinfo {author} {\bibfnamefont {W.~A.}\ \bibnamefont {Curtin}},\ }\bibfield
   {title} {\bibinfo {title} {Comprehensive first-principles study of stable
  stacking faults in hcp metals},\ }\href
  {https://doi.org/10.1016/j.actamat.2016.10.042} {\bibfield  {journal}
  {\bibinfo  {journal} {Acta Materialia}\ }\textbf {\bibinfo {volume} {123}},\
  \bibinfo {pages} {223} (\bibinfo {year} {2017})}\BibitemShut {NoStop}%
\bibitem [{\citenamefont {Clouet}(2021)}]{clouet_2021_babel}%
  \BibitemOpen
  \bibfield  {author} {\bibinfo {author} {\bibfnamefont {E.}~\bibnamefont
  {Clouet}},\ }\href {http://emmanuel.clouet.free.fr/Programs/Babel/index.html}
  {\bibinfo {title} {Babel package}} (\bibinfo {year} {2021})\BibitemShut
  {NoStop}%
\bibitem [{\citenamefont {V{\'\i}tek}\ \emph {et~al.}(1970)\citenamefont
  {V{\'\i}tek}, \citenamefont {Perrin},\ and\ \citenamefont
  {Bowen}}]{vitek_1970_pma}%
  \BibitemOpen
  \bibfield  {author} {\bibinfo {author} {\bibfnamefont {V.}~\bibnamefont
  {V{\'\i}tek}}, \bibinfo {author} {\bibfnamefont {R.~C.}\ \bibnamefont
  {Perrin}},\ and\ \bibinfo {author} {\bibfnamefont {D.~K.}\ \bibnamefont
  {Bowen}},\ }\bibfield  {title} {\bibinfo {title} {The core structure of
  $1/2\langle 111 \rangle$ screw dislocations in b.c.c. crystals},\ }\href
  {https://doi.org/10.1080/14786437008238490} {\bibfield  {journal} {\bibinfo
  {journal} {Philosophical Magazine A}\ }\textbf {\bibinfo {volume} {21}},\
  \bibinfo {pages} {1049} (\bibinfo {year} {1970})}\BibitemShut {NoStop}%
\bibitem [{\citenamefont {Stukowski}(2009)}]{stukowski_2009_msmse}%
  \BibitemOpen
  \bibfield  {author} {\bibinfo {author} {\bibfnamefont {A.}~\bibnamefont
  {Stukowski}},\ }\bibfield  {title} {\bibinfo {title} {Visualization and
  analysis of atomistic simulation data with {OVITO--}the {Open Visualization
  Tool}},\ }\href {https://doi.org/10.1088/0965-0393/18/1/015012} {\bibfield
  {journal} {\bibinfo  {journal} {Modelling and Simulation in Materials Science
  and Engineering}\ }\textbf {\bibinfo {volume} {18}},\ \bibinfo {pages}
  {015012} (\bibinfo {year} {2009})}\BibitemShut {NoStop}%
\bibitem [{\citenamefont {Lejaeghere}\ \emph {et~al.}(2016)\citenamefont
  {Lejaeghere}, \citenamefont {Bihlmayer}, \citenamefont {Björkman},
  \citenamefont {Blaha}, \citenamefont {Blügel}, \citenamefont {Blum},
  \citenamefont {Caliste}, \citenamefont {Castelli}, \citenamefont {Clark},
  \citenamefont {{Dal Corso}}, \citenamefont {de~Gironcoli}, \citenamefont
  {Deutsch}, \citenamefont {Dewhurst}, \citenamefont {{Di Marco}},
  \citenamefont {Draxl}, \citenamefont {Dułak}, \citenamefont {Eriksson},
  \citenamefont {Flores-Livas}, \citenamefont {Garrity}, \citenamefont
  {Genovese}, \citenamefont {Giannozzi}, \citenamefont {Giantomassi},
  \citenamefont {Goedecker}, \citenamefont {Gonze}, \citenamefont {Grånäs},
  \citenamefont {Gross}, \citenamefont {Gulans}, \citenamefont {Gygi},
  \citenamefont {Hamann}, \citenamefont {Hasnip}, \citenamefont {Holzwarth},
  \citenamefont {Iuşan}, \citenamefont {Jochym}, \citenamefont {Jollet},
  \citenamefont {Jones}, \citenamefont {Kresse}, \citenamefont {Koepernik},
  \citenamefont {Küçükbenli}, \citenamefont {Kvashnin}, \citenamefont
  {Locht}, \citenamefont {Lubeck}, \citenamefont {Marsman}, \citenamefont
  {Marzari}, \citenamefont {Nitzsche}, \citenamefont {Nordström},
  \citenamefont {Ozaki}, \citenamefont {Paulatto}, \citenamefont {Pickard},
  \citenamefont {Poelmans}, \citenamefont {Probert}, \citenamefont {Refson},
  \citenamefont {Richter}, \citenamefont {Rignanese}, \citenamefont {Saha},
  \citenamefont {Scheffler}, \citenamefont {Schlipf}, \citenamefont {Schwarz},
  \citenamefont {Sharma}, \citenamefont {Tavazza}, \citenamefont {Thunström},
  \citenamefont {Tkatchenko}, \citenamefont {Torrent}, \citenamefont
  {Vanderbilt}, \citenamefont {van Setten}, \citenamefont {{Van Speybroeck}},
  \citenamefont {Wills}, \citenamefont {Yates}, \citenamefont {Zhang},\ and\
  \citenamefont {Cottenier}}]{lejaeghere_2016_science}%
  \BibitemOpen
  \bibfield  {author} {\bibinfo {author} {\bibfnamefont {K.}~\bibnamefont
  {Lejaeghere}}, \bibinfo {author} {\bibfnamefont {G.}~\bibnamefont
  {Bihlmayer}}, \bibinfo {author} {\bibfnamefont {T.}~\bibnamefont
  {Björkman}}, \bibinfo {author} {\bibfnamefont {P.}~\bibnamefont {Blaha}},
  \bibinfo {author} {\bibfnamefont {S.}~\bibnamefont {Blügel}}, \bibinfo
  {author} {\bibfnamefont {V.}~\bibnamefont {Blum}}, \bibinfo {author}
  {\bibfnamefont {D.}~\bibnamefont {Caliste}}, \bibinfo {author} {\bibfnamefont
  {I.~E.}\ \bibnamefont {Castelli}}, \bibinfo {author} {\bibfnamefont {S.~J.}\
  \bibnamefont {Clark}}, \bibinfo {author} {\bibfnamefont {A.}~\bibnamefont
  {{Dal Corso}}}, \bibinfo {author} {\bibfnamefont {S.}~\bibnamefont
  {de~Gironcoli}}, \bibinfo {author} {\bibfnamefont {T.}~\bibnamefont
  {Deutsch}}, \bibinfo {author} {\bibfnamefont {J.~K.}\ \bibnamefont
  {Dewhurst}}, \bibinfo {author} {\bibfnamefont {I.}~\bibnamefont {{Di
  Marco}}}, \bibinfo {author} {\bibfnamefont {C.}~\bibnamefont {Draxl}},
  \bibinfo {author} {\bibfnamefont {M.}~\bibnamefont {Dułak}}, \bibinfo
  {author} {\bibfnamefont {O.}~\bibnamefont {Eriksson}}, \bibinfo {author}
  {\bibfnamefont {J.}~\bibnamefont {Flores-Livas}}, \bibinfo {author}
  {\bibfnamefont {K.~F.}\ \bibnamefont {Garrity}}, \bibinfo {author}
  {\bibfnamefont {L.}~\bibnamefont {Genovese}}, \bibinfo {author}
  {\bibfnamefont {P.}~\bibnamefont {Giannozzi}}, \bibinfo {author}
  {\bibfnamefont {M.}~\bibnamefont {Giantomassi}}, \bibinfo {author}
  {\bibfnamefont {S.}~\bibnamefont {Goedecker}}, \bibinfo {author}
  {\bibfnamefont {X.}~\bibnamefont {Gonze}}, \bibinfo {author} {\bibfnamefont
  {O.}~\bibnamefont {Grånäs}}, \bibinfo {author} {\bibfnamefont {E.~K.~U.}\
  \bibnamefont {Gross}}, \bibinfo {author} {\bibfnamefont {A.}~\bibnamefont
  {Gulans}}, \bibinfo {author} {\bibfnamefont {F.}~\bibnamefont {Gygi}},
  \bibinfo {author} {\bibfnamefont {D.~R.}\ \bibnamefont {Hamann}}, \bibinfo
  {author} {\bibfnamefont {P.~J.}\ \bibnamefont {Hasnip}}, \bibinfo {author}
  {\bibfnamefont {N.~A.~W.}\ \bibnamefont {Holzwarth}}, \bibinfo {author}
  {\bibfnamefont {D.}~\bibnamefont {Iuşan}}, \bibinfo {author} {\bibfnamefont
  {D.~B.}\ \bibnamefont {Jochym}}, \bibinfo {author} {\bibfnamefont
  {F.}~\bibnamefont {Jollet}}, \bibinfo {author} {\bibfnamefont
  {D.}~\bibnamefont {Jones}}, \bibinfo {author} {\bibfnamefont
  {G.}~\bibnamefont {Kresse}}, \bibinfo {author} {\bibfnamefont
  {K.}~\bibnamefont {Koepernik}}, \bibinfo {author} {\bibfnamefont
  {E.}~\bibnamefont {Küçükbenli}}, \bibinfo {author} {\bibfnamefont {Y.~O.}\
  \bibnamefont {Kvashnin}}, \bibinfo {author} {\bibfnamefont {I.~L.~M.}\
  \bibnamefont {Locht}}, \bibinfo {author} {\bibfnamefont {S.}~\bibnamefont
  {Lubeck}}, \bibinfo {author} {\bibfnamefont {M.}~\bibnamefont {Marsman}},
  \bibinfo {author} {\bibfnamefont {N.}~\bibnamefont {Marzari}}, \bibinfo
  {author} {\bibfnamefont {U.}~\bibnamefont {Nitzsche}}, \bibinfo {author}
  {\bibfnamefont {L.}~\bibnamefont {Nordström}}, \bibinfo {author}
  {\bibfnamefont {T.}~\bibnamefont {Ozaki}}, \bibinfo {author} {\bibfnamefont
  {L.}~\bibnamefont {Paulatto}}, \bibinfo {author} {\bibfnamefont {C.~J.}\
  \bibnamefont {Pickard}}, \bibinfo {author} {\bibfnamefont {W.}~\bibnamefont
  {Poelmans}}, \bibinfo {author} {\bibfnamefont {M.~I.~J.}\ \bibnamefont
  {Probert}}, \bibinfo {author} {\bibfnamefont {K.}~\bibnamefont {Refson}},
  \bibinfo {author} {\bibfnamefont {M.}~\bibnamefont {Richter}}, \bibinfo
  {author} {\bibfnamefont {G.-M.}\ \bibnamefont {Rignanese}}, \bibinfo {author}
  {\bibfnamefont {S.}~\bibnamefont {Saha}}, \bibinfo {author} {\bibfnamefont
  {M.}~\bibnamefont {Scheffler}}, \bibinfo {author} {\bibfnamefont
  {M.}~\bibnamefont {Schlipf}}, \bibinfo {author} {\bibfnamefont
  {K.}~\bibnamefont {Schwarz}}, \bibinfo {author} {\bibfnamefont
  {S.}~\bibnamefont {Sharma}}, \bibinfo {author} {\bibfnamefont
  {F.}~\bibnamefont {Tavazza}}, \bibinfo {author} {\bibfnamefont
  {P.}~\bibnamefont {Thunström}}, \bibinfo {author} {\bibfnamefont
  {A.}~\bibnamefont {Tkatchenko}}, \bibinfo {author} {\bibfnamefont
  {M.}~\bibnamefont {Torrent}}, \bibinfo {author} {\bibfnamefont
  {D.}~\bibnamefont {Vanderbilt}}, \bibinfo {author} {\bibfnamefont {M.~J.}\
  \bibnamefont {van Setten}}, \bibinfo {author} {\bibfnamefont
  {V.}~\bibnamefont {{Van Speybroeck}}}, \bibinfo {author} {\bibfnamefont
  {J.~M.}\ \bibnamefont {Wills}}, \bibinfo {author} {\bibfnamefont {J.~R.}\
  \bibnamefont {Yates}}, \bibinfo {author} {\bibfnamefont {G.-X.}\ \bibnamefont
  {Zhang}},\ and\ \bibinfo {author} {\bibfnamefont {S.}~\bibnamefont
  {Cottenier}},\ }\bibfield  {title} {\bibinfo {title} {Reproducibility in
  density functional theory calculations of solids},\ }\href
  {https://doi.org/10.1126/science.aad3000} {\bibfield  {journal} {\bibinfo
  {journal} {Science}\ }\textbf {\bibinfo {volume} {351}},\ \bibinfo {pages}
  {aad3000} (\bibinfo {year} {2016})}\BibitemShut {NoStop}%
\bibitem [{\citenamefont {Henkelman}\ and\ \citenamefont
  {J{\'o}nsson}(2000)}]{henkelman_2000_11_jcp}%
  \BibitemOpen
  \bibfield  {author} {\bibinfo {author} {\bibfnamefont {G.}~\bibnamefont
  {Henkelman}}\ and\ \bibinfo {author} {\bibfnamefont {H.}~\bibnamefont
  {J{\'o}nsson}},\ }\bibfield  {title} {\bibinfo {title} {Improved tangent
  estimate in the nudged elastic band method for finding minimum energy paths
  and saddle points},\ }\href {https://doi.org/10.1063/1.1323224} {\bibfield
  {journal} {\bibinfo  {journal} {The Journal of Chemical Physics}\ }\textbf
  {\bibinfo {volume} {113}},\ \bibinfo {pages} {9978} (\bibinfo {year}
  {2000})}\BibitemShut {NoStop}%
\bibitem [{\citenamefont {Samolyuk}\ \emph {et~al.}(2012)\citenamefont
  {Samolyuk}, \citenamefont {Osetsky},\ and\ \citenamefont
  {Stoller}}]{samolyuk_2013_jpcm}%
  \BibitemOpen
  \bibfield  {author} {\bibinfo {author} {\bibfnamefont {G.~D.}\ \bibnamefont
  {Samolyuk}}, \bibinfo {author} {\bibfnamefont {Y.~N.}\ \bibnamefont
  {Osetsky}},\ and\ \bibinfo {author} {\bibfnamefont {R.~E.}\ \bibnamefont
  {Stoller}},\ }\bibfield  {title} {\bibinfo {title} {The influence of
  transition metal solutes on the dislocation core structure and values of the
  {Peierls} stress and barrier in tungsten},\ }\href
  {https://doi.org/10.1088/0953-8984/25/2/025403} {\bibfield  {journal}
  {\bibinfo  {journal} {Journal of Physics: Condensed Matter}\ }\textbf
  {\bibinfo {volume} {25}},\ \bibinfo {pages} {025403} (\bibinfo {year}
  {2012})}\BibitemShut {NoStop}%
\bibitem [{\citenamefont {Xu}\ and\ \citenamefont
  {Moriarty}(1998)}]{xu_1998_cms}%
  \BibitemOpen
  \bibfield  {author} {\bibinfo {author} {\bibfnamefont {W.}~\bibnamefont
  {Xu}}\ and\ \bibinfo {author} {\bibfnamefont {J.~A.}\ \bibnamefont
  {Moriarty}},\ }\bibfield  {title} {\bibinfo {title} {Accurate atomistic
  simulations of the {Peierls} barrier and kink-pair formation energy for
  $\langle 111 \rangle$ screw dislocations in bcc {Mo}},\ }\href
  {https://doi.org/10.1016/s0927-0256(97)00161-4} {\bibfield  {journal}
  {\bibinfo  {journal} {Computational Materials Science}\ }\textbf {\bibinfo
  {volume} {9}},\ \bibinfo {pages} {348} (\bibinfo {year} {1998})}\BibitemShut
  {NoStop}%
\bibitem [{\citenamefont {Sun}\ and\ \citenamefont
  {Ceder}(2013)}]{sun_2013_ss}%
  \BibitemOpen
  \bibfield  {author} {\bibinfo {author} {\bibfnamefont {W.}~\bibnamefont
  {Sun}}\ and\ \bibinfo {author} {\bibfnamefont {G.}~\bibnamefont {Ceder}},\
  }\bibfield  {title} {\bibinfo {title} {Efficient creation and convergence of
  surface slabs},\ }\href {https://doi.org/10.1016/j.susc.2013.05.016}
  {\bibfield  {journal} {\bibinfo  {journal} {Surface Science}\ }\textbf
  {\bibinfo {volume} {617}},\ \bibinfo {pages} {53} (\bibinfo {year}
  {2013})}\BibitemShut {NoStop}%
\bibitem [{\citenamefont {Yang}\ and\ \citenamefont
  {Qi}(2018)}]{yang_2018_prb}%
  \BibitemOpen
  \bibfield  {author} {\bibinfo {author} {\bibfnamefont {C.}~\bibnamefont
  {Yang}}\ and\ \bibinfo {author} {\bibfnamefont {L.}~\bibnamefont {Qi}},\
  }\bibfield  {title} {\bibinfo {title} {\textit{Ab initio} calculations of
  ideal strength and lattice instability in {W-Ta} and {W-Re} alloys},\ }\href
  {https://doi.org/10.1103/physrevb.97.014107} {\bibfield  {journal} {\bibinfo
  {journal} {Physical Review B}\ }\textbf {\bibinfo {volume} {97}},\ \bibinfo
  {pages} {014107} (\bibinfo {year} {2018})}\BibitemShut {NoStop}%
\bibitem [{\citenamefont {Hu}\ \emph {et~al.}(2021)\citenamefont {Hu},
  \citenamefont {Sundar}, \citenamefont {Ogata},\ and\ \citenamefont
  {Qi}}]{hu_2021_am}%
  \BibitemOpen
  \bibfield  {author} {\bibinfo {author} {\bibfnamefont {Y.-J.}\ \bibnamefont
  {Hu}}, \bibinfo {author} {\bibfnamefont {A.}~\bibnamefont {Sundar}}, \bibinfo
  {author} {\bibfnamefont {S.}~\bibnamefont {Ogata}},\ and\ \bibinfo {author}
  {\bibfnamefont {L.}~\bibnamefont {Qi}},\ }\bibfield  {title} {\bibinfo
  {title} {Screening of generalized stacking fault energies, surface energies
  and intrinsic ductile potency of refractory multicomponent alloys},\ }\href
  {https://doi.org/10.1016/j.actamat.2021.116800} {\bibfield  {journal}
  {\bibinfo  {journal} {Acta Materialia}\ }\textbf {\bibinfo {volume} {210}},\
  \bibinfo {pages} {116800} (\bibinfo {year} {2021})}\BibitemShut {NoStop}%
\bibitem [{\citenamefont {Rice}(1992)}]{rice_1992_jmps}%
  \BibitemOpen
  \bibfield  {author} {\bibinfo {author} {\bibfnamefont {J.~R.}\ \bibnamefont
  {Rice}},\ }\bibfield  {title} {\bibinfo {title} {Dislocation nucleation from
  a crack tip{: A}n analysis based on the {P}eierls concept},\ }\href
  {https://doi.org/10.1016/S0022-5096(05)80012-2} {\bibfield  {journal}
  {\bibinfo  {journal} {Journal of the Mechanics and Physics of Solids}\
  }\textbf {\bibinfo {volume} {40}},\ \bibinfo {pages} {239} (\bibinfo {year}
  {1992})}\BibitemShut {NoStop}%
\bibitem [{\citenamefont {Mak}\ \emph {et~al.}(2021)\citenamefont {Mak},
  \citenamefont {Yin},\ and\ \citenamefont {Curtin}}]{mak_2021_jmps}%
  \BibitemOpen
  \bibfield  {author} {\bibinfo {author} {\bibfnamefont {E.}~\bibnamefont
  {Mak}}, \bibinfo {author} {\bibfnamefont {B.}~\bibnamefont {Yin}},\ and\
  \bibinfo {author} {\bibfnamefont {W.}~\bibnamefont {Curtin}},\ }\bibfield
  {title} {\bibinfo {title} {A ductility criterion for bcc high entropy
  alloys},\ }\href {https://doi.org/10.1016/j.jmps.2021.104389} {\bibfield
  {journal} {\bibinfo  {journal} {Journal of the Mechanics and Physics of
  Solids}\ }\textbf {\bibinfo {volume} {152}},\ \bibinfo {pages} {104389}
  (\bibinfo {year} {2021})}\BibitemShut {NoStop}%
\bibitem [{\citenamefont {Vanderbilt}(1990)}]{vanderbilt_1990_prb}%
  \BibitemOpen
  \bibfield  {author} {\bibinfo {author} {\bibfnamefont {D.}~\bibnamefont
  {Vanderbilt}},\ }\bibfield  {title} {\bibinfo {title} {Soft self-consistent
  pseudopotentials in a generalized eigenvalue formalism},\ }\href
  {http://link.aps.org/doi/10.1103/PhysRevB.41.7892} {\bibfield  {journal}
  {\bibinfo  {journal} {Physical Review B}\ }\textbf {\bibinfo {volume} {41}},\
  \bibinfo {pages} {7892} (\bibinfo {year} {1990})}\BibitemShut {NoStop}%
\bibitem [{\citenamefont {Rappe}\ \emph {et~al.}(1990)\citenamefont {Rappe},
  \citenamefont {Rabe}, \citenamefont {Kaxiras},\ and\ \citenamefont
  {Joannopoulos}}]{rappe_1990_prb}%
  \BibitemOpen
  \bibfield  {author} {\bibinfo {author} {\bibfnamefont {A.~M.}\ \bibnamefont
  {Rappe}}, \bibinfo {author} {\bibfnamefont {K.~M.}\ \bibnamefont {Rabe}},
  \bibinfo {author} {\bibfnamefont {E.}~\bibnamefont {Kaxiras}},\ and\ \bibinfo
  {author} {\bibfnamefont {J.~D.}\ \bibnamefont {Joannopoulos}},\ }\bibfield
  {title} {\bibinfo {title} {Optimized pseudopotentials},\ }\href
  {https://doi.org/10.1103/physrevb.41.1227} {\bibfield  {journal} {\bibinfo
  {journal} {Physical Review B}\ }\textbf {\bibinfo {volume} {41}},\ \bibinfo
  {pages} {1227} (\bibinfo {year} {1990})}\BibitemShut {NoStop}%
\bibitem [{\citenamefont {{Dal Corso}}(2014)}]{corso_2014_cms}%
  \BibitemOpen
  \bibfield  {author} {\bibinfo {author} {\bibfnamefont {A.}~\bibnamefont {{Dal
  Corso}}},\ }\bibfield  {title} {\bibinfo {title} {Pseudopotentials periodic
  table{: From H to Pu}},\ }\href
  {https://doi.org/10.1016/j.commatsci.2014.07.043} {\bibfield  {journal}
  {\bibinfo  {journal} {Computational Materials Science}\ }\textbf {\bibinfo
  {volume} {95}},\ \bibinfo {pages} {337} (\bibinfo {year} {2014})}\BibitemShut
  {NoStop}%
\bibitem [{\citenamefont {Marzari}\ \emph {et~al.}(1999)\citenamefont
  {Marzari}, \citenamefont {Vanderbilt}, \citenamefont {{De Vita}},\ and\
  \citenamefont {Payne}}]{marzari_1999_prl}%
  \BibitemOpen
  \bibfield  {author} {\bibinfo {author} {\bibfnamefont {N.}~\bibnamefont
  {Marzari}}, \bibinfo {author} {\bibfnamefont {D.}~\bibnamefont {Vanderbilt}},
  \bibinfo {author} {\bibfnamefont {A.}~\bibnamefont {{De Vita}}},\ and\
  \bibinfo {author} {\bibfnamefont {M.~C.}\ \bibnamefont {Payne}},\ }\bibfield
  {title} {\bibinfo {title} {Thermal contraction and disordering of the al(110)
  surface},\ }\href {https://doi.org/10.1103/PhysRevLett.82.3296} {\bibfield
  {journal} {\bibinfo  {journal} {Physical Review Letters}\ }\textbf {\bibinfo
  {volume} {82}},\ \bibinfo {pages} {3296} (\bibinfo {year}
  {1999})}\BibitemShut {NoStop}%
\bibitem [{\citenamefont {Li}\ \emph {et~al.}(2020)\citenamefont {Li},
  \citenamefont {Chen}, \citenamefont {Zheng}, \citenamefont {Zuo},\ and\
  \citenamefont {Ong}}]{zheng_2020_npjcm}%
  \BibitemOpen
  \bibfield  {author} {\bibinfo {author} {\bibfnamefont {X.-G.}\ \bibnamefont
  {Li}}, \bibinfo {author} {\bibfnamefont {C.}~\bibnamefont {Chen}}, \bibinfo
  {author} {\bibfnamefont {H.}~\bibnamefont {Zheng}}, \bibinfo {author}
  {\bibfnamefont {Y.}~\bibnamefont {Zuo}},\ and\ \bibinfo {author}
  {\bibfnamefont {S.~P.}\ \bibnamefont {Ong}},\ }\bibfield  {title} {\bibinfo
  {title} {Complex strengthening mechanisms in the {NbMoTaW} multi-principal
  element alloy},\ }\href {https://doi.org/10.1038/s41524-020-0339-0}
  {\bibfield  {journal} {\bibinfo  {journal} {npj Computational Materials}\
  }\textbf {\bibinfo {volume} {6}},\ \bibinfo {pages} {70} (\bibinfo {year}
  {2020})}\BibitemShut {NoStop}%
\bibitem [{\citenamefont {Mendelev}(2020)}]{mendelev_2020_openkim}%
  \BibitemOpen
  \bibfield  {author} {\bibinfo {author} {\bibfnamefont {M.}~\bibnamefont
  {Mendelev}},\ }\bibfield  {title} {\bibinfo {title} {{Finnis--Sinclair}
  potential for the {Fe--Ni--Cr} system developed by {M}endelev et al. (2020)
  v000},\ }\bibfield  {journal} {\bibinfo  {journal} {OpenKIM}\ }\href
  {https://doi.org/10.25950/538764D4} {10.25950/538764D4} (\bibinfo {year}
  {2020})\BibitemShut {NoStop}%
\bibitem [{\citenamefont {Starikov}\ \emph {et~al.}(2018)\citenamefont
  {Starikov}, \citenamefont {Kolotova}, \citenamefont {Kuksin}, \citenamefont
  {Smirnova},\ and\ \citenamefont {Tseplyaev}}]{starikov_2017_jnm}%
  \BibitemOpen
  \bibfield  {author} {\bibinfo {author} {\bibfnamefont {S.}~\bibnamefont
  {Starikov}}, \bibinfo {author} {\bibfnamefont {L.}~\bibnamefont {Kolotova}},
  \bibinfo {author} {\bibfnamefont {A.}~\bibnamefont {Kuksin}}, \bibinfo
  {author} {\bibfnamefont {D.}~\bibnamefont {Smirnova}},\ and\ \bibinfo
  {author} {\bibfnamefont {V.}~\bibnamefont {Tseplyaev}},\ }\bibfield  {title}
  {\bibinfo {title} {Atomistic simulation of cubic and tetragonal phases of
  {U-Mo} alloy: {S}tructure and thermodynamic properties},\ }\href
  {https://doi.org/10.1016/j.jnucmat.2017.11.047} {\bibfield  {journal}
  {\bibinfo  {journal} {Journal of Nuclear Materials}\ }\textbf {\bibinfo
  {volume} {499}},\ \bibinfo {pages} {451} (\bibinfo {year}
  {2018})}\BibitemShut {NoStop}%
\bibitem [{\citenamefont {Pink}\ and\ \citenamefont
  {Arsenault}(1980)}]{pink_1980_pms}%
  \BibitemOpen
  \bibfield  {author} {\bibinfo {author} {\bibfnamefont {E.}~\bibnamefont
  {Pink}}\ and\ \bibinfo {author} {\bibfnamefont {R.~J.}\ \bibnamefont
  {Arsenault}},\ }\bibfield  {title} {\bibinfo {title} {Low-temperature
  softening in body-centered cubic alloys},\ }\href
  {https://doi.org/10.1016/0079-6425(79)90003-3} {\bibfield  {journal}
  {\bibinfo  {journal} {Progress in Materials Science}\ }\textbf {\bibinfo
  {volume} {24}},\ \bibinfo {pages} {1} (\bibinfo {year} {1980})}\BibitemShut
  {NoStop}%
\bibitem [{\citenamefont {Chiesa}\ \emph {et~al.}(2009)\citenamefont {Chiesa},
  \citenamefont {Gilbert}, \citenamefont {Dudarev}, \citenamefont {Derlet},\
  and\ \citenamefont {Swygenhoven}}]{chiesa_2009_pm}%
  \BibitemOpen
  \bibfield  {author} {\bibinfo {author} {\bibfnamefont {S.}~\bibnamefont
  {Chiesa}}, \bibinfo {author} {\bibfnamefont {M.}~\bibnamefont {Gilbert}},
  \bibinfo {author} {\bibfnamefont {S.}~\bibnamefont {Dudarev}}, \bibinfo
  {author} {\bibfnamefont {P.}~\bibnamefont {Derlet}},\ and\ \bibinfo {author}
  {\bibfnamefont {H.~V.}\ \bibnamefont {Swygenhoven}},\ }\bibfield  {title}
  {\bibinfo {title} {The non-degenerate core structure of a $1/2\langle 111
  \rangle$ screw dislocation in bcc transition metals modelled using
  {F}innis{--}{S}inclair potentials: The necessary and sufficient conditions},\
  }\href {https://doi.org/10.1080/14786430903250835} {\bibfield  {journal}
  {\bibinfo  {journal} {Philosophical Magazine}\ }\textbf {\bibinfo {volume}
  {89}},\ \bibinfo {pages} {3235} (\bibinfo {year} {2009})}\BibitemShut
  {NoStop}%
\bibitem [{\citenamefont {Mendelev}\ \emph {et~al.}(2003)\citenamefont
  {Mendelev}, \citenamefont {Han}, \citenamefont {Srolovitz}, \citenamefont
  {Ackland}, \citenamefont {Sun},\ and\ \citenamefont
  {Asta}}]{mendelev_2003_pm}%
  \BibitemOpen
  \bibfield  {author} {\bibinfo {author} {\bibfnamefont {M.~I.}\ \bibnamefont
  {Mendelev}}, \bibinfo {author} {\bibfnamefont {S.}~\bibnamefont {Han}},
  \bibinfo {author} {\bibfnamefont {D.~J.}\ \bibnamefont {Srolovitz}}, \bibinfo
  {author} {\bibfnamefont {G.~J.}\ \bibnamefont {Ackland}}, \bibinfo {author}
  {\bibfnamefont {D.~Y.}\ \bibnamefont {Sun}},\ and\ \bibinfo {author}
  {\bibfnamefont {M.}~\bibnamefont {Asta}},\ }\bibfield  {title} {\bibinfo
  {title} {Development of new interatomic potentials appropriate for
  crystalline and liquid iron},\ }\href
  {https://doi.org/10.1080/14786430310001613264} {\bibfield  {journal}
  {\bibinfo  {journal} {Philosophical Magazine}\ }\textbf {\bibinfo {volume}
  {83}},\ \bibinfo {pages} {3977} (\bibinfo {year} {2003})}\BibitemShut
  {NoStop}%
\bibitem [{\citenamefont {Zuo}\ \emph {et~al.}(2020)\citenamefont {Zuo},
  \citenamefont {Chen}, \citenamefont {Li}, \citenamefont {Deng}, \citenamefont
  {Chen}, \citenamefont {Behler}, \citenamefont {Cs{\'a}nyi}, \citenamefont
  {Shapeev}, \citenamefont {Thompson}, \citenamefont {Wood},\ and\
  \citenamefont {et~al.}}]{zuo_2020_jpca}%
  \BibitemOpen
  \bibfield  {author} {\bibinfo {author} {\bibfnamefont {Y.}~\bibnamefont
  {Zuo}}, \bibinfo {author} {\bibfnamefont {C.}~\bibnamefont {Chen}}, \bibinfo
  {author} {\bibfnamefont {X.}~\bibnamefont {Li}}, \bibinfo {author}
  {\bibfnamefont {Z.}~\bibnamefont {Deng}}, \bibinfo {author} {\bibfnamefont
  {Y.}~\bibnamefont {Chen}}, \bibinfo {author} {\bibfnamefont {J.}~\bibnamefont
  {Behler}}, \bibinfo {author} {\bibfnamefont {G.}~\bibnamefont {Cs{\'a}nyi}},
  \bibinfo {author} {\bibfnamefont {A.~V.}\ \bibnamefont {Shapeev}}, \bibinfo
  {author} {\bibfnamefont {A.~P.}\ \bibnamefont {Thompson}}, \bibinfo {author}
  {\bibfnamefont {M.~A.}\ \bibnamefont {Wood}},\ and\ \bibinfo {author}
  {\bibnamefont {et~al.}},\ }\bibfield  {title} {\bibinfo {title} {Performance
  and cost assessment of machine learning interatomic potentials},\ }\href
  {https://doi.org/10.1021/acs.jpca.9b08723} {\bibfield  {journal} {\bibinfo
  {journal} {The Journal of Physical Chemistry A}\ }\textbf {\bibinfo {volume}
  {124}},\ \bibinfo {pages} {731} (\bibinfo {year} {2020})}\BibitemShut
  {NoStop}%
\bibitem [{\citenamefont {Alam}\ and\ \citenamefont
  {Groh}(2015)}]{alam_2015_jpcs}%
  \BibitemOpen
  \bibfield  {author} {\bibinfo {author} {\bibfnamefont {M.}~\bibnamefont
  {Alam}}\ and\ \bibinfo {author} {\bibfnamefont {S.}~\bibnamefont {Groh}},\
  }\bibfield  {title} {\bibinfo {title} {Dislocation modeling in bcc lithium:
  {A} comparison between continuum and atomistic predictions in the modified
  embedded atoms method},\ }\href {https://doi.org/10.1016/j.jpcs.2015.02.007}
  {\bibfield  {journal} {\bibinfo  {journal} {Journal of Physics and Chemistry
  of Solids}\ }\textbf {\bibinfo {volume} {82}},\ \bibinfo {pages} {1}
  (\bibinfo {year} {2015})}\BibitemShut {NoStop}%
\bibitem [{\citenamefont {Groh}\ and\ \citenamefont
  {Alam}(2015)}]{groh_2015_msmse}%
  \BibitemOpen
  \bibfield  {author} {\bibinfo {author} {\bibfnamefont {S.}~\bibnamefont
  {Groh}}\ and\ \bibinfo {author} {\bibfnamefont {M.}~\bibnamefont {Alam}},\
  }\bibfield  {title} {\bibinfo {title} {Fracture behavior of lithium single
  crystal in the framework of (semi-)empirical force field derived from
  first-principles},\ }\href {http://stacks.iop.org/0965-0393/23/i=4/a=045008}
  {\bibfield  {journal} {\bibinfo  {journal} {Modelling and Simulation in
  Materials Science and Engineering}\ }\textbf {\bibinfo {volume} {23}},\
  \bibinfo {pages} {045008} (\bibinfo {year} {2015})}\BibitemShut {NoStop}%
\bibitem [{\citenamefont {Zhou}(2010)}]{zhou_2010_lmp}%
  \BibitemOpen
  \bibfield  {author} {\bibinfo {author} {\bibfnamefont {X.}~\bibnamefont
  {Zhou}},\ }\href {https://doi.org/10.25950/b2223d98} {\bibinfo {title}
  {{LAMMPS EIM} potential for the {Br-Cl-Cs-F-I-K-Li-Na-Rb} system}} (\bibinfo
  {year} {2010})\BibitemShut {NoStop}%
\bibitem [{\citenamefont {Foxall}\ and\ \citenamefont
  {Statham}(1970)}]{foxall_1970_am}%
  \BibitemOpen
  \bibfield  {author} {\bibinfo {author} {\bibfnamefont {R.}~\bibnamefont
  {Foxall}}\ and\ \bibinfo {author} {\bibfnamefont {C.}~\bibnamefont
  {Statham}},\ }\bibfield  {title} {\bibinfo {title} {Dislocation arrangements
  in deformed single crystals of niobium-molybdenum alloys and niobium-9
  {at.\%} rhenium},\ }\href {https://doi.org/10.1016/0001-6160(70)90105-7}
  {\bibfield  {journal} {\bibinfo  {journal} {Acta Metallurgica}\ }\textbf
  {\bibinfo {volume} {18}},\ \bibinfo {pages} {1147} (\bibinfo {year}
  {1970})}\BibitemShut {NoStop}%
\bibitem [{\citenamefont {Peters}\ and\ \citenamefont
  {Hendrickson}(1970)}]{peters_1970_mt}%
  \BibitemOpen
  \bibfield  {author} {\bibinfo {author} {\bibfnamefont {B.~C.}\ \bibnamefont
  {Peters}}\ and\ \bibinfo {author} {\bibfnamefont {A.~A.}\ \bibnamefont
  {Hendrickson}},\ }\bibfield  {title} {\bibinfo {title} {Solid solution
  strengthening in {Nb-Ta} and {Nb-Mo} alloy single crystals},\ }\href
  {https://doi.org/10.1007/bf02643445} {\bibfield  {journal} {\bibinfo
  {journal} {Metallurgical Transactions}\ }\textbf {\bibinfo {volume} {1}},\
  \bibinfo {pages} {2271} (\bibinfo {year} {1970})}\BibitemShut {NoStop}%
\bibitem [{\citenamefont {Ulitchny}\ \emph {et~al.}(1971)\citenamefont
  {Ulitchny}, \citenamefont {Sagues},\ and\ \citenamefont
  {Gibala}}]{ulitchny_1971_iaea}%
  \BibitemOpen
  \bibfield  {author} {\bibinfo {author} {\bibfnamefont {M.~G.}\ \bibnamefont
  {Ulitchny}}, \bibinfo {author} {\bibfnamefont {A.~A.}\ \bibnamefont
  {Sagues}},\ and\ \bibinfo {author} {\bibfnamefont {R.}~\bibnamefont
  {Gibala}},\ }\bibfield  {title} {\bibinfo {title} {Alloy softening in
  niobium-and tantalum-base solid solutions}\ }(\bibinfo {year}
  {1971})\BibitemShut {NoStop}%
\bibitem [{\citenamefont {Statham}\ and\ \citenamefont
  {Christian}(1971)}]{statham_1971_sm}%
  \BibitemOpen
  \bibfield  {author} {\bibinfo {author} {\bibfnamefont {C.}~\bibnamefont
  {Statham}}\ and\ \bibinfo {author} {\bibfnamefont {J.}~\bibnamefont
  {Christian}},\ }\bibfield  {title} {\bibinfo {title} {Solution hardening and
  softening in niobium-molybdenum and niobium-rhenium alloys},\ }\href
  {https://doi.org/10.1016/0036-9748(71)90237-7} {\bibfield  {journal}
  {\bibinfo  {journal} {Scripta Metallurgica}\ }\textbf {\bibinfo {volume}
  {5}},\ \bibinfo {pages} {399} (\bibinfo {year} {1971})}\BibitemShut {NoStop}%
\bibitem [{\citenamefont {Stephens}\ and\ \citenamefont
  {Witzke}(1975)}]{stephens_1975_jlcm}%
  \BibitemOpen
  \bibfield  {author} {\bibinfo {author} {\bibfnamefont {J.~R.}\ \bibnamefont
  {Stephens}}\ and\ \bibinfo {author} {\bibfnamefont {W.~R.}\ \bibnamefont
  {Witzke}},\ }\bibfield  {title} {\bibinfo {title} {Hardness behavior of
  binary and ternary niobium alloys at 77 and 300 {K}},\ }\href
  {https://doi.org/10.1016/0022-5088(75)90061-2} {\bibfield  {journal}
  {\bibinfo  {journal} {Journal of the Less Common Metals}\ }\textbf {\bibinfo
  {volume} {40}},\ \bibinfo {pages} {195} (\bibinfo {year} {1975})}\BibitemShut
  {NoStop}%
\bibitem [{\citenamefont {Botta}\ \emph {et~al.}(1988)\citenamefont {Botta},
  \citenamefont {Christian},\ and\ \citenamefont {Taylor}}]{botta_1988_pma}%
  \BibitemOpen
  \bibfield  {author} {\bibinfo {author} {\bibfnamefont {W.~J.}\ \bibnamefont
  {Botta}}, \bibinfo {author} {\bibfnamefont {J.~W.}\ \bibnamefont
  {Christian}},\ and\ \bibinfo {author} {\bibfnamefont {G.}~\bibnamefont
  {Taylor}},\ }\bibfield  {title} {\bibinfo {title} {Solution hardening and
  softening of {Nb-Zr} single crystals},\ }\href
  {https://doi.org/10.1080/01418618808209915} {\bibfield  {journal} {\bibinfo
  {journal} {Philosophical Magazine A}\ }\textbf {\bibinfo {volume} {57}},\
  \bibinfo {pages} {703} (\bibinfo {year} {1988})}\BibitemShut {NoStop}%
\bibitem [{\citenamefont {Arsenault}(1966)}]{arsenault_1966_am}%
  \BibitemOpen
  \bibfield  {author} {\bibinfo {author} {\bibfnamefont {R.}~\bibnamefont
  {Arsenault}},\ }\bibfield  {title} {\bibinfo {title} {An investigation of the
  mechanism of thermally activated deformation in tantalum and tantalum-base
  alloys},\ }\href {https://doi.org/10.1016/0001-6160(66)90003-4} {\bibfield
  {journal} {\bibinfo  {journal} {Acta Metallurgica}\ }\textbf {\bibinfo
  {volume} {14}},\ \bibinfo {pages} {831} (\bibinfo {year} {1966})}\BibitemShut
  {NoStop}%
\bibitem [{\citenamefont {Mordike}\ \emph {et~al.}(1970)\citenamefont
  {Mordike}, \citenamefont {Rogausch},\ and\ \citenamefont
  {Braithwaite}}]{mordike_1970_msj}%
  \BibitemOpen
  \bibfield  {author} {\bibinfo {author} {\bibfnamefont {B.~L.}\ \bibnamefont
  {Mordike}}, \bibinfo {author} {\bibfnamefont {K.~D.}\ \bibnamefont
  {Rogausch}},\ and\ \bibinfo {author} {\bibfnamefont {A.~A.}\ \bibnamefont
  {Braithwaite}},\ }\bibfield  {title} {\bibinfo {title}
  {Solid-solution-hardening of tantalum-base alloys},\ }\href
  {https://doi.org/10.1179/msc.1970.4.1.37} {\bibfield  {journal} {\bibinfo
  {journal} {Metal Science Journal}\ }\textbf {\bibinfo {volume} {4}},\
  \bibinfo {pages} {37} (\bibinfo {year} {1970})}\BibitemShut {NoStop}%
\bibitem [{\citenamefont {Gypen}\ and\ \citenamefont
  {Deruyttere}(1982)}]{gypen_1982_jlcm}%
  \BibitemOpen
  \bibfield  {author} {\bibinfo {author} {\bibfnamefont {L.}~\bibnamefont
  {Gypen}}\ and\ \bibinfo {author} {\bibfnamefont {A.}~\bibnamefont
  {Deruyttere}},\ }\bibfield  {title} {\bibinfo {title} {Thermally activated
  deformation in tantalum-base solid solutions},\ }\href
  {https://doi.org/10.1016/0022-5088(82)90208-9} {\bibfield  {journal}
  {\bibinfo  {journal} {Journal of the Less Common Metals}\ }\textbf {\bibinfo
  {volume} {86}},\ \bibinfo {pages} {219} (\bibinfo {year} {1982})}\BibitemShut
  {NoStop}%
\bibitem [{\citenamefont {Pink}\ and\ \citenamefont
  {Arsenault}(1972)}]{pink_1972_msj}%
  \BibitemOpen
  \bibfield  {author} {\bibinfo {author} {\bibfnamefont {E.}~\bibnamefont
  {Pink}}\ and\ \bibinfo {author} {\bibfnamefont {R.~J.}\ \bibnamefont
  {Arsenault}},\ }\bibfield  {title} {\bibinfo {title} {{Solid-Solution
  Strengthening and Weakening of Vanadium--Titanium Alloys}},\ }\href
  {https://doi.org/10.1179/030634572790445777} {\bibfield  {journal} {\bibinfo
  {journal} {Metal Science Journal}\ }\textbf {\bibinfo {volume} {6}},\
  \bibinfo {pages} {1} (\bibinfo {year} {1972})}\BibitemShut {NoStop}%
\bibitem [{\citenamefont {Owen}\ \emph {et~al.}(1987)\citenamefont {Owen},
  \citenamefont {Spitzig},\ and\ \citenamefont {Buck}}]{owen_1987_mta}%
  \BibitemOpen
  \bibfield  {author} {\bibinfo {author} {\bibfnamefont {C.~V.}\ \bibnamefont
  {Owen}}, \bibinfo {author} {\bibfnamefont {W.~A.}\ \bibnamefont {Spitzig}},\
  and\ \bibinfo {author} {\bibfnamefont {O.}~\bibnamefont {Buck}},\ }\bibfield
  {title} {\bibinfo {title} {{effects of hydrogen on low temperature hardening
  and embrittlement of V-Cr alloys}},\ }\href
  {https://doi.org/10.1007/bf02646143} {\bibfield  {journal} {\bibinfo
  {journal} {Metallurgical Transactions A}\ }\textbf {\bibinfo {volume} {18}},\
  \bibinfo {pages} {1593} (\bibinfo {year} {1987})}\BibitemShut {NoStop}%
\bibitem [{\citenamefont {Owen}\ \emph {et~al.}(1985)\citenamefont {Owen},
  \citenamefont {Rowland},\ and\ \citenamefont {Buck}}]{owen_1985_mmta}%
  \BibitemOpen
  \bibfield  {author} {\bibinfo {author} {\bibfnamefont {C.~V.}\ \bibnamefont
  {Owen}}, \bibinfo {author} {\bibfnamefont {T.~J.}\ \bibnamefont {Rowland}},\
  and\ \bibinfo {author} {\bibfnamefont {O.}~\bibnamefont {Buck}},\ }\bibfield
  {title} {\bibinfo {title} {{Effects of hydrogen on some mechanical properties
  of vanadium-titanium alloys}},\ }\href {https://doi.org/10.1007/bf02656712}
  {\bibfield  {journal} {\bibinfo  {journal} {Metallurgical Transactions A}\
  }\textbf {\bibinfo {volume} {16}},\ \bibinfo {pages} {59} (\bibinfo {year}
  {1985})}\BibitemShut {NoStop}%
\bibitem [{\citenamefont {Smialek}\ \emph {et~al.}(1970)\citenamefont
  {Smialek}, \citenamefont {Webb},\ and\ \citenamefont
  {Mitchell}}]{smialek_1970_sm}%
  \BibitemOpen
  \bibfield  {author} {\bibinfo {author} {\bibfnamefont {R.}~\bibnamefont
  {Smialek}}, \bibinfo {author} {\bibfnamefont {G.}~\bibnamefont {Webb}},\ and\
  \bibinfo {author} {\bibfnamefont {T.}~\bibnamefont {Mitchell}},\ }\bibfield
  {title} {\bibinfo {title} {Solid solution softening in bcc metal alloys},\
  }\href {https://doi.org/10.1016/0036-9748(70)90139-0} {\bibfield  {journal}
  {\bibinfo  {journal} {Scripta Metallurgica}\ }\textbf {\bibinfo {volume}
  {4}},\ \bibinfo {pages} {33} (\bibinfo {year} {1970})}\BibitemShut {NoStop}%
\bibitem [{\citenamefont {Stephens}\ and\ \citenamefont
  {Witzke}(1971)}]{stephens_1971_jlcm}%
  \BibitemOpen
  \bibfield  {author} {\bibinfo {author} {\bibfnamefont {J.}~\bibnamefont
  {Stephens}}\ and\ \bibinfo {author} {\bibfnamefont {W.}~\bibnamefont
  {Witzke}},\ }\bibfield  {title} {\bibinfo {title} {Alloy softening in group
  via metals alloyed with rhenium},\ }\href
  {https://doi.org/10.1016/0022-5088(71)90043-9} {\bibfield  {journal}
  {\bibinfo  {journal} {Journal of the Less Common Metals}\ }\textbf {\bibinfo
  {volume} {23}},\ \bibinfo {pages} {325} (\bibinfo {year} {1971})}\BibitemShut
  {NoStop}%
\bibitem [{\citenamefont {Raffo}(1969)}]{raffo_1969_jlcm}%
  \BibitemOpen
  \bibfield  {author} {\bibinfo {author} {\bibfnamefont {P.}~\bibnamefont
  {Raffo}},\ }\bibfield  {title} {\bibinfo {title} {Yielding and fracture in
  tungsten and tungsten-rhenium alloys},\ }\href
  {https://doi.org/10.1016/0022-5088(69)90047-2} {\bibfield  {journal}
  {\bibinfo  {journal} {Journal of the Less Common Metals}\ }\textbf {\bibinfo
  {volume} {17}},\ \bibinfo {pages} {133} (\bibinfo {year} {1969})}\BibitemShut
  {NoStop}%
\bibitem [{\citenamefont {Davidson}\ and\ \citenamefont
  {Brotzen}(1970)}]{davidson_1970_am}%
  \BibitemOpen
  \bibfield  {author} {\bibinfo {author} {\bibfnamefont {D.}~\bibnamefont
  {Davidson}}\ and\ \bibinfo {author} {\bibfnamefont {F.}~\bibnamefont
  {Brotzen}},\ }\bibfield  {title} {\bibinfo {title} {Plastic deformation of
  molybdenum-rhenium alloy crystals},\ }\href
  {https://doi.org/10.1016/0001-6160(70)90132-x} {\bibfield  {journal}
  {\bibinfo  {journal} {Acta Metallurgica}\ }\textbf {\bibinfo {volume} {18}},\
  \bibinfo {pages} {463} (\bibinfo {year} {1970})}\BibitemShut {NoStop}%
\bibitem [{\citenamefont {P{\"o}hl}\ \emph {et~al.}(2013)\citenamefont
  {P{\"o}hl}, \citenamefont {Schatte},\ and\ \citenamefont
  {Leitner}}]{pohl_2013_jac}%
  \BibitemOpen
  \bibfield  {author} {\bibinfo {author} {\bibfnamefont {C.}~\bibnamefont
  {P{\"o}hl}}, \bibinfo {author} {\bibfnamefont {J.}~\bibnamefont {Schatte}},\
  and\ \bibinfo {author} {\bibfnamefont {H.}~\bibnamefont {Leitner}},\
  }\bibfield  {title} {\bibinfo {title} {Solid solution softening of
  polycrystalline molybdenum--hafnium alloys},\ }\href
  {https://doi.org/10.1016/j.jallcom.2013.04.138} {\bibfield  {journal}
  {\bibinfo  {journal} {Journal of Alloys and Compounds}\ }\textbf {\bibinfo
  {volume} {576}},\ \bibinfo {pages} {250} (\bibinfo {year}
  {2013})}\BibitemShut {NoStop}%
\bibitem [{\citenamefont {Klopp}(1968)}]{klopp_1968_nasa}%
  \BibitemOpen
  \bibfield  {author} {\bibinfo {author} {\bibfnamefont {W.~D.}\ \bibnamefont
  {Klopp}},\ }\href@noop {} {\emph {\bibinfo {title} {Review of Ductilizing of
  Group {VIA} Elements by Rhenium and other Solutes}}},\ \bibinfo {number}
  {D-4955}\ (\bibinfo  {publisher} {NASA},\ \bibinfo {year} {1968})\BibitemShut
  {NoStop}%
\bibitem [{\citenamefont {Mendelev}\ \emph {et~al.}(2007)\citenamefont
  {Mendelev}, \citenamefont {Han}, \citenamefont {Son}, \citenamefont
  {Ackland},\ and\ \citenamefont {J.}}]{mendelev_2007_prb}%
  \BibitemOpen
  \bibfield  {author} {\bibinfo {author} {\bibfnamefont {M.~I.}\ \bibnamefont
  {Mendelev}}, \bibinfo {author} {\bibfnamefont {S.}~\bibnamefont {Han}},
  \bibinfo {author} {\bibfnamefont {W.-J.}\ \bibnamefont {Son}}, \bibinfo
  {author} {\bibfnamefont {G.~J.}\ \bibnamefont {Ackland}},\ and\ \bibinfo
  {author} {\bibfnamefont {D.}~\bibnamefont {J.}, \bibfnamefont {Srolovitz}},\
  }\bibfield  {title} {\bibinfo {title} {Simulation of the interaction between
  {Fe} impurities and point defects in {V}},\ }\href
  {https://doi.org/10.1103/physrevb.76.214105} {\bibfield  {journal} {\bibinfo
  {journal} {Physical Review B}\ }\textbf {\bibinfo {volume} {76}},\ \bibinfo
  {pages} {214105} (\bibinfo {year} {2007})}\BibitemShut {NoStop}%
\bibitem [{\citenamefont {Lee}\ \emph {et~al.}(2001)\citenamefont {Lee},
  \citenamefont {Baskes}, \citenamefont {Kim},\ and\ \citenamefont {{Koo
  Cho}}}]{lee_2001_prb}%
  \BibitemOpen
  \bibfield  {author} {\bibinfo {author} {\bibfnamefont {B.-J.}\ \bibnamefont
  {Lee}}, \bibinfo {author} {\bibfnamefont {M.~I.}\ \bibnamefont {Baskes}},
  \bibinfo {author} {\bibfnamefont {H.}~\bibnamefont {Kim}},\ and\ \bibinfo
  {author} {\bibfnamefont {Y.}~\bibnamefont {{Koo Cho}}},\ }\bibfield  {title}
  {\bibinfo {title} {Second nearest-neighbor modified embedded atom method
  potentials for bcc transition metals},\ }\href
  {https://doi.org/10.1103/PhysRevB.64.184102} {\bibfield  {journal} {\bibinfo
  {journal} {Physical Review B}\ }\textbf {\bibinfo {volume} {64}},\ \bibinfo
  {pages} {184102} (\bibinfo {year} {2001})}\BibitemShut {NoStop}%
\bibitem [{\citenamefont {Maisel}\ \emph {et~al.}(2017)\citenamefont {Maisel},
  \citenamefont {Ko}, \citenamefont {Zhang}, \citenamefont {Grabowski},\ and\
  \citenamefont {Neugebauer}}]{maisel_2017_prm}%
  \BibitemOpen
  \bibfield  {author} {\bibinfo {author} {\bibfnamefont {S.~B.}\ \bibnamefont
  {Maisel}}, \bibinfo {author} {\bibfnamefont {W.-S.}\ \bibnamefont {Ko}},
  \bibinfo {author} {\bibfnamefont {J.-L.}\ \bibnamefont {Zhang}}, \bibinfo
  {author} {\bibfnamefont {B.}~\bibnamefont {Grabowski}},\ and\ \bibinfo
  {author} {\bibfnamefont {J.}~\bibnamefont {Neugebauer}},\ }\bibfield  {title}
  {\bibinfo {title} {Thermomechanical response of {NiTi} shape-memory
  nanoprecipitates in {TiV} alloys},\ }\href
  {https://doi.org/10.1103/physrevmaterials.1.033610} {\bibfield  {journal}
  {\bibinfo  {journal} {Physical Review Materials}\ }\textbf {\bibinfo {volume}
  {1}},\ \bibinfo {pages} {033610} (\bibinfo {year} {2017})}\BibitemShut
  {NoStop}%
\bibitem [{\citenamefont {Lin}\ \emph {et~al.}(2014{\natexlab{a}})\citenamefont
  {Lin}, \citenamefont {Mrovec},\ and\ \citenamefont
  {Vitek}}]{vitek_2014_msmse}%
  \BibitemOpen
  \bibfield  {author} {\bibinfo {author} {\bibfnamefont {Y.-S.}\ \bibnamefont
  {Lin}}, \bibinfo {author} {\bibfnamefont {M.}~\bibnamefont {Mrovec}},\ and\
  \bibinfo {author} {\bibfnamefont {V.}~\bibnamefont {Vitek}},\ }\bibfield
  {title} {\bibinfo {title} {A new method for development of bond-order
  potentials for transition bcc metals},\ }\href
  {https://doi.org/10.1088/0965-0393/22/3/034002} {\bibfield  {journal}
  {\bibinfo  {journal} {Modelling and Simulation in Materials Science and
  Engineering}\ }\textbf {\bibinfo {volume} {22}},\ \bibinfo {pages} {034002}
  (\bibinfo {year} {2014}{\natexlab{a}})}\BibitemShut {NoStop}%
\bibitem [{\citenamefont {Byggm{\"a}star}\ \emph {et~al.}(2020)\citenamefont
  {Byggm{\"a}star}, \citenamefont {Nordlund},\ and\ \citenamefont
  {Djurabekova}}]{byggmastar_2020_prm}%
  \BibitemOpen
  \bibfield  {author} {\bibinfo {author} {\bibfnamefont {J.}~\bibnamefont
  {Byggm{\"a}star}}, \bibinfo {author} {\bibfnamefont {K.}~\bibnamefont
  {Nordlund}},\ and\ \bibinfo {author} {\bibfnamefont {F.}~\bibnamefont
  {Djurabekova}},\ }\bibfield  {title} {\bibinfo {title} {Gaussian
  approximation potentials for body-centered-cubic transition metals},\ }\href
  {https://doi.org/10.1103/physrevmaterials.4.093802} {\bibfield  {journal}
  {\bibinfo  {journal} {Physical Review Materials}\ }\textbf {\bibinfo {volume}
  {4}},\ \bibinfo {pages} {093802} (\bibinfo {year} {2020})}\BibitemShut
  {NoStop}%
\bibitem [{\citenamefont {Zhang}\ \emph {et~al.}(2016)\citenamefont {Zhang},
  \citenamefont {Ashcraft}, \citenamefont {Mendelev}, \citenamefont {Wang},\
  and\ \citenamefont {Kelton}}]{zhang_2016_jcp}%
  \BibitemOpen
  \bibfield  {author} {\bibinfo {author} {\bibfnamefont {Y.}~\bibnamefont
  {Zhang}}, \bibinfo {author} {\bibfnamefont {R.}~\bibnamefont {Ashcraft}},
  \bibinfo {author} {\bibfnamefont {M.}~\bibnamefont {Mendelev}}, \bibinfo
  {author} {\bibfnamefont {C.~Z.}\ \bibnamefont {Wang}},\ and\ \bibinfo
  {author} {\bibfnamefont {K.~F.}\ \bibnamefont {Kelton}},\ }\bibfield  {title}
  {\bibinfo {title} {Experimental and molecular dynamics simulation study of
  structure of liquid and amorphous {Ni}$_{62}${Nb}$_{38}$ alloy},\ }\href
  {https://doi.org/10.1063/1.4968212} {\bibfield  {journal} {\bibinfo
  {journal} {The Journal of Chemical Physics}\ }\textbf {\bibinfo {volume}
  {145}},\ \bibinfo {pages} {204505} (\bibinfo {year} {2016})}\BibitemShut
  {NoStop}%
\bibitem [{\citenamefont {Yang}\ and\ \citenamefont
  {Qi}(2019)}]{yang_2019_cms}%
  \BibitemOpen
  \bibfield  {author} {\bibinfo {author} {\bibfnamefont {C.}~\bibnamefont
  {Yang}}\ and\ \bibinfo {author} {\bibfnamefont {L.}~\bibnamefont {Qi}},\
  }\bibfield  {title} {\bibinfo {title} {Modified embedded-atom method
  potential of niobium for studies on mechanical properties},\ }\href
  {https://doi.org/10.1016/j.commatsci.2019.01.047} {\bibfield  {journal}
  {\bibinfo  {journal} {Computational Materials Science}\ }\textbf {\bibinfo
  {volume} {161}},\ \bibinfo {pages} {351} (\bibinfo {year}
  {2019})}\BibitemShut {NoStop}%
\bibitem [{\citenamefont {Smirnova}\ and\ \citenamefont
  {Starikov}(2017)}]{starikov_2017_cms}%
  \BibitemOpen
  \bibfield  {author} {\bibinfo {author} {\bibfnamefont {D.}~\bibnamefont
  {Smirnova}}\ and\ \bibinfo {author} {\bibfnamefont {S.}~\bibnamefont
  {Starikov}},\ }\bibfield  {title} {\bibinfo {title} {An interatomic potential
  for simulation of {Zr-Nb} system},\ }\href
  {https://doi.org/10.1016/j.commatsci.2016.12.016} {\bibfield  {journal}
  {\bibinfo  {journal} {Computational Materials Science}\ }\textbf {\bibinfo
  {volume} {129}},\ \bibinfo {pages} {259} (\bibinfo {year}
  {2017})}\BibitemShut {NoStop}%
\bibitem [{\citenamefont {Ravelo}\ \emph {et~al.}(2013)\citenamefont {Ravelo},
  \citenamefont {Germann}, \citenamefont {Guerrero}, \citenamefont {An},\ and\
  \citenamefont {Holian}}]{ravelo_2013_prb}%
  \BibitemOpen
  \bibfield  {author} {\bibinfo {author} {\bibfnamefont {R.}~\bibnamefont
  {Ravelo}}, \bibinfo {author} {\bibfnamefont {T.~C.}\ \bibnamefont {Germann}},
  \bibinfo {author} {\bibfnamefont {O.}~\bibnamefont {Guerrero}}, \bibinfo
  {author} {\bibfnamefont {Q.}~\bibnamefont {An}},\ and\ \bibinfo {author}
  {\bibfnamefont {B.~L.}\ \bibnamefont {Holian}},\ }\bibfield  {title}
  {\bibinfo {title} {Shock-induced plasticity in tantalum single crystals{:
  I}nteratomic potentials and large-scale molecular-dynamics simulations},\
  }\href {https://doi.org/10.1103/physrevb.88.134101} {\bibfield  {journal}
  {\bibinfo  {journal} {Physical Review B}\ }\textbf {\bibinfo {volume} {88}},\
  \bibinfo {pages} {134101} (\bibinfo {year} {2013})}\BibitemShut {NoStop}%
\bibitem [{\citenamefont {Li}\ \emph {et~al.}(2003)\citenamefont {Li},
  \citenamefont {Siegel}, \citenamefont {Adams},\ and\ \citenamefont
  {Liu}}]{li_2003_prb}%
  \BibitemOpen
  \bibfield  {author} {\bibinfo {author} {\bibfnamefont {Y.}~\bibnamefont
  {Li}}, \bibinfo {author} {\bibfnamefont {D.~J.}\ \bibnamefont {Siegel}},
  \bibinfo {author} {\bibfnamefont {J.~B.}\ \bibnamefont {Adams}},\ and\
  \bibinfo {author} {\bibfnamefont {X.-Y.}\ \bibnamefont {Liu}},\ }\bibfield
  {title} {\bibinfo {title} {Embedded-atom-method tantalum potential developed
  by the force-matching method},\ }\href
  {https://doi.org/10.1103/PhysRevB.67.125101} {\bibfield  {journal} {\bibinfo
  {journal} {Physical Review B}\ }\textbf {\bibinfo {volume} {67}},\ \bibinfo
  {pages} {125101} (\bibinfo {year} {2003})}\BibitemShut {NoStop}%
\bibitem [{\citenamefont {Zhou}\ \emph {et~al.}(2004)\citenamefont {Zhou},
  \citenamefont {Johnson},\ and\ \citenamefont {Wadley}}]{zhou_2004_prb}%
  \BibitemOpen
  \bibfield  {author} {\bibinfo {author} {\bibfnamefont {X.~W.}\ \bibnamefont
  {Zhou}}, \bibinfo {author} {\bibfnamefont {R.~A.}\ \bibnamefont {Johnson}},\
  and\ \bibinfo {author} {\bibfnamefont {H.~N.~G.}\ \bibnamefont {Wadley}},\
  }\bibfield  {title} {\bibinfo {title} {Misfit-energy-increasing dislocations
  in vapor-deposited {CoFe}/{NiFe} multilayers},\ }\href
  {https://doi.org/10.1103/physrevb.69.144113} {\bibfield  {journal} {\bibinfo
  {journal} {Physical Review B}\ }\textbf {\bibinfo {volume} {69}},\ \bibinfo
  {pages} {144113} (\bibinfo {year} {2004})}\BibitemShut {NoStop}%
\bibitem [{\citenamefont {Chen}\ \emph {et~al.}(2019)\citenamefont {Chen},
  \citenamefont {Fang}, \citenamefont {Liu}, \citenamefont {Hu}, \citenamefont
  {Gao}, \citenamefont {Gao},\ and\ \citenamefont {Deng}}]{chen_2019_cms}%
  \BibitemOpen
  \bibfield  {author} {\bibinfo {author} {\bibfnamefont {Y.}~\bibnamefont
  {Chen}}, \bibinfo {author} {\bibfnamefont {J.}~\bibnamefont {Fang}}, \bibinfo
  {author} {\bibfnamefont {L.}~\bibnamefont {Liu}}, \bibinfo {author}
  {\bibfnamefont {W.}~\bibnamefont {Hu}}, \bibinfo {author} {\bibfnamefont
  {N.}~\bibnamefont {Gao}}, \bibinfo {author} {\bibfnamefont {F.}~\bibnamefont
  {Gao}},\ and\ \bibinfo {author} {\bibfnamefont {H.}~\bibnamefont {Deng}},\
  }\bibfield  {title} {\bibinfo {title} {Development of the interatomic
  potentials for {W-Ta} system},\ }\href
  {https://doi.org/10.1016/j.commatsci.2019.03.021} {\bibfield  {journal}
  {\bibinfo  {journal} {Computational Materials Science}\ }\textbf {\bibinfo
  {volume} {163}},\ \bibinfo {pages} {91} (\bibinfo {year} {2019})}\BibitemShut
  {NoStop}%
\bibitem [{\citenamefont {Pun}\ \emph {et~al.}(2015)\citenamefont {Pun},
  \citenamefont {Darling}, \citenamefont {Kecskes},\ and\ \citenamefont
  {Mishin}}]{pun_2015_am}%
  \BibitemOpen
  \bibfield  {author} {\bibinfo {author} {\bibfnamefont {G.~P.}\ \bibnamefont
  {Pun}}, \bibinfo {author} {\bibfnamefont {K.}~\bibnamefont {Darling}},
  \bibinfo {author} {\bibfnamefont {L.}~\bibnamefont {Kecskes}},\ and\ \bibinfo
  {author} {\bibfnamefont {Y.}~\bibnamefont {Mishin}},\ }\bibfield  {title}
  {\bibinfo {title} {Angular-dependent interatomic potential for the {Cu--Ta}
  system and its application to structural stability of nano-crystalline
  alloys},\ }\href {https://doi.org/10.1016/j.actamat.2015.08.052} {\bibfield
  {journal} {\bibinfo  {journal} {Acta Materialia}\ }\textbf {\bibinfo {volume}
  {100}},\ \bibinfo {pages} {377} (\bibinfo {year} {2015})}\BibitemShut
  {NoStop}%
\bibitem [{\citenamefont {Thompson}\ \emph {et~al.}(2015)\citenamefont
  {Thompson}, \citenamefont {Swiler}, \citenamefont {Trott}, \citenamefont
  {Foiles},\ and\ \citenamefont {Tucker}}]{thompson_2015_jcp}%
  \BibitemOpen
  \bibfield  {author} {\bibinfo {author} {\bibfnamefont {A.~P.}\ \bibnamefont
  {Thompson}}, \bibinfo {author} {\bibfnamefont {L.~P.}\ \bibnamefont
  {Swiler}}, \bibinfo {author} {\bibfnamefont {C.~R.}\ \bibnamefont {Trott}},
  \bibinfo {author} {\bibfnamefont {S.~M.}\ \bibnamefont {Foiles}},\ and\
  \bibinfo {author} {\bibfnamefont {G.~J.}\ \bibnamefont {Tucker}},\ }\bibfield
   {title} {\bibinfo {title} {Spectral neighbor analysis method for automated
  generation of quantum-accurate interatomic potentials},\ }\href
  {https://doi.org/10.1016/j.jcp.2014.12.018} {\bibfield  {journal} {\bibinfo
  {journal} {Journal of Computational Physics}\ }\textbf {\bibinfo {volume}
  {285}},\ \bibinfo {pages} {316} (\bibinfo {year} {2015})}\BibitemShut
  {NoStop}%
\bibitem [{\citenamefont {Choi}\ \emph {et~al.}(2017)\citenamefont {Choi},
  \citenamefont {Kim}, \citenamefont {Seol},\ and\ \citenamefont
  {Lee}}]{choi_2017_cms}%
  \BibitemOpen
  \bibfield  {author} {\bibinfo {author} {\bibfnamefont {W.-M.}\ \bibnamefont
  {Choi}}, \bibinfo {author} {\bibfnamefont {Y.}~\bibnamefont {Kim}}, \bibinfo
  {author} {\bibfnamefont {D.}~\bibnamefont {Seol}},\ and\ \bibinfo {author}
  {\bibfnamefont {B.-J.}\ \bibnamefont {Lee}},\ }\bibfield  {title} {\bibinfo
  {title} {Modified embedded-atom method interatomic potentials for the {Co-Cr,
  Co-Fe, Co-Mn, Cr-Mn and Mn-Ni} binary systems},\ }\href
  {https://doi.org/10.1016/j.commatsci.2017.01.002} {\bibfield  {journal}
  {\bibinfo  {journal} {Computational Materials Science}\ }\textbf {\bibinfo
  {volume} {130}},\ \bibinfo {pages} {121} (\bibinfo {year}
  {2017})}\BibitemShut {NoStop}%
\bibitem [{\citenamefont {Howells}\ and\ \citenamefont
  {Mishin}(2018)}]{mishin_2018_msmse}%
  \BibitemOpen
  \bibfield  {author} {\bibinfo {author} {\bibfnamefont {C.~A.}\ \bibnamefont
  {Howells}}\ and\ \bibinfo {author} {\bibfnamefont {Y.}~\bibnamefont
  {Mishin}},\ }\bibfield  {title} {\bibinfo {title} {Angular-dependent
  interatomic potential for the binary {Ni--Cr} system},\ }\href
  {https://doi.org/10.1088/1361-651x/aae400} {\bibfield  {journal} {\bibinfo
  {journal} {Modelling and Simulation in Materials Science and Engineering}\
  }\textbf {\bibinfo {volume} {26}},\ \bibinfo {pages} {085008} (\bibinfo
  {year} {2018})}\BibitemShut {NoStop}%
\bibitem [{\citenamefont {Ackland}\ \emph {et~al.}(1987)\citenamefont
  {Ackland}, \citenamefont {Tichy}, \citenamefont {Vitek},\ and\ \citenamefont
  {Finnis}}]{ackland_1987_pma}%
  \BibitemOpen
  \bibfield  {author} {\bibinfo {author} {\bibfnamefont {G.~J.}\ \bibnamefont
  {Ackland}}, \bibinfo {author} {\bibfnamefont {G.}~\bibnamefont {Tichy}},
  \bibinfo {author} {\bibfnamefont {V.}~\bibnamefont {Vitek}},\ and\ \bibinfo
  {author} {\bibfnamefont {M.~W.}\ \bibnamefont {Finnis}},\ }\bibfield  {title}
  {\bibinfo {title} {Simple {N}-body potentials for the noble metals and
  nickel},\ }\href {https://doi.org/10.1080/01418618708204485} {\bibfield
  {journal} {\bibinfo  {journal} {Philosophical Magazine A}\ }\textbf {\bibinfo
  {volume} {56}},\ \bibinfo {pages} {735} (\bibinfo {year} {1987})}\BibitemShut
  {NoStop}%
\bibitem [{\citenamefont {Park}\ \emph {et~al.}(2012)\citenamefont {Park},
  \citenamefont {Fellinger}, \citenamefont {Lenosky}, \citenamefont {Tipton},
  \citenamefont {Trinkle}, \citenamefont {Rudin}, \citenamefont {Woodward},
  \citenamefont {Wilkins},\ and\ \citenamefont {Hennig}}]{park_2012_prb}%
  \BibitemOpen
  \bibfield  {author} {\bibinfo {author} {\bibfnamefont {H.}~\bibnamefont
  {Park}}, \bibinfo {author} {\bibfnamefont {M.~R.}\ \bibnamefont {Fellinger}},
  \bibinfo {author} {\bibfnamefont {T.~J.}\ \bibnamefont {Lenosky}}, \bibinfo
  {author} {\bibfnamefont {W.~W.}\ \bibnamefont {Tipton}}, \bibinfo {author}
  {\bibfnamefont {D.~R.}\ \bibnamefont {Trinkle}}, \bibinfo {author}
  {\bibfnamefont {S.~P.}\ \bibnamefont {Rudin}}, \bibinfo {author}
  {\bibfnamefont {C.}~\bibnamefont {Woodward}}, \bibinfo {author}
  {\bibfnamefont {J.~W.}\ \bibnamefont {Wilkins}},\ and\ \bibinfo {author}
  {\bibfnamefont {R.~G.}\ \bibnamefont {Hennig}},\ }\bibfield  {title}
  {\bibinfo {title} {{\it Ab initio} based empirical potential used to study
  the mechanical properties of molybdenum},\ }\href
  {https://doi.org/10.1103/PhysRevB.85.214121} {\bibfield  {journal} {\bibinfo
  {journal} {Physical Review B}\ }\textbf {\bibinfo {volume} {85}},\ \bibinfo
  {pages} {214121} (\bibinfo {year} {2012})}\BibitemShut {NoStop}%
\bibitem [{\citenamefont {Kim}\ \emph {et~al.}(2017)\citenamefont {Kim},
  \citenamefont {Seol}, \citenamefont {Ji}, \citenamefont {Jang}, \citenamefont
  {Kim},\ and\ \citenamefont {Lee}}]{kim_2017_calphad}%
  \BibitemOpen
  \bibfield  {author} {\bibinfo {author} {\bibfnamefont {J.-S.}\ \bibnamefont
  {Kim}}, \bibinfo {author} {\bibfnamefont {D.}~\bibnamefont {Seol}}, \bibinfo
  {author} {\bibfnamefont {J.}~\bibnamefont {Ji}}, \bibinfo {author}
  {\bibfnamefont {H.-S.}\ \bibnamefont {Jang}}, \bibinfo {author}
  {\bibfnamefont {Y.}~\bibnamefont {Kim}},\ and\ \bibinfo {author}
  {\bibfnamefont {B.-J.}\ \bibnamefont {Lee}},\ }\bibfield  {title} {\bibinfo
  {title} {Second nearest-neighbor modified embedded-atom method interatomic
  potentials for the {Pt-M (M = Al, Co, Cu, Mo, Ni, Ti, V)} binary systems},\
  }\href {https://doi.org/10.1016/j.calphad.2017.09.005} {\bibfield  {journal}
  {\bibinfo  {journal} {Calphad}\ }\textbf {\bibinfo {volume} {59}},\ \bibinfo
  {pages} {131} (\bibinfo {year} {2017})}\BibitemShut {NoStop}%
\bibitem [{\citenamefont {Chen}\ \emph {et~al.}(2017)\citenamefont {Chen},
  \citenamefont {Deng}, \citenamefont {Tran}, \citenamefont {Tang},
  \citenamefont {Chu},\ and\ \citenamefont {Ong}}]{chen_2017_prm}%
  \BibitemOpen
  \bibfield  {author} {\bibinfo {author} {\bibfnamefont {C.}~\bibnamefont
  {Chen}}, \bibinfo {author} {\bibfnamefont {Z.}~\bibnamefont {Deng}}, \bibinfo
  {author} {\bibfnamefont {R.}~\bibnamefont {Tran}}, \bibinfo {author}
  {\bibfnamefont {H.}~\bibnamefont {Tang}}, \bibinfo {author} {\bibfnamefont
  {I.-H.}\ \bibnamefont {Chu}},\ and\ \bibinfo {author} {\bibfnamefont {S.~P.}\
  \bibnamefont {Ong}},\ }\bibfield  {title} {\bibinfo {title} {Accurate force
  field for molybdenum by machine learning large materials data},\ }\href
  {https://doi.org/10.1103/PhysRevMaterials.1.043603} {\bibfield  {journal}
  {\bibinfo  {journal} {Physical Review Materials}\ }\textbf {\bibinfo {volume}
  {1}},\ \bibinfo {pages} {043603} (\bibinfo {year} {2017})}\BibitemShut
  {NoStop}%
\bibitem [{\citenamefont {Olsson}(2009)}]{olsson_2009_cms}%
  \BibitemOpen
  \bibfield  {author} {\bibinfo {author} {\bibfnamefont {P.~A.}\ \bibnamefont
  {Olsson}},\ }\bibfield  {title} {\bibinfo {title} {Semi-empirical atomistic
  study of point defect properties in {BCC} transition metals},\ }\href
  {https://doi.org/10.1016/j.commatsci.2009.06.025} {\bibfield  {journal}
  {\bibinfo  {journal} {Computational Materials Science}\ }\textbf {\bibinfo
  {volume} {47}},\ \bibinfo {pages} {135} (\bibinfo {year} {2009})}\BibitemShut
  {NoStop}%
\bibitem [{\citenamefont {Marinica}\ \emph {et~al.}(2013)\citenamefont
  {Marinica}, \citenamefont {Ventelon}, \citenamefont {Gilbert}, \citenamefont
  {Proville}, \citenamefont {Dudarev}, \citenamefont {Marian}, \citenamefont
  {Bencteux},\ and\ \citenamefont {Willaime}}]{marinica_2013_jpcm}%
  \BibitemOpen
  \bibfield  {author} {\bibinfo {author} {\bibfnamefont {M.-C.}\ \bibnamefont
  {Marinica}}, \bibinfo {author} {\bibfnamefont {L.}~\bibnamefont {Ventelon}},
  \bibinfo {author} {\bibfnamefont {M.~R.}\ \bibnamefont {Gilbert}}, \bibinfo
  {author} {\bibfnamefont {L.}~\bibnamefont {Proville}}, \bibinfo {author}
  {\bibfnamefont {S.~L.}\ \bibnamefont {Dudarev}}, \bibinfo {author}
  {\bibfnamefont {J.}~\bibnamefont {Marian}}, \bibinfo {author} {\bibfnamefont
  {G.}~\bibnamefont {Bencteux}},\ and\ \bibinfo {author} {\bibfnamefont
  {F.}~\bibnamefont {Willaime}},\ }\bibfield  {title} {\bibinfo {title}
  {Interatomic potentials for modelling radiation defects and dislocations in
  tungsten},\ }\href {https://doi.org/10.1088/0953-8984/25/39/395502}
  {\bibfield  {journal} {\bibinfo  {journal} {Journal of Physics: Condensed
  Matter}\ }\textbf {\bibinfo {volume} {25}},\ \bibinfo {pages} {395502}
  (\bibinfo {year} {2013})}\BibitemShut {NoStop}%
\bibitem [{\citenamefont {Bonny}\ \emph {et~al.}(2017)\citenamefont {Bonny},
  \citenamefont {Bakaev}, \citenamefont {Terentyev},\ and\ \citenamefont
  {Mastrikov}}]{bonny_2017_jap}%
  \BibitemOpen
  \bibfield  {author} {\bibinfo {author} {\bibfnamefont {G.}~\bibnamefont
  {Bonny}}, \bibinfo {author} {\bibfnamefont {A.}~\bibnamefont {Bakaev}},
  \bibinfo {author} {\bibfnamefont {D.}~\bibnamefont {Terentyev}},\ and\
  \bibinfo {author} {\bibfnamefont {Y.~A.}\ \bibnamefont {Mastrikov}},\
  }\bibfield  {title} {\bibinfo {title} {Interatomic potential to study plastic
  deformation in tungsten-rhenium alloys},\ }\href
  {https://doi.org/10.1063/1.4982361} {\bibfield  {journal} {\bibinfo
  {journal} {Journal of Applied Physics}\ }\textbf {\bibinfo {volume} {121}},\
  \bibinfo {pages} {165107} (\bibinfo {year} {2017})}\BibitemShut {NoStop}%
\bibitem [{\citenamefont {Lenosky}(2017)}]{lenosky_2017_openkim}%
  \BibitemOpen
  \bibfield  {author} {\bibinfo {author} {\bibfnamefont {T.}~\bibnamefont
  {Lenosky}},\ }\bibfield  {title} {\bibinfo {title} {{LAMMPS MEAM P}otential
  for {W} developed by {L}enosky (2017) v000},\ }\bibfield  {journal} {\bibinfo
   {journal} {OpenKIM}\ }\href {https://doi.org/10.25950/264A7B3E}
  {10.25950/264A7B3E} (\bibinfo {year} {2017})\BibitemShut {NoStop}%
\bibitem [{\citenamefont {Park}\ \emph {et~al.}(2021)\citenamefont {Park},
  \citenamefont {Fellinger}, \citenamefont {Lenosky}, \citenamefont {Tipton},
  \citenamefont {Trinkle}, \citenamefont {Rudin}, \citenamefont {Woodward},
  \citenamefont {Wilkins},\ and\ \citenamefont {Hennig}}]{park_2012_openkim}%
  \BibitemOpen
  \bibfield  {author} {\bibinfo {author} {\bibfnamefont {H.}~\bibnamefont
  {Park}}, \bibinfo {author} {\bibfnamefont {M.}~\bibnamefont {Fellinger}},
  \bibinfo {author} {\bibfnamefont {T.}~\bibnamefont {Lenosky}}, \bibinfo
  {author} {\bibfnamefont {W.}~\bibnamefont {Tipton}}, \bibinfo {author}
  {\bibfnamefont {D.}~\bibnamefont {Trinkle}}, \bibinfo {author} {\bibfnamefont
  {S.~P.}\ \bibnamefont {Rudin}}, \bibinfo {author} {\bibfnamefont
  {C.}~\bibnamefont {Woodward}}, \bibinfo {author} {\bibfnamefont
  {J.}~\bibnamefont {Wilkins}},\ and\ \bibinfo {author} {\bibfnamefont
  {R.}~\bibnamefont {Hennig}},\ }\href {https://doi.org/10.25950/A4A622CF}
  {\bibinfo {title} {{MEAM }potential for {W} developed by {P}ark et al. (2012)
  v001}} (\bibinfo {year} {2021})\BibitemShut {NoStop}%
\bibitem [{\citenamefont {Morris}\ \emph {et~al.}(2008)\citenamefont {Morris},
  \citenamefont {Aga}, \citenamefont {Levashov},\ and\ \citenamefont
  {Egami}}]{morris_2008_prb}%
  \BibitemOpen
  \bibfield  {author} {\bibinfo {author} {\bibfnamefont {J.~R.}\ \bibnamefont
  {Morris}}, \bibinfo {author} {\bibfnamefont {R.~S.}\ \bibnamefont {Aga}},
  \bibinfo {author} {\bibfnamefont {V.}~\bibnamefont {Levashov}},\ and\
  \bibinfo {author} {\bibfnamefont {T.}~\bibnamefont {Egami}},\ }\bibfield
  {title} {\bibinfo {title} {Many-body effects in bcc metals{: A}n embedded
  atom model extension of the modified {Johnson} pair potential for iron},\
  }\href {https://doi.org/10.1103/physrevb.77.174201} {\bibfield  {journal}
  {\bibinfo  {journal} {Physical Review B}\ }\textbf {\bibinfo {volume} {77}},\
  \bibinfo {pages} {174201} (\bibinfo {year} {2008})}\BibitemShut {NoStop}%
\bibitem [{\citenamefont {Proville}\ \emph {et~al.}(2012)\citenamefont
  {Proville}, \citenamefont {Rodney},\ and\ \citenamefont
  {Marinica}}]{proville_2012_nm}%
  \BibitemOpen
  \bibfield  {author} {\bibinfo {author} {\bibfnamefont {L.}~\bibnamefont
  {Proville}}, \bibinfo {author} {\bibfnamefont {D.}~\bibnamefont {Rodney}},\
  and\ \bibinfo {author} {\bibfnamefont {M.-C.}\ \bibnamefont {Marinica}},\
  }\bibfield  {title} {\bibinfo {title} {Quantum effect on thermally activated
  glide of dislocations},\ }\href {https://doi.org/10.1038/nmat3401} {\bibfield
   {journal} {\bibinfo  {journal} {Nature Materials}\ }\textbf {\bibinfo
  {volume} {11}},\ \bibinfo {pages} {845} (\bibinfo {year} {2012})}\BibitemShut
  {NoStop}%
\bibitem [{\citenamefont {Bonny}\ \emph {et~al.}(2009)\citenamefont {Bonny},
  \citenamefont {Pasianot},\ and\ \citenamefont {Malerba}}]{bonny_2009_pm}%
  \BibitemOpen
  \bibfield  {author} {\bibinfo {author} {\bibfnamefont {G.}~\bibnamefont
  {Bonny}}, \bibinfo {author} {\bibfnamefont {R.~C.}\ \bibnamefont
  {Pasianot}},\ and\ \bibinfo {author} {\bibfnamefont {L.}~\bibnamefont
  {Malerba}},\ }\bibfield  {title} {\bibinfo {title} {Fitting interatomic
  potentials consistent with thermodynamics: {Fe, Cu, Ni} and their alloys},\
  }\href {https://doi.org/10.1080/14786430903299337} {\bibfield  {journal}
  {\bibinfo  {journal} {Philosophical Magazine}\ }\textbf {\bibinfo {volume}
  {89}},\ \bibinfo {pages} {3451} (\bibinfo {year} {2009})}\BibitemShut
  {NoStop}%
\bibitem [{\citenamefont {Chiesa}\ \emph {et~al.}(2011)\citenamefont {Chiesa},
  \citenamefont {Derlet}, \citenamefont {Dudarev},\ and\ \citenamefont
  {Swygenhoven}}]{chiesa_2011_jpcm}%
  \BibitemOpen
  \bibfield  {author} {\bibinfo {author} {\bibfnamefont {S.}~\bibnamefont
  {Chiesa}}, \bibinfo {author} {\bibfnamefont {P.~M.}\ \bibnamefont {Derlet}},
  \bibinfo {author} {\bibfnamefont {S.~L.}\ \bibnamefont {Dudarev}},\ and\
  \bibinfo {author} {\bibfnamefont {H.~V.}\ \bibnamefont {Swygenhoven}},\
  }\bibfield  {title} {\bibinfo {title} {Optimization of the magnetic potential
  for $\alpha${-Fe}},\ }\href {https://doi.org/10.1088/0953-8984/23/20/206001}
  {\bibfield  {journal} {\bibinfo  {journal} {Journal of Physics: Condensed
  Matter}\ }\textbf {\bibinfo {volume} {23}},\ \bibinfo {pages} {206001}
  (\bibinfo {year} {2011})}\BibitemShut {NoStop}%
\bibitem [{\citenamefont {Chamati}\ \emph {et~al.}(2006)\citenamefont
  {Chamati}, \citenamefont {Papanicolaou}, \citenamefont {Mishin},\ and\
  \citenamefont {Papaconstantopoulos}}]{chamati_2006_ss}%
  \BibitemOpen
  \bibfield  {author} {\bibinfo {author} {\bibfnamefont {H.}~\bibnamefont
  {Chamati}}, \bibinfo {author} {\bibfnamefont {N.}~\bibnamefont
  {Papanicolaou}}, \bibinfo {author} {\bibfnamefont {Y.}~\bibnamefont
  {Mishin}},\ and\ \bibinfo {author} {\bibfnamefont {D.}~\bibnamefont
  {Papaconstantopoulos}},\ }\bibfield  {title} {\bibinfo {title} {Embedded-atom
  potential for {F}e and its application to self-diffusion on {F}e(100)},\
  }\href {https://doi.org/10.1016/j.susc.2006.02.010} {\bibfield  {journal}
  {\bibinfo  {journal} {Surface Science}\ }\textbf {\bibinfo {volume} {600}},\
  \bibinfo {pages} {1793} (\bibinfo {year} {2006})}\BibitemShut {NoStop}%
\bibitem [{\citenamefont {Meyer}\ and\ \citenamefont
  {Entel}(1998)}]{meyer_1998_prb}%
  \BibitemOpen
  \bibfield  {author} {\bibinfo {author} {\bibfnamefont {R.}~\bibnamefont
  {Meyer}}\ and\ \bibinfo {author} {\bibfnamefont {P.}~\bibnamefont {Entel}},\
  }\bibfield  {title} {\bibinfo {title} {Martensite-austenite transition and
  phonon dispersion curves of {Fe}$_{1-x}${Ni}$_x$ studied by
  molecular-dynamics simulations},\ }\href
  {https://doi.org/10.1103/physrevb.57.5140} {\bibfield  {journal} {\bibinfo
  {journal} {Physical Review B}\ }\textbf {\bibinfo {volume} {57}},\ \bibinfo
  {pages} {5140} (\bibinfo {year} {1998})}\BibitemShut {NoStop}%
\bibitem [{\citenamefont {Ackland}\ \emph {et~al.}(1997)\citenamefont
  {Ackland}, \citenamefont {Bacon}, \citenamefont {Calder},\ and\ \citenamefont
  {Harry}}]{ackland_1997_pma}%
  \BibitemOpen
  \bibfield  {author} {\bibinfo {author} {\bibfnamefont {G.~J.}\ \bibnamefont
  {Ackland}}, \bibinfo {author} {\bibfnamefont {D.~J.}\ \bibnamefont {Bacon}},
  \bibinfo {author} {\bibfnamefont {A.~F.}\ \bibnamefont {Calder}},\ and\
  \bibinfo {author} {\bibfnamefont {T.}~\bibnamefont {Harry}},\ }\bibfield
  {title} {\bibinfo {title} {Computer simulation of point defect properties in
  dilute {Fe--Cu} alloy using a many-body interatomic potential},\ }\href
  {https://doi.org/10.1080/01418619708207198} {\bibfield  {journal} {\bibinfo
  {journal} {Philosophical Magazine A}\ }\textbf {\bibinfo {volume} {75}},\
  \bibinfo {pages} {713} (\bibinfo {year} {1997})}\BibitemShut {NoStop}%
\bibitem [{\citenamefont {Malerba}\ \emph {et~al.}(2010)\citenamefont
  {Malerba}, \citenamefont {Marinica}, \citenamefont {Anento}, \citenamefont
  {Bj{\"o}rkas}, \citenamefont {Nguyen}, \citenamefont {Domain}, \citenamefont
  {Djurabekova}, \citenamefont {Olsson}, \citenamefont {Nordlund},
  \citenamefont {Serra}, \citenamefont {Terentyev}, \citenamefont {Willaime},\
  and\ \citenamefont {Becquart}}]{malerba_2010_jnm}%
  \BibitemOpen
  \bibfield  {author} {\bibinfo {author} {\bibfnamefont {L.}~\bibnamefont
  {Malerba}}, \bibinfo {author} {\bibfnamefont {M.}~\bibnamefont {Marinica}},
  \bibinfo {author} {\bibfnamefont {N.}~\bibnamefont {Anento}}, \bibinfo
  {author} {\bibfnamefont {C.}~\bibnamefont {Bj{\"o}rkas}}, \bibinfo {author}
  {\bibfnamefont {H.}~\bibnamefont {Nguyen}}, \bibinfo {author} {\bibfnamefont
  {C.}~\bibnamefont {Domain}}, \bibinfo {author} {\bibfnamefont
  {F.}~\bibnamefont {Djurabekova}}, \bibinfo {author} {\bibfnamefont
  {P.}~\bibnamefont {Olsson}}, \bibinfo {author} {\bibfnamefont
  {K.}~\bibnamefont {Nordlund}}, \bibinfo {author} {\bibfnamefont
  {A.}~\bibnamefont {Serra}}, \bibinfo {author} {\bibfnamefont
  {D.}~\bibnamefont {Terentyev}}, \bibinfo {author} {\bibfnamefont
  {F.}~\bibnamefont {Willaime}},\ and\ \bibinfo {author} {\bibfnamefont
  {C.}~\bibnamefont {Becquart}},\ }\bibfield  {title} {\bibinfo {title}
  {Comparison of empirical interatomic potentials for iron applied to radiation
  damage studies},\ }\href {https://doi.org/10.1016/j.jnucmat.2010.05.017}
  {\bibfield  {journal} {\bibinfo  {journal} {Journal of Nuclear Materials}\
  }\textbf {\bibinfo {volume} {406}},\ \bibinfo {pages} {19} (\bibinfo {year}
  {2010})}\BibitemShut {NoStop}%
\bibitem [{\citenamefont {Starikov}\ \emph {et~al.}(2021)\citenamefont
  {Starikov}, \citenamefont {Smirnova}, \citenamefont {Pradhan}, \citenamefont
  {Lysogorskiy}, \citenamefont {Chapman}, \citenamefont {Mrovec},\ and\
  \citenamefont {Drautz}}]{starikov_2021_prm}%
  \BibitemOpen
  \bibfield  {author} {\bibinfo {author} {\bibfnamefont {S.}~\bibnamefont
  {Starikov}}, \bibinfo {author} {\bibfnamefont {D.}~\bibnamefont {Smirnova}},
  \bibinfo {author} {\bibfnamefont {T.}~\bibnamefont {Pradhan}}, \bibinfo
  {author} {\bibfnamefont {Y.}~\bibnamefont {Lysogorskiy}}, \bibinfo {author}
  {\bibfnamefont {H.}~\bibnamefont {Chapman}}, \bibinfo {author} {\bibfnamefont
  {M.}~\bibnamefont {Mrovec}},\ and\ \bibinfo {author} {\bibfnamefont
  {R.}~\bibnamefont {Drautz}},\ }\bibfield  {title} {\bibinfo {title}
  {Angular-dependent interatomic potential for large-scale atomistic simulation
  of iron: {D}evelopment and comprehensive comparison with existing interatomic
  models},\ }\href {https://doi.org/10.1103/physrevmaterials.5.063607}
  {\bibfield  {journal} {\bibinfo  {journal} {Physical Review Materials}\
  }\textbf {\bibinfo {volume} {5}},\ \bibinfo {pages} {063607} (\bibinfo {year}
  {2021})}\BibitemShut {NoStop}%
\bibitem [{\citenamefont {Kim}\ \emph {et~al.}(2009)\citenamefont {Kim},
  \citenamefont {Jung},\ and\ \citenamefont {Lee}}]{kim_2009_am}%
  \BibitemOpen
  \bibfield  {author} {\bibinfo {author} {\bibfnamefont {H.-K.}\ \bibnamefont
  {Kim}}, \bibinfo {author} {\bibfnamefont {W.-S.}\ \bibnamefont {Jung}},\ and\
  \bibinfo {author} {\bibfnamefont {B.-J.}\ \bibnamefont {Lee}},\ }\bibfield
  {title} {\bibinfo {title} {Modified embedded-atom method interatomic
  potentials for the {Fe--Ti--C} and {Fe--Ti--N} ternary systems},\ }\href
  {https://doi.org/10.1016/j.actamat.2009.03.019} {\bibfield  {journal}
  {\bibinfo  {journal} {Acta Materialia}\ }\textbf {\bibinfo {volume} {57}},\
  \bibinfo {pages} {3140} (\bibinfo {year} {2009})}\BibitemShut {NoStop}%
\bibitem [{\citenamefont {Liyanage}\ \emph {et~al.}(2014)\citenamefont
  {Liyanage}, \citenamefont {Kim}, \citenamefont {Houze}, \citenamefont {Kim},
  \citenamefont {Tschopp}, \citenamefont {Baskes},\ and\ \citenamefont
  {Horstemeyer}}]{liyanage_2014_prb}%
  \BibitemOpen
  \bibfield  {author} {\bibinfo {author} {\bibfnamefont {L.~S.~I.}\
  \bibnamefont {Liyanage}}, \bibinfo {author} {\bibfnamefont {S.-G.}\
  \bibnamefont {Kim}}, \bibinfo {author} {\bibfnamefont {J.}~\bibnamefont
  {Houze}}, \bibinfo {author} {\bibfnamefont {S.}~\bibnamefont {Kim}}, \bibinfo
  {author} {\bibfnamefont {M.~A.}\ \bibnamefont {Tschopp}}, \bibinfo {author}
  {\bibfnamefont {M.~I.}\ \bibnamefont {Baskes}},\ and\ \bibinfo {author}
  {\bibfnamefont {M.~F.}\ \bibnamefont {Horstemeyer}},\ }\bibfield  {title}
  {\bibinfo {title} {Structural, elastic, and thermal properties of cementite
  ({Fe}$_3${C}) calculated using a modified embedded atom method},\ }\href
  {https://doi.org/10.1103/physrevb.89.094102} {\bibfield  {journal} {\bibinfo
  {journal} {Physical Review B}\ }\textbf {\bibinfo {volume} {89}},\ \bibinfo
  {pages} {094102} (\bibinfo {year} {2014})}\BibitemShut {NoStop}%
\bibitem [{\citenamefont {Byggm{\"a}star}\ and\ \citenamefont
  {Granberg}(2020)}]{byggmastar_2020_jnm}%
  \BibitemOpen
  \bibfield  {author} {\bibinfo {author} {\bibfnamefont {J.}~\bibnamefont
  {Byggm{\"a}star}}\ and\ \bibinfo {author} {\bibfnamefont {F.}~\bibnamefont
  {Granberg}},\ }\bibfield  {title} {\bibinfo {title} {Dynamical stability of
  radiation-induced {C15} clusters in iron},\ }\href
  {https://doi.org/10.1016/j.jnucmat.2019.151893} {\bibfield  {journal}
  {\bibinfo  {journal} {Journal of Nuclear Materials}\ }\textbf {\bibinfo
  {volume} {528}},\ \bibinfo {pages} {151893} (\bibinfo {year}
  {2020})}\BibitemShut {NoStop}%
\bibitem [{\citenamefont {Aryanpour}\ \emph {et~al.}(2010)\citenamefont
  {Aryanpour}, \citenamefont {van Duin},\ and\ \citenamefont
  {Kubicki}}]{aryanpour_2010_jpca}%
  \BibitemOpen
  \bibfield  {author} {\bibinfo {author} {\bibfnamefont {M.}~\bibnamefont
  {Aryanpour}}, \bibinfo {author} {\bibfnamefont {A.~C.~T.}\ \bibnamefont {van
  Duin}},\ and\ \bibinfo {author} {\bibfnamefont {J.~D.}\ \bibnamefont
  {Kubicki}},\ }\bibfield  {title} {\bibinfo {title} {Development of a reactive
  force field for iron{--}oxyhydroxide systems},\ }\href
  {https://doi.org/10.1021/jp101332k} {\bibfield  {journal} {\bibinfo
  {journal} {The Journal of Physical Chemistry A}\ }\textbf {\bibinfo {volume}
  {114}},\ \bibinfo {pages} {6298} (\bibinfo {year} {2010})}\BibitemShut
  {NoStop}%
\bibitem [{\citenamefont {Mori}\ and\ \citenamefont
  {Ozaki}(2020)}]{mori_2020_prm}%
  \BibitemOpen
  \bibfield  {author} {\bibinfo {author} {\bibfnamefont {H.}~\bibnamefont
  {Mori}}\ and\ \bibinfo {author} {\bibfnamefont {T.}~\bibnamefont {Ozaki}},\
  }\bibfield  {title} {\bibinfo {title} {Neural network atomic potential to
  investigate the dislocation dynamics in bcc iron},\ }\href
  {https://doi.org/10.1103/physrevmaterials.4.040601} {\bibfield  {journal}
  {\bibinfo  {journal} {Physical Review Materials}\ }\textbf {\bibinfo {volume}
  {4}},\ \bibinfo {pages} {040601} (\bibinfo {year} {2020})}\BibitemShut
  {NoStop}%
\bibitem [{\citenamefont {Lin}\ \emph {et~al.}(2014{\natexlab{b}})\citenamefont
  {Lin}, \citenamefont {Mrovec},\ and\ \citenamefont {Vitek}}]{lin_2014_msmse}%
  \BibitemOpen
  \bibfield  {author} {\bibinfo {author} {\bibfnamefont {Y.-S.}\ \bibnamefont
  {Lin}}, \bibinfo {author} {\bibfnamefont {M.}~\bibnamefont {Mrovec}},\ and\
  \bibinfo {author} {\bibfnamefont {V.}~\bibnamefont {Vitek}},\ }\bibfield
  {title} {\bibinfo {title} {A new method for development of bond-order
  potentials for transition bcc metals},\ }\href
  {https://doi.org/10.1088/0965-0393/22/3/034002} {\bibfield  {journal}
  {\bibinfo  {journal} {Modelling and Simulation in Materials Science and
  Engineering}\ }\textbf {\bibinfo {volume} {22}},\ \bibinfo {pages} {034002}
  (\bibinfo {year} {2014}{\natexlab{b}})}\BibitemShut {NoStop}%
\bibitem [{\citenamefont {Shang}\ \emph {et~al.}(2010)\citenamefont {Shang},
  \citenamefont {Saengdeejing}, \citenamefont {Mei}, \citenamefont {Kim},
  \citenamefont {Zhang}, \citenamefont {Ganeshan}, \citenamefont {Wang},\ and\
  \citenamefont {Liu}}]{shang_2010_cms}%
  \BibitemOpen
  \bibfield  {author} {\bibinfo {author} {\bibfnamefont {S.}~\bibnamefont
  {Shang}}, \bibinfo {author} {\bibfnamefont {A.}~\bibnamefont {Saengdeejing}},
  \bibinfo {author} {\bibfnamefont {Z.}~\bibnamefont {Mei}}, \bibinfo {author}
  {\bibfnamefont {D.}~\bibnamefont {Kim}}, \bibinfo {author} {\bibfnamefont
  {H.}~\bibnamefont {Zhang}}, \bibinfo {author} {\bibfnamefont
  {S.}~\bibnamefont {Ganeshan}}, \bibinfo {author} {\bibfnamefont
  {Y.}~\bibnamefont {Wang}},\ and\ \bibinfo {author} {\bibfnamefont
  {Z.}~\bibnamefont {Liu}},\ }\bibfield  {title} {\bibinfo {title}
  {First-principles calculations of pure elements: Equations of state and
  elastic stiffness constants},\ }\href
  {https://doi.org/10.1016/j.commatsci.2010.03.041} {\bibfield  {journal}
  {\bibinfo  {journal} {Computational Materials Science}\ }\textbf {\bibinfo
  {volume} {48}},\ \bibinfo {pages} {813} (\bibinfo {year} {2010})}\BibitemShut
  {NoStop}%
\bibitem [{\citenamefont {Hutcheon}\ and\ \citenamefont
  {Needs}(2019)}]{hutcheon_2019_prb}%
  \BibitemOpen
  \bibfield  {author} {\bibinfo {author} {\bibfnamefont {M.}~\bibnamefont
  {Hutcheon}}\ and\ \bibinfo {author} {\bibfnamefont {R.}~\bibnamefont
  {Needs}},\ }\bibfield  {title} {\bibinfo {title} {Structural and vibrational
  properties of lithium under ambient conditions within density functional
  theory},\ }\bibfield  {journal} {\bibinfo  {journal} {Physical Review B}\
  }\textbf {\bibinfo {volume} {99}},\ \href
  {https://doi.org/10.1103/physrevb.99.014111} {10.1103/physrevb.99.014111}
  (\bibinfo {year} {2019})\BibitemShut {NoStop}%
\bibitem [{\citenamefont {Xie}\ \emph {et~al.}(2008)\citenamefont {Xie},
  \citenamefont {Ma}, \citenamefont {Cui}, \citenamefont {Li}, \citenamefont
  {Qiu},\ and\ \citenamefont {Zou}}]{xie_2008_njp}%
  \BibitemOpen
  \bibfield  {author} {\bibinfo {author} {\bibfnamefont {Y.}~\bibnamefont
  {Xie}}, \bibinfo {author} {\bibfnamefont {Y.~M.}\ \bibnamefont {Ma}},
  \bibinfo {author} {\bibfnamefont {T.}~\bibnamefont {Cui}}, \bibinfo {author}
  {\bibfnamefont {Y.}~\bibnamefont {Li}}, \bibinfo {author} {\bibfnamefont
  {J.}~\bibnamefont {Qiu}},\ and\ \bibinfo {author} {\bibfnamefont {G.~T.}\
  \bibnamefont {Zou}},\ }\bibfield  {title} {\bibinfo {title} {Origin of bcc to
  fcc phase transition under pressure in alkali metals},\ }\href
  {https://doi.org/10.1088/1367-2630/10/6/063022} {\bibfield  {journal}
  {\bibinfo  {journal} {New Journal of Physics}\ }\textbf {\bibinfo {volume}
  {10}},\ \bibinfo {pages} {063022} (\bibinfo {year} {2008})}\BibitemShut
  {NoStop}%
\end{thebibliography}%

\clearpage
\section*{ACKNOWLEDGEMENTS}
\subsection*{Funding}
This work is supported by the Research Grants Council, Hong Kong SAR through the Early Career Scheme Fund,  the General Research Fund, and the Collaborative Research Fund under project numbers 21205019, 11211019 and 8730054, respectively.  

\subsection*{Author contributions}
RW developed interatomic potentials and performed atomistic simulations. LZ calculated core structures using DFT. SP and RW performed VCA calculations. ZW designed and directed the research. ZW and DJS wrote the paper.  All authors analysed the data and discussed the results.

\subsection*{Competing interests}
The authors declare no competing interests.

\subsection*{Data and materials availability}
All data, computational methods and models are contained in the Supplementary Information.

\section*{SUPPLEMENTARY INFORMATION}

\indent Supplementary text \\
\indent Supplementary Figures 1-19\\
\indent Supplementary Tables 1-7\\
\indent References \textit{(71-145)}


\balancecolsandclearpage
\newpage

\clearpage
\newcommand{\beginextenddata}{%
  \setcounter{table}{0}
  \setcounter{figure}{0}
  \setcounter{equation}{0}
  \setcounter{page}{1}
  \renewcommand{\thesection}{\arabic{section}.}
  \setcounter{section}{0}
  \renewcommand{\thesubsection}{\thesection\arabic{subsection}}
  \renewcommand{\thesubsubsection}{\thesubsection.\arabic{subsubsection}}
}

\beginextenddata

\newgeometry{left=2.5cm, right=2.5cm, top=2.5cm, bottom=2.5cm}

\fancypagestyle{plain}{
\fancyfoot[R]{\thepage}}

\pagestyle{plain}
\renewcommand\headrulewidth{0pt}
\lhead{}\chead{}\rhead{}
\cfoot{}

\large

\baselineskip15pt

\begin{center}
  \vspace{0.5cm}
  {\huge Supplementary Information for} \\
  \vspace{0.5cm}
  \textbf{\Large The Taming of the Screw: Dislocation Cores in BCC Metals and Alloys} \\
  \vspace{1cm}
  {\large Rui Wang et al.} \\
  \vspace{1cm}
  \vspace{2cm}
\end{center}

\makeatletter

\titleformat{\subsubsection}
{\large}{\thesubsubsection}{1em}{{#1}}

\titleformat{\subsection}
{\bfseries\large}{\thesubsection}{1em}{{#1}}

\titleformat{\section}
{\bfseries\Large}{\thesection}{1em}{{#1}}

\clearpage

\newpage

\section{Supplementary Text}

\subsection{Cross-validation of \(\chi\) in binary W-TM alloys with DFT VCA in QE}
\label{sec:chi_vca_qe}

In VASP, pseudopotentials for Y, Zr and Nb are only available with valence states Y-sv: \(4s^2 4p^6 4d^1 5s^2\), Zr-sv: \(4s^2 4p^6 4d^2 5s^2\) and Nb-pv/Nb-sv: \(4p^6 4d^4 5s^1 / 4s^2 4p^6 4d^4 5s^1 \), which are different from other elements in their respective columns.  This may yield some inconsistency when comparing the effects of solutes within the same Group/column in the periodic table.   Therefore, we have also calculated \(\chi\) for 28 binary W-TM alloys using VCA as implemented in the open source plane wave DFT package, Quantum ESPRESSO (QE~\cite{giannozzi_2009_jpcm}). In QE, the exchange and correlation functional is describled by the GGA with the PBE~\cite{perdew_1996_prl} parameterization.  Core electrons are replaced by pseudopotentials constructed using the weighting scheme in Eqn.~\ref{eqn:vca_weighted_core_ve}.  The scalar relativistic ultrasoft pseudo-potentials~\cite{vanderbilt_1990_prb,rappe_1990_prb}, as generated by the PSlibrary~\cite{corso_2014_cms}, are used for the core electrons of the constituent pure elements~\cite{bellaiche_2000_prb}.  The valence states of all elements are shown in Supplementary Table~\ref{tab:vs_state_vca_vasp_qe}.

For all calculations, the plane-wave function kinetic energy cutoffs and charge density cutoffs are taken as 2041 eV (150 Ry) and 24490 eV (1800 Ry) respectively. The Marzari-Vanderbilt cold smearing~\cite{marzari_1999_prl} with a width of 0.03 eV (0.002 Ry) is employed. Energy differences between the BCC and FCC structure are calculated using primitive cells with one atom per cell and a $\Gamma$-centered $24 \times 24 \times 24$ Monkhorst-Pack k-point mesh~\cite{monkhorst_1976_prb}.  Note that the energy difference is an per-atom quantity in perfect crystals and is insensitive to the size of the supercell.

Structure optimization is carried out by allowing variations of primitive cell shape, volume, and ion positions.  Convergence is assumed when total energy difference between two consecutive steps is less than 1.36 $\times 10^{-4}$ eV (1 $\times 10^{-5}$ Ry) and force components on the ion are below 2.57 $\times 10^{-3}$ eV/\AA \ (1 $\times 10^{-4}$ Ry/bohr). The convergence criterion for electronic self-consistency is 1.36 $\times 10^{-7}$ eV (1 $\times 10^{-8}$ Ry).

Supplementary Fig.~\ref{fig:vca_w_qe} shows the \(\chi\) values as a function of solute concentration for 28 W-TM alloys calculated by QE.  The general trend of solute effects on \(\chi\) is similar to that in Fig.~\ref{fig:vca} calculated by VASP; both results show that solutes on the left of W increase \(\chi\) and solutes on the right of W decrease \(\chi\) at low concentrations, consistent with the change in valence electron concentrations.  The QE results show that solutes in the same Group exhibit nearly the same effects on changing \(\chi\) in W (note that they are obtained by using the same valence electron state for solutes in the same Group).  Nevertheless, the overall trends can be robustly predicted in VASP or QE, since \(\chi\) only depends on the energy difference between two elementary structures and can be obtained efficiently.

\subsection{Three special cases}
Three potentials (Nb-SNAP~\cite{zheng_2020_npjcm}, Cr-EAM~\cite{mendelev_2020_openkim} and Mo-ADP~\cite{starikov_2017_jnm}) examined in the current work do not follow the \(\chi\) model prediction.  Nb-SNAP has \(\Delta E = 111\) meV/atom and BCC elastic constants \(C_{11} = 253\)  GPa, \(C_{12} = 136\) GPa and \(C_{44} = 14\) GPa, whilte DFT shows \(\Delta E = 322\) meV/atom and BCC elastic constants are \(C_{11} = 253\) GPa, \(C_{12} = 133\) GPa and \(C_{44} = 31\) GPa from the experiments. Cr-EAM has \(\Delta E = 10\) meV/atom and a meta stable FCC structure with \(C_{11} = 585\) GPa, \(C_{12} = 302\) GPa, \(C_{44} = 212\) GPa, while DFT predicts \(\Delta E = 412\) meV/atom and the FCC structure is unstable. Mo-ADP has \(\Delta E = 363\) meV/atom and BCC elastic constants \(C_{11} = 547\) GPa, \(C_{12} = 168\) GPa, \(C_{44} = 160\) GPa, while DFT shows \(\Delta E = 424\) meV/atom and BCC elastic constants from experiments are \(C_{11} = 472\) GPa, \(C_{12} = 158\) GPa, \(C_{44} = 106\) GPa.  The three potentials thus have basic properties largely different from DFT calculations or experimental measurements.  These interatomic potentials appear to be less physical and can not reasonably represent the elementary metals, respectively, which may be responsible for their unusual behaviours.

\subsection{Solute Softening and Hardening in BCC Binary Alloys}
\label{sec:sssh_exp}
In the current work, we introduced the material index \(\chi\) based on a geometric model and \(\chi\) is defined using \(\Delta E\) at 0 K.  The Peierls barrier of the screw dislocation is shown to be proportional to \(\chi\) in VCA DFT calculations.  To further extend the validity of the \(\chi\)-model, we compare our prediction with experimental data on yield strength or hardness measured in binary BCC transition metal alloys, which were well documented in classical works~\cite{pink_1980_pms}.  Since yield strength and hardness are temperature and strain-rate dependent, we choose to focus on data measured at low temperature and in particular, 77-78 K where sufficient experiments were conducted and ductility/toughness is most needed.  Experimental data could also be influenced by extrinsic factors such as sample size and impurity. We thus choose to normalize yield strength or hardness of alloys by their corresponding values of pure metals. Supplementary Fig.~\ref{fig:hs_gp_v_exp} and~\ref{fig:hs_gp_vi_exp} show the normalized strength/hardness as a function of solute concentrations.  In Group V (V, Nb, Ta)-based BCC alloys, solutes on their right in the periodic table show hardening effects, while solutes on the left exhibit softening in general.  An opposite trend is seen in Group VI (Cr, Mo, W)-based transition metal alloys, despite some scattering in the data from different experiments.  In addition, the potency of solute softening/hardening is also related to the solute position on the periodic table (or valence electron concentrations); solutes further away from the base metal have stronger hardening/softening effects, fully consistent with the \(\chi\)-model prediction shown in Figs.~\ref{fig:vca},~\ref{fig:chi_prop}, Supplementary Figs.~\ref{fig:vca_mo_vasp},~\ref{fig:vca_w_qe},~\ref{fig:vca_v_vasp} and~\ref{fig:vca_wtms_pe}.  While the general trends on solute effects have been studied in some alloys in previous studies~\cite{trinkle_2005_science,medvedeva_2007_prb,hu_2017_am}, the \(\chi\) material index provides a univeral, simple model rationalising all BCC elements across the periodic table.  We note that many solutes also exhibit softening effects in BCC Fe. The importance of Fe and its magnetism warrant a separate, dedicated study.

\subsection{Inter-string interaction energy}
The inter-string interaction energy (ISIE) captures energy variation during atomic displacements similar to that in the core center of the screw dislocation, i.e., displacement in the screw dislocation Burgers vector \(\langle 111 \rangle\) direction~\cite{chiesa_2009_pm,li_2012_am}. Chiesa et al.~\cite{chiesa_2009_pm} proposed a necessary and sufficient criterion relating the core structure to ISIE profile in BCC structures.  Using FS-EAM interatomic potentials, they demonstrated that the ND core and D core are obtained when the ISIE profile has a double-hump and single-hump, respectively.  In this work, we examine this criterion using both DFT and a wide range of interatomic potentials.  In DFT, we first construct a fully periodic supercell (24 atoms) with supercell vectors \((\mathbf{c}_1,\mathbf{c}_2,\mathbf{c}_3) = ([11\bar{2}], 2[\bar{1}10], [111])\), as shown in Supplementary Fig.~\ref{fig:interstring}.  The supercell used in molecular static calculations (LAMMPS) has the same orientation, but larger size (144 atoms) with supercell vectors \((\mathbf{c}_1,\mathbf{c}_2,\mathbf{c}_3) = (2[11\bar{2}], 3[\bar{1}10], 2[111])\).  The ISIE is calculated by choosing a single column of atoms aligned in the \(z\)-axis (\(\langle  111 \rangle\) direction) (Supplementary Fig.~\ref{fig:interstring}a) and displacing them in the same direction, while other atoms are fixed. Three elements (Fe, Ta and W) are selected for the comparison between DFT and four interatomic potentials (all predict ND cores) in Supplementary Fig.~\ref{fig:interstring}b. DFT results show a double-hump in the ISIE profile in Fe and W, while a single-hump is seen in Ta, suggesting no clear trend in the shape of ISIE among the BCC transition metals. In contrast, Fe-GAP~\cite{dragoni_2018_prm}, Ta-SNAP~\cite{zheng_2020_npjcm} and W-EAM~\cite{setyawan_2018_jap} have a single hump in the ISIE profile, while Fe-EAM~\cite{mendelev_2003_pm} has a double-hump. A broad examination on all considered interatomic potentials (Supplementary Table~\ref{tab:potential}) is shown in Supplementary Fig.~\ref{fig:interstring_pot}, which gives no correlation between the core structure and ISIE profile in general. Thus, both DFT and interatomic potential results suggest that the core structure is not solely controlled by the profile of ISIE.

\newpage



\newcommand{\beginsupplementaryfigure}{%
  \pagenumbering{arabic}
  \setcounter{page}{1}
  \setcounter{table}{0}
  \renewcommand{\thetable}{\arabic{table}}%
  \setcounter{figure}{0}
}

\captionsetup[figure]{justification=raggedright,labelfont={bf},labelformat={default},labelsep=space,name={Supplementary Fig.~\!}}
\captionsetup[table]{justification=raggedright,labelfont={bf},labelformat={default},labelsep=space,name={Supplementary Table~\!}}

\section*{Supplementary Figures}
\beginsupplementaryfigure

\begin{figure*}[!htbp]
  \centering
  \includegraphics[width=0.7\textwidth]{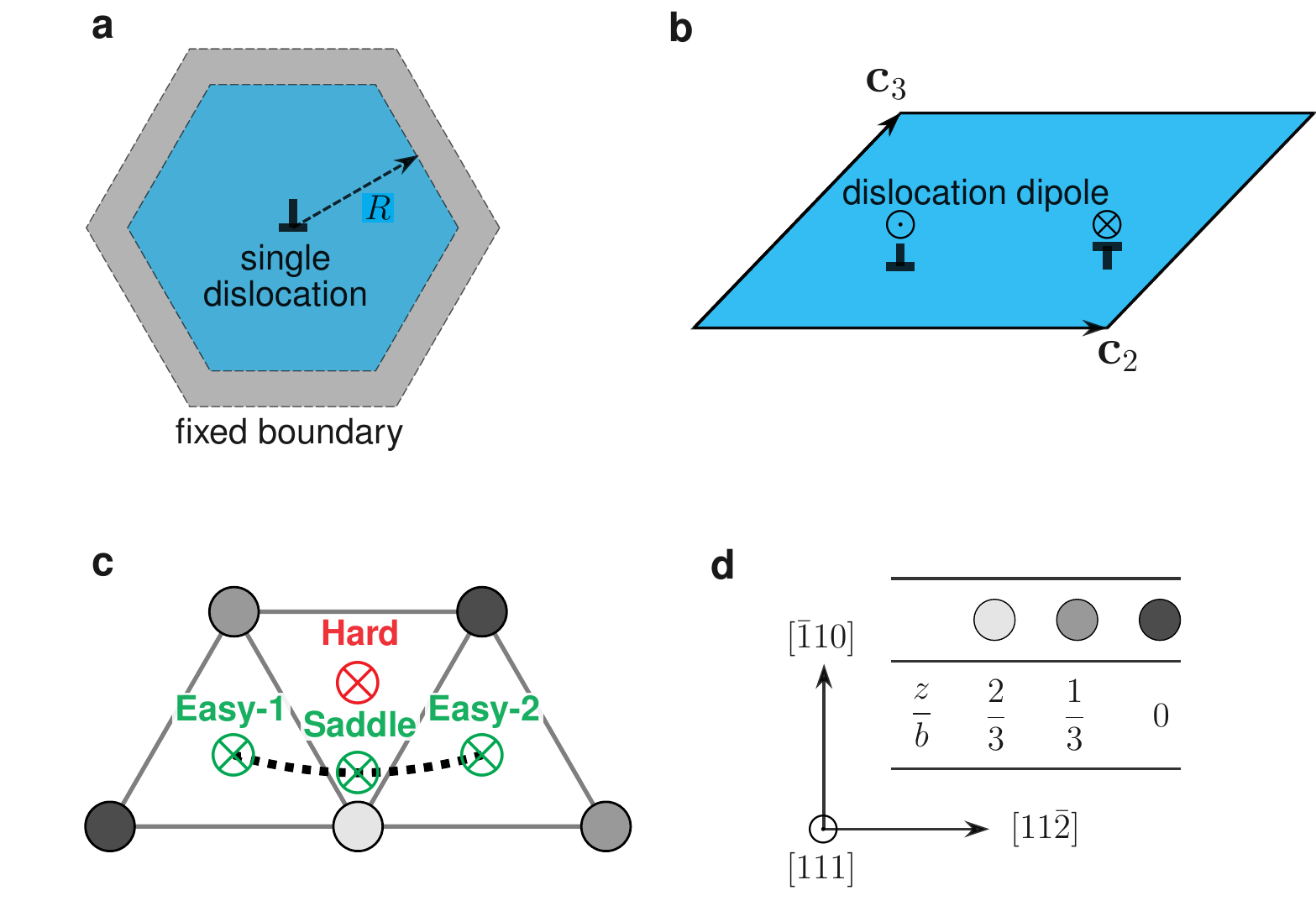}
  \caption{\label{fig:dis_sche}\(\vert\) \textbf{Supercell for dislocation core simulations.} \textbf{a} A hexagonal prism with a single dislocation at the center and fixed boundary conditions. \textbf{b} Dislocation dipole configuration. \textbf{c} Dislocation core center at easy core, hard core and saddle core positions. \textbf{d} Atom colour scheme and crystallographic orientations. In (\textbf{c}), atoms are coloured based on their positions in the \(\langle  111 \rangle\) directions shown in (\textbf{d}). }
\end{figure*}

\begin{figure}[!htbp]
  \centering
  \includegraphics[width=0.9\textwidth]{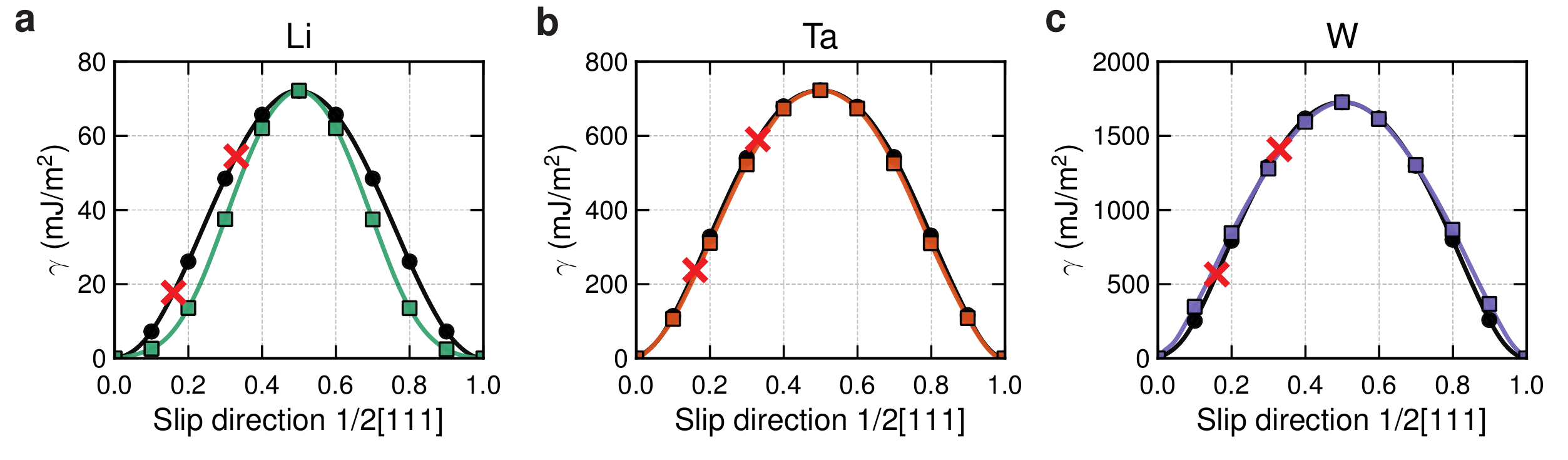}
  \caption{\label{fig:gamma_line_li_ta_w}\(\vert\) \textbf{The generalized stacking fault \(\gamma\) lines along the \(\langle 111 \rangle\) direction and minimum energy path (MEP) on the \(\{110\}\) plane. } \textbf{a} Li. \textbf{b} Ta. \textbf{c} W. The black lines are \(\gamma\)-lines calculated using the classical method and the coloured lines are the MEP on the \(\gamma\)-surface shown in Fig.~\ref{fig:gamma_disl_Li_Ta_W}\textbf{d}-\textbf{f}.  The crosses (``\(\times\)'') denote slips at \(\mathbf{b}/6\) and \(\mathbf{b}/3\), respectively.  All results are calculated using DFT with parameters in Supplementary Table.~\ref{tab:dft_details}.}
\end{figure}

\begin{figure}[!htbp]
  \centering
  \includegraphics[width=0.5\textwidth]{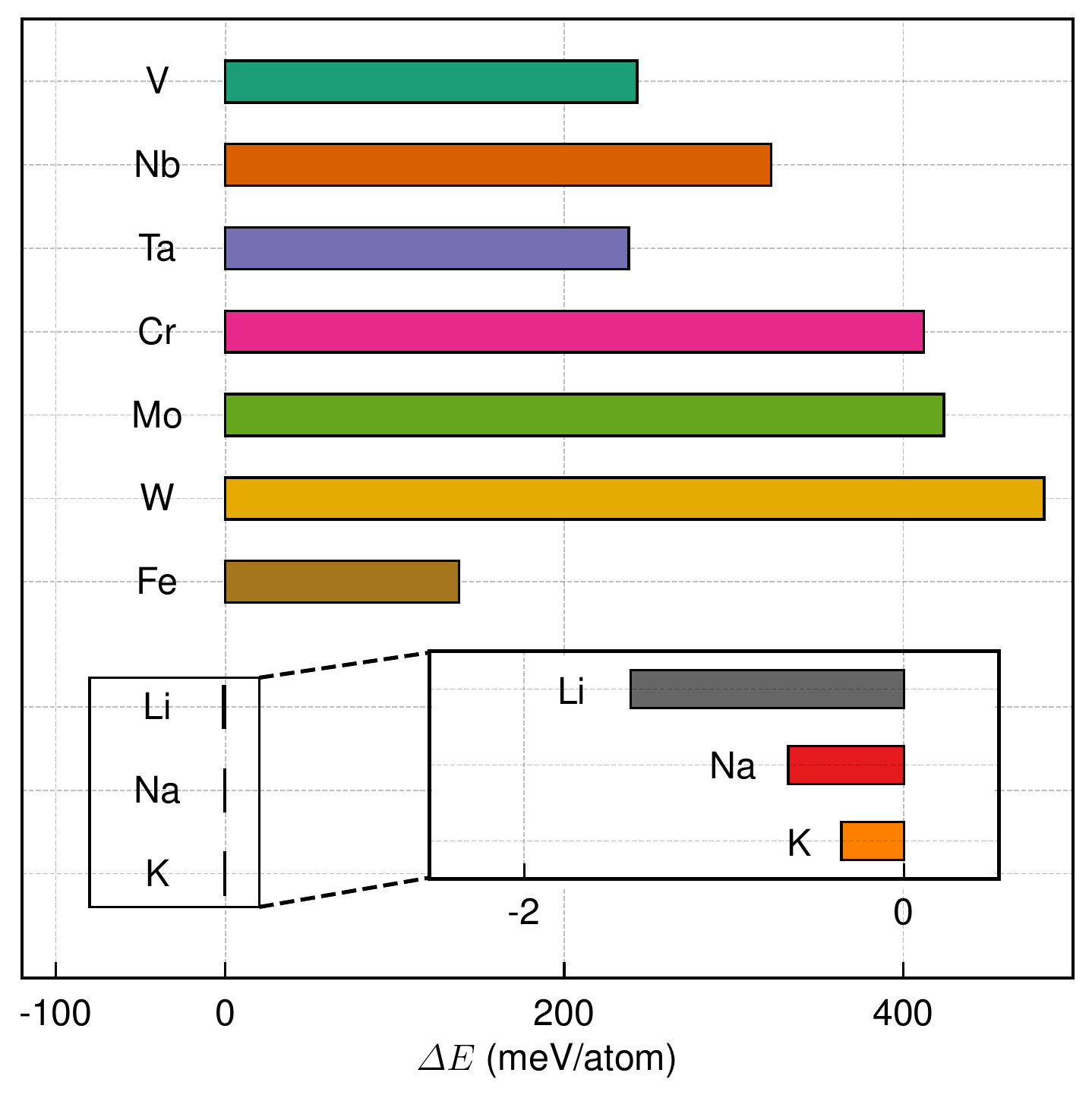}
  \caption{\label{fig:deltaE}\(\vert\) \textbf{Energy difference \(\Delta E\) between the BCC and FCC structure for 7 transition and 3 alkaline metals calculated by DFT in VASP.} Transition metals have large energy differences \(\Delta E> 138\) meV/atom and alkaline metals have vanishingly small energy differences \(\Delta E \approx -1 \) meV/atom.  The calculation parameters are shown in Supplementary Table~\ref{tab:dft_details}. Exact values of \(\Delta E\) are shown in Supplementary Table~\ref{tab:delta_E_bcc_fcc}.}
\end{figure}

\begin{figure}[!htbp]
  \centering
  \includegraphics[width=0.95\textwidth]{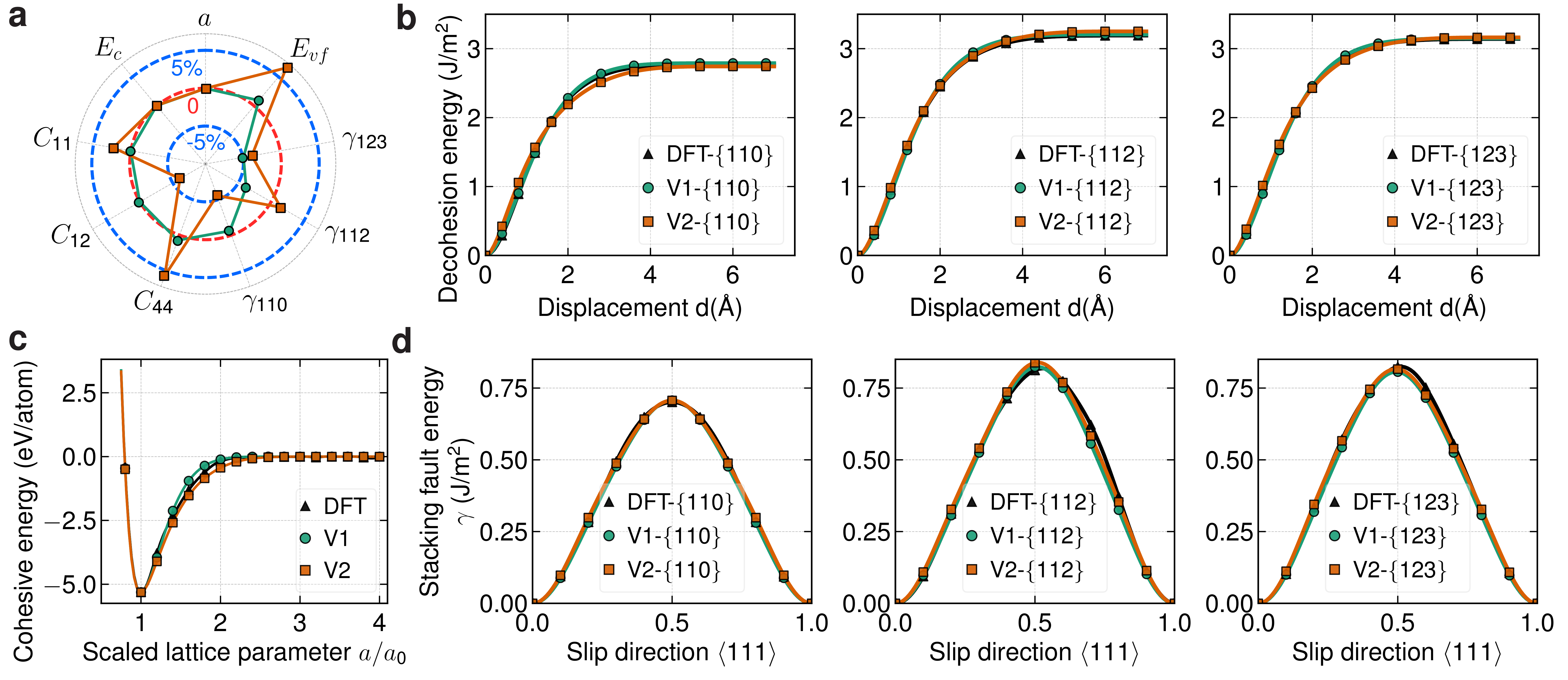}
  \caption{\label{fig:meam_prop_v1_v2}\(\vert\) \textbf{Properties of two MEAM potentials for V1 and V2.} \textbf{a} Fitting errors for the lattice parameter, cohesive energy, elastic constants, surface energies and vacancy formation energy for our MEAM potentials V1 and V2. Blue circles are separared by 5\% errors. Points closer to or farther from the center than the target means negative or positive errors. \textbf{b} Decohesion energy curves for \{110\}, \{112\} and \{123\} planes. \textbf{c} Cohesive energy curves. \textbf{d} Generalized stacking fault energy curves in the slip direction \(\langle 111 \rangle\) for \{110\}, \{112\} and \{123\}. All properties are included in the training set.}
\end{figure}

\begin{figure}[!htbp]
  \centering
  \includegraphics[width=0.6\textwidth]{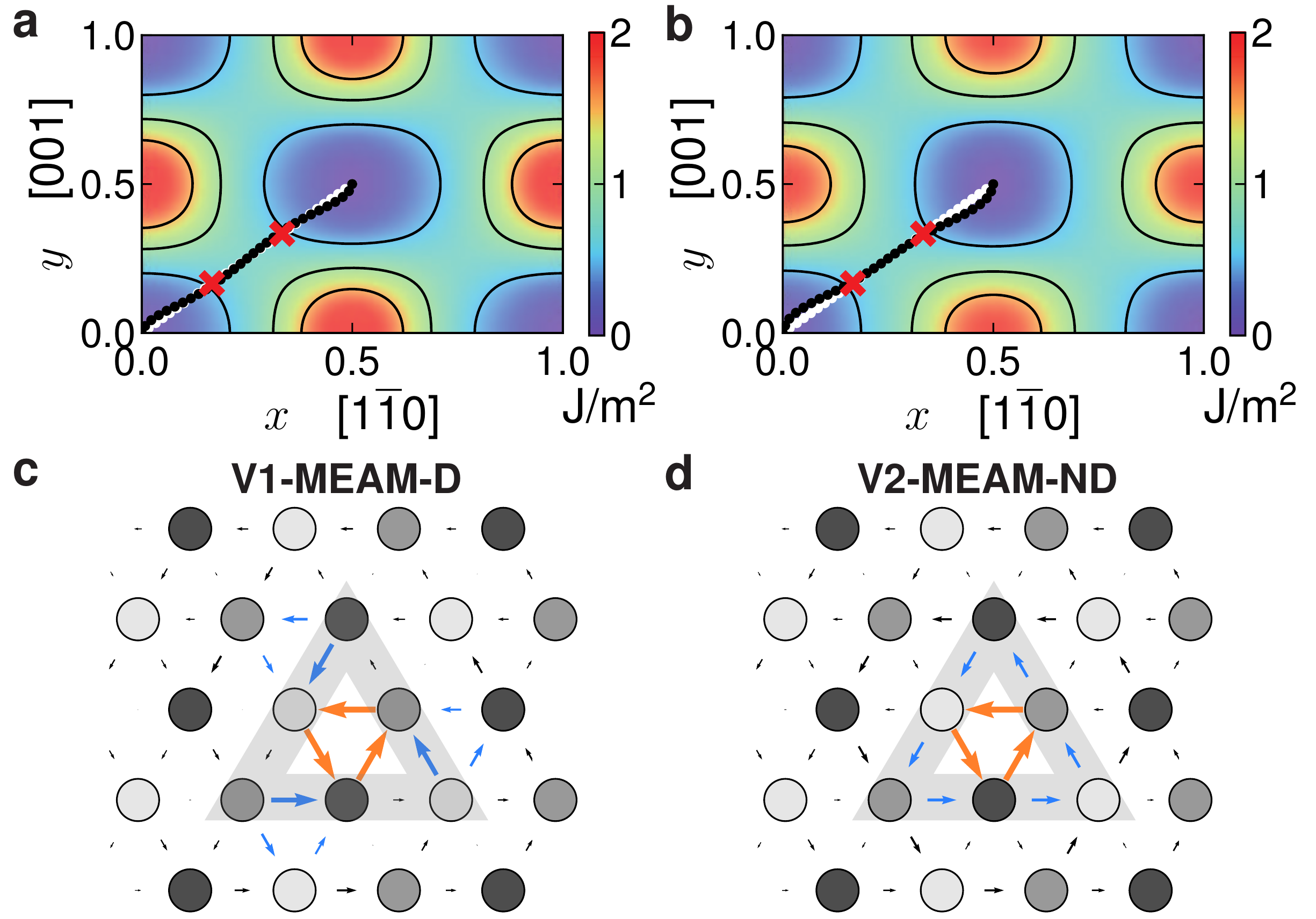}
  \caption{\label{fig:prop_core_type_v}\(\vert\) \textbf{Generalized stacking fault energy surface and dislocation core structure for two interatomic potentials}. \textbf{a-b} Two potentials with nearly identical \(\gamma\)-surfaces. \textbf{c-d} A degenerate (D) core and a non-degenerate (ND) core.}
\end{figure}

\begin{figure}[!htbp]
  \centering
  \includegraphics[width=0.6\textwidth]{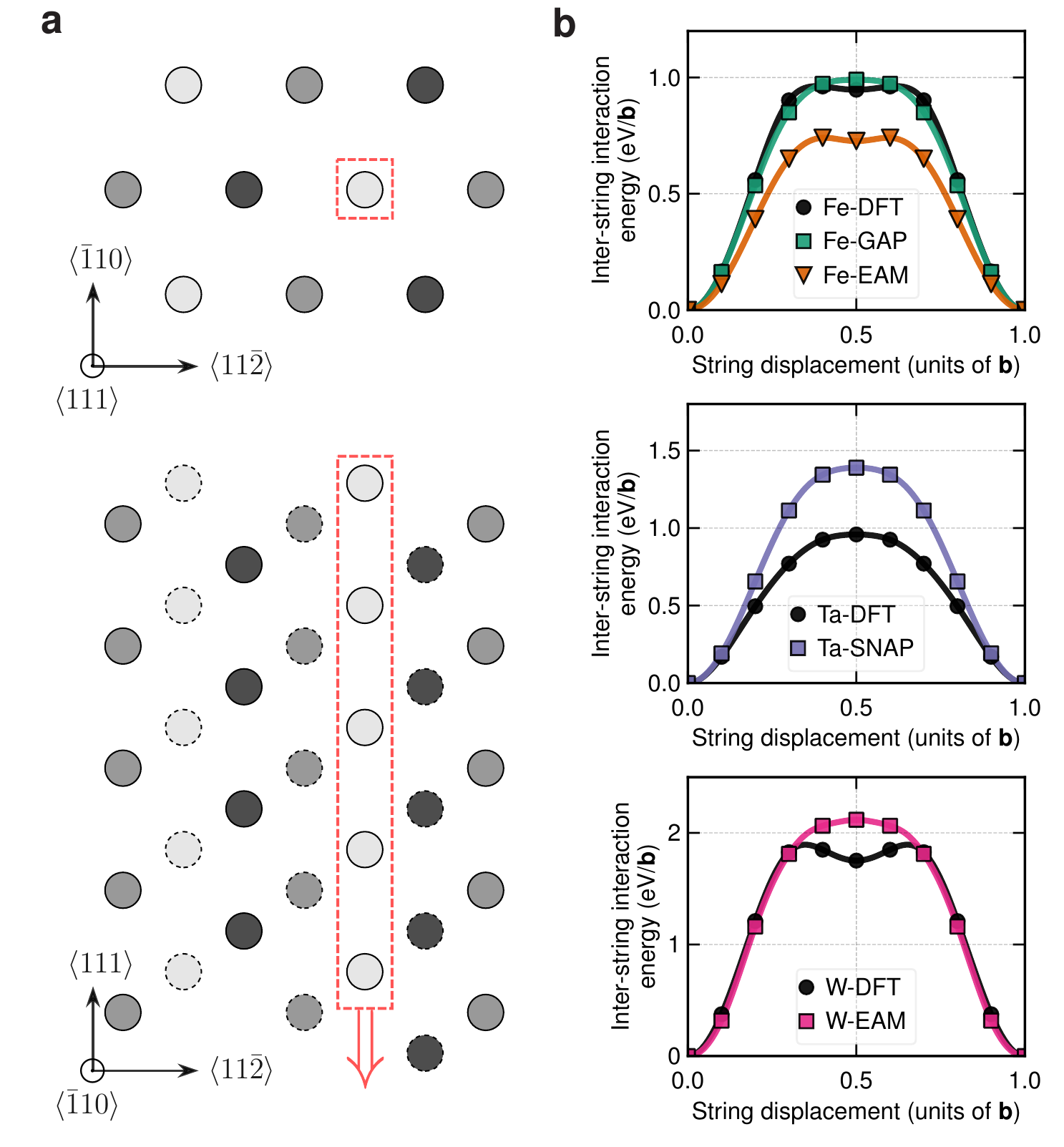}
  \caption{\label{fig:interstring}\(\vert\) \textbf{Supercell model and calculation of inter-string interaction energy.} \textbf{a} The inter-string model for BCC lattice. A \(\langle  111 \rangle\) string atom (in red dashed box) is displaced along \(\langle  111 \rangle\) direction while other atoms are fixed. \textbf{b} The inter-string interaction energy with displacements along  \(\langle  111 \rangle\) for Fe, Ta, W calculated by DFT and in comparison with four interatomic potentials (Fe-GAP~\cite{dragoni_2018_prm}, Fe-EAM~\cite{mendelev_2003_pm}, Ta-SNAP~\cite{zheng_2020_npjcm}, W-EAM~\cite{setyawan_2018_jap}). All potentials predict ND-core and thus the inter-string criterion is not general.}
\end{figure}

\begin{figure}[!htbp]
  \centering
  \includegraphics[width=0.5\textwidth]{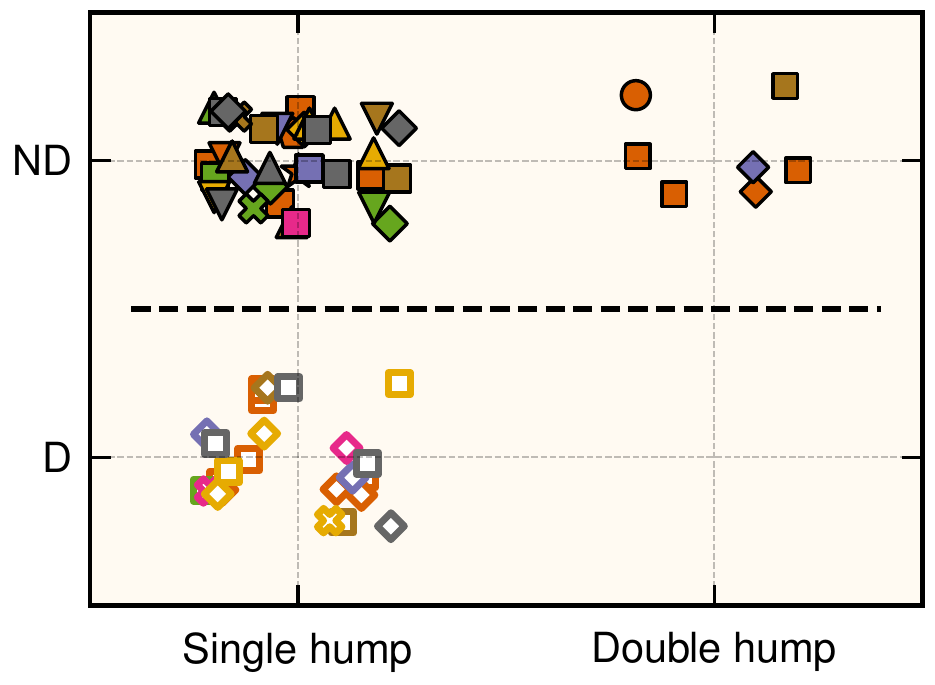}
  \caption{\label{fig:interstring_pot}\(\vert\) \textbf{Dislocation core structure versus the inter-string interaction energy (ISIE) profile.} Symbol shapes and colours denote potential forms and element types as in Fig.~\ref{fig:prop_core_type}. Generally, no clear trend is seen between the ND-/D-cores and the ISIE profile (single/double hump).}
\end{figure}


\begin{figure}[!htbp]
  \centering
  \includegraphics[width=0.6\textwidth]{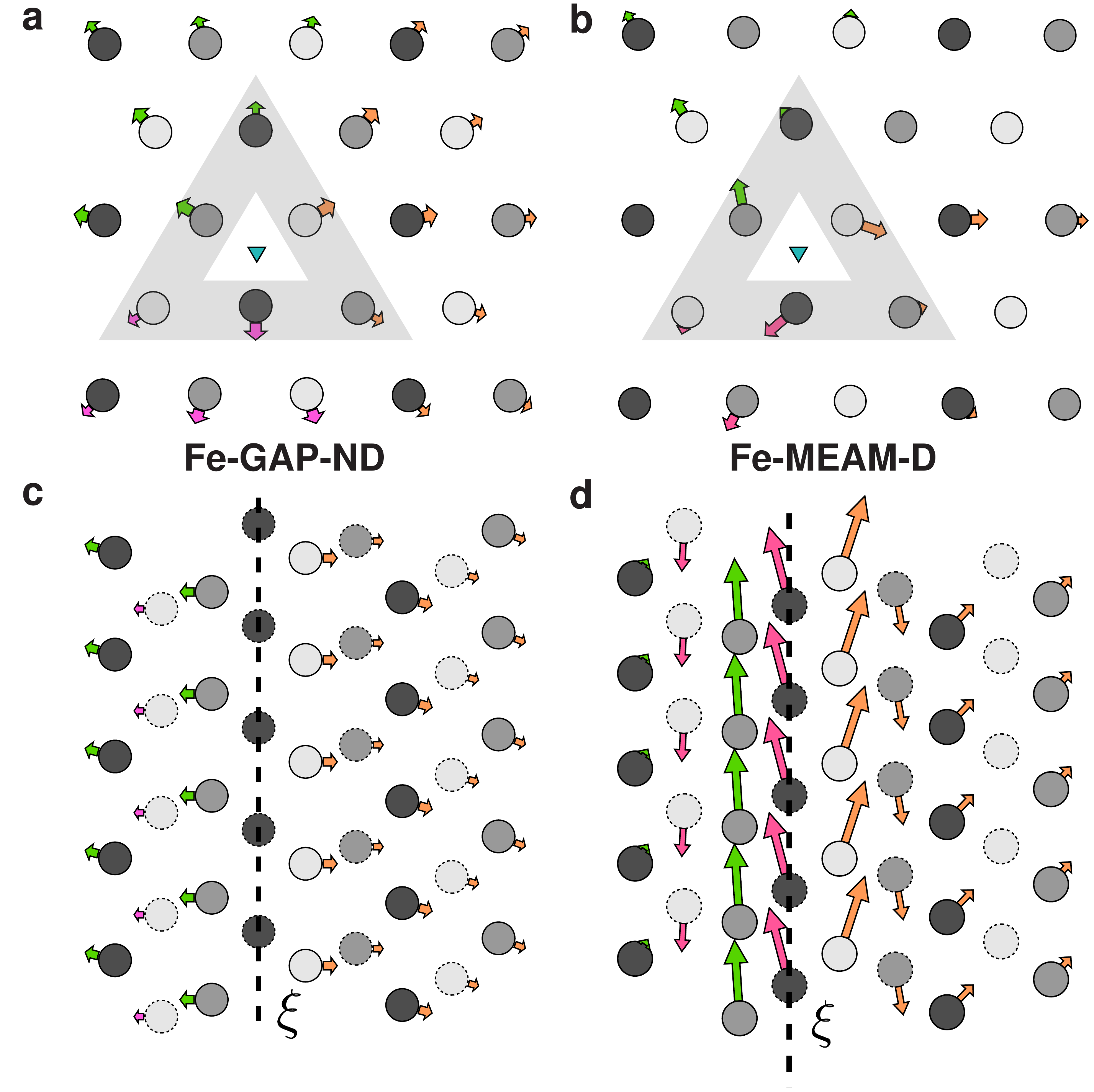}
  \caption{\label{fig:fe_core_eecd}\(\vert\) \textbf{Excessive atomic displacements (EADs) of the screw dislocation calculated by two interatomic potentials.} \textbf{a-b} Viewed in the \(\langle 111 \rangle\) direction. \textbf{c-d} Viewed in the \(\langle 110 \rangle\) direction. The crystallographic orientation is shown in Fig.~\ref{fig:bcc_fcc}. The arrows represent EADs and are magnified by \textbf{6} times for comparisons.  The two potentials, Fe-GAP~\cite{dragoni_2018_prm} and Fe-MEAM~\cite{asadi_2015_prb}, give \(\Delta E = 159\) meV/atom (\(\chi=1.15\)) and \(\Delta E = 44\) meV/atom (\(\chi=0.32\)), respectively. The Fe-GAP potential exhibits a ND-core with small EADs and Fe-MEAM has a D-core with substantial EADs in \(\langle  110 \rangle\) direction on \(\{110\}\) planes, consistent with the \(\chi\)-model prediction.}
\end{figure}

\begin{figure}[!htbp]
  \centering
  \includegraphics[width=0.7\textwidth]{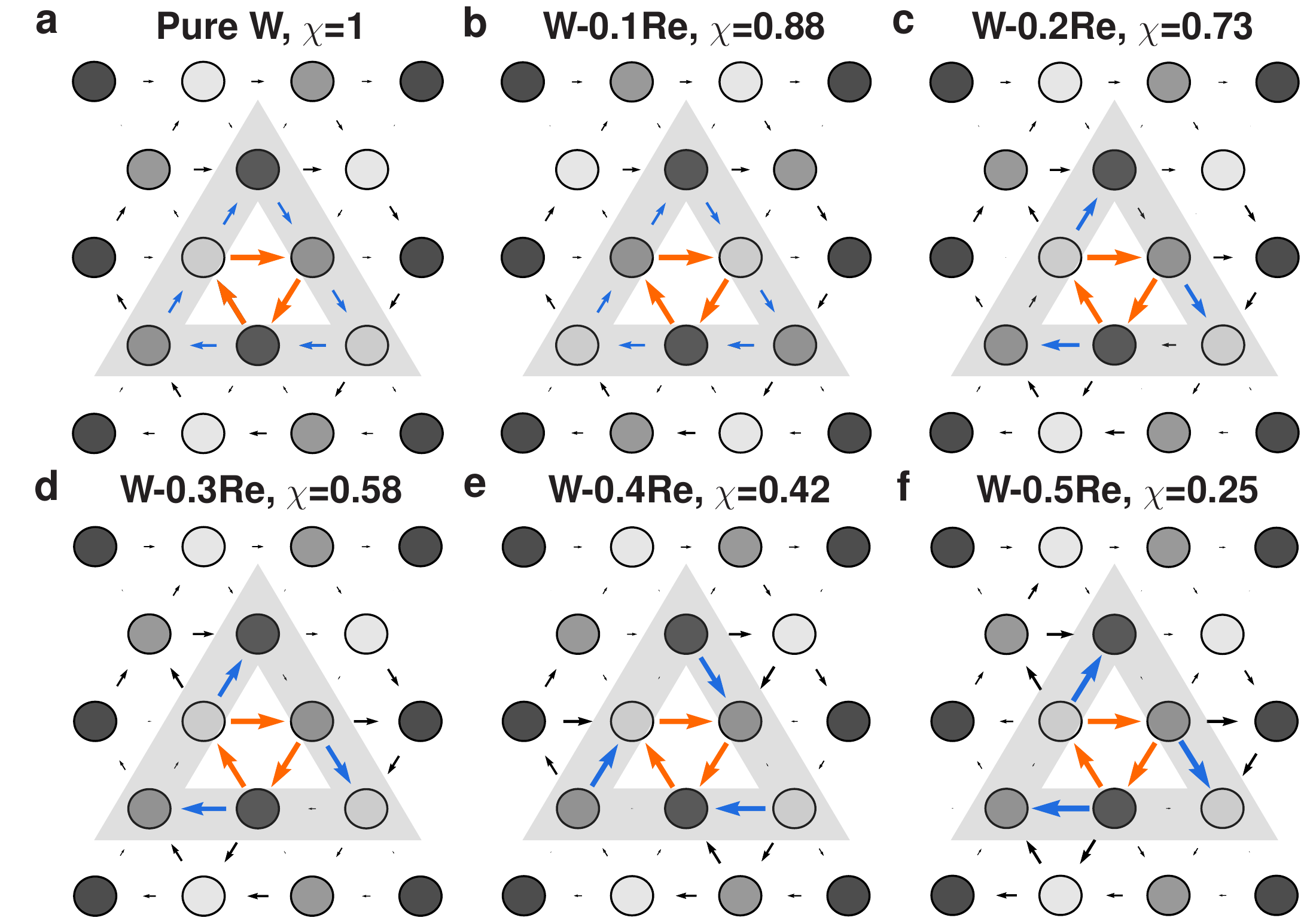}
  \caption{\label{fig:wre_core}\(\vert\) \textbf{Screw dislocation core structures of binary W-Re alloys as a function of Re concentration.}  \textbf{a} Pure W. \textbf{b} W-10at.\%Re. \textbf{c} W-20at.\%Re. \textbf{d} W-30at.\%Re. \textbf{e} W-40at.\%Re. \textbf{f} W-50at.\%Re. All core structures are calculated using VCA as implemented in VASP. ND-cores are seen in pure W and W-10at.\%Re. D-cores are seen in W-20at.\%Re, W-30at.\%Re, W-40at.\%Re and W-50at.\%Re.  Core transition from ND to D occurs at \(\chi\approx0.8\) between 10at.\%Re and 20at.\%Re.}
\end{figure}

\begin{figure}[!htbp]
  \centering
  \includegraphics[width=0.65\textwidth]{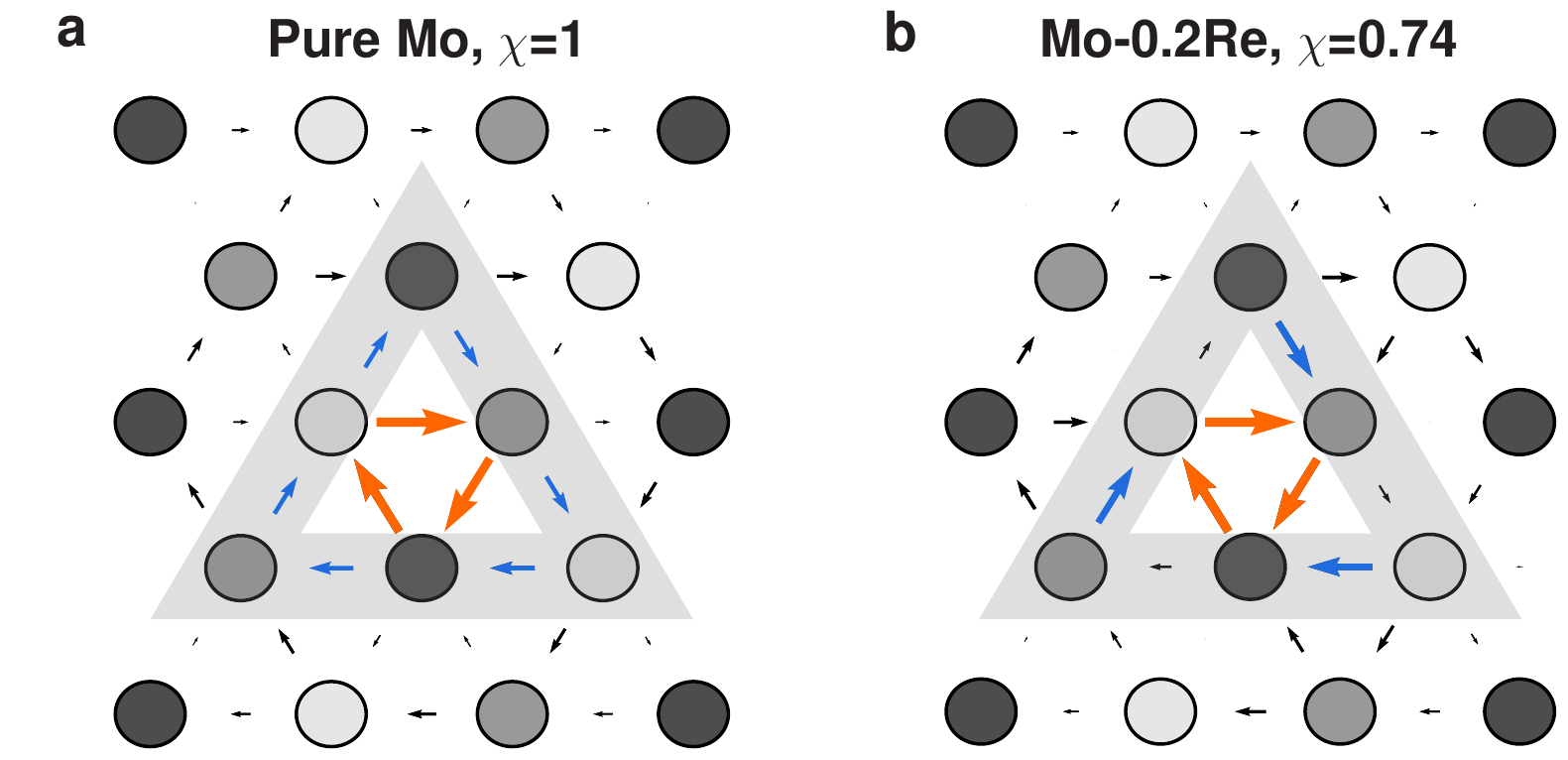}
  \caption{\label{fig:more_core}\(\vert\) \textbf{Screw dislocation core structures of pure Mo and Mo-20at.\%Re alloy.} \textbf{a} Pure Mo. \textbf{b} Mo-20at.\%Re. Pure Mo adopts the ND core while D core is preferred in Mo-20at.\%Re. The core transition happens at 20at.\%Re, as in the case of W-Re alloys.}
\end{figure}

\begin{figure}[!htbp]
  \centering
  \includegraphics[width=0.7\textwidth]{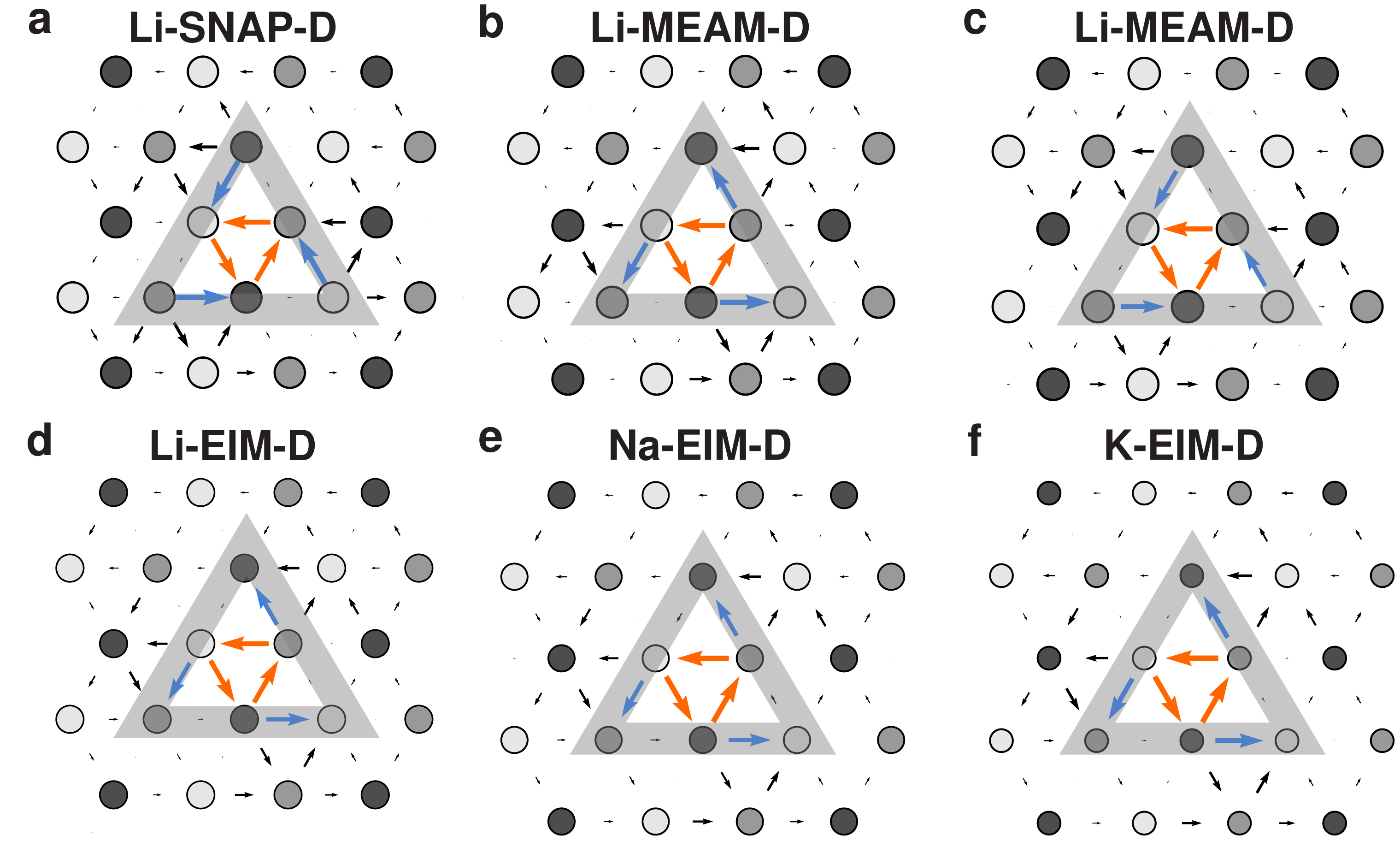}
  \caption{\label{fig:alkali_core}\(\vert\) \textbf{Core structures of alkaline metals calculated by the SNAP, EIM and MEAM interatomic potentials and visualised by the differential displacement map.} \textbf{a} Li-SNAP~\cite{zuo_2020_jpca}, \(\Delta E = 2.3\) meV/atom. \textbf{b} Li-MEAM~\cite{alam_2015_jpcs}, \(\Delta E = 1.4\) meV/atom. \textbf{c} Li-MEAM~\cite{groh_2015_msmse},  \(\Delta E = 2.1\) meV/atom. \textbf{d} Li-EIM, \(\Delta E = -23.5\) meV/atom. \textbf{e} Na, \(\Delta E = -21.4\) meV/atom. \textbf{f} K, \(\Delta E = -15.5\) meV/atom. All potentials exhibit D-cores. The potentials in (\textbf{d}-\textbf{f}) are developed by X. Zhou~\cite{zhou_2010_lmp}.}
\end{figure}

\begin{figure}[!htbp]
  \centering
  \includegraphics[width=0.80\textwidth]{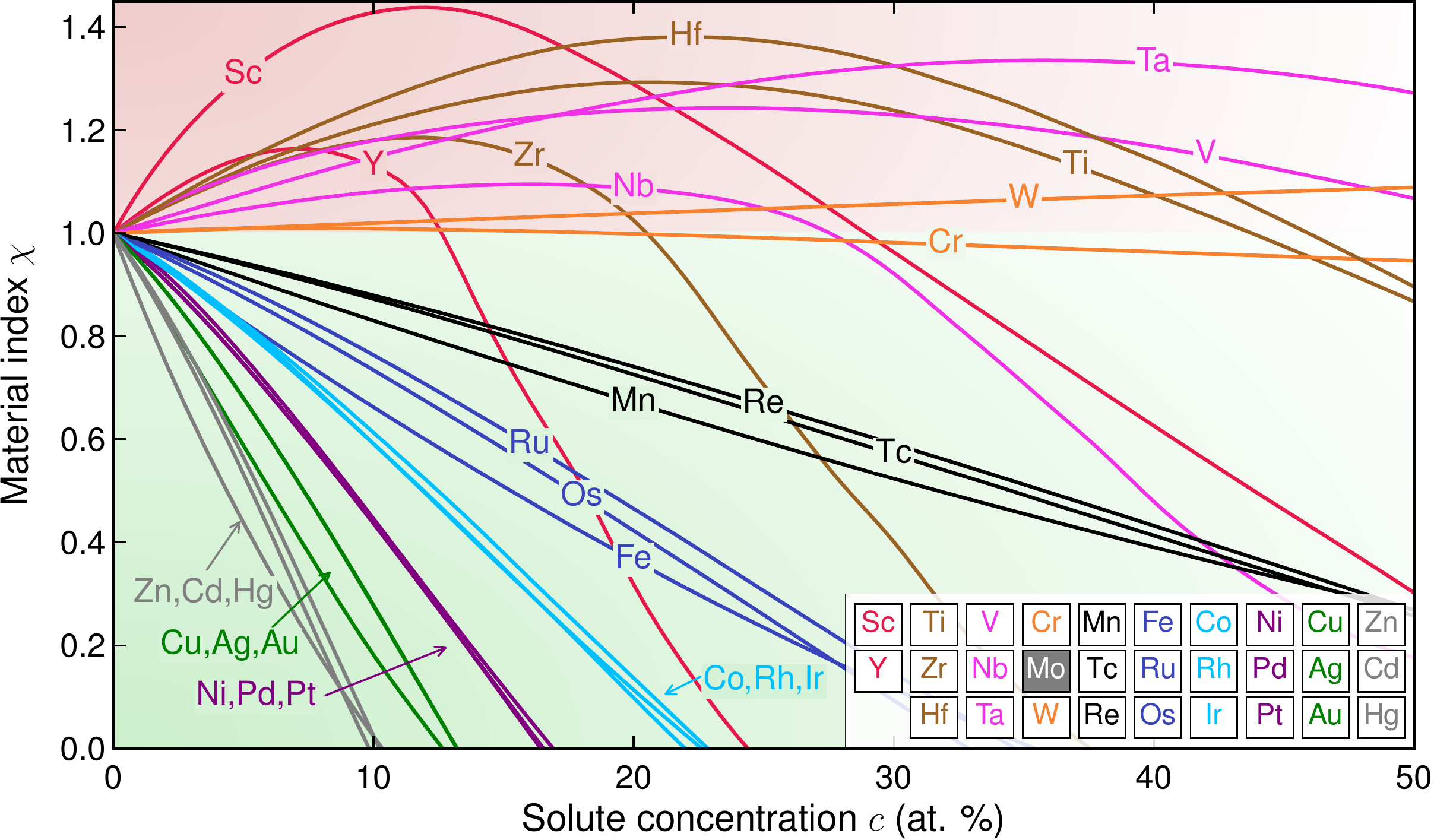}
  \caption{\label{fig:vca_mo_vasp}\(\vert\) \textbf{Prediction of \(\chi\) as a function of solute concentration for binary Mo-TM alloys using VCA as implemented in VASP.} }
\end{figure}

\begin{figure}[!htbp]
  \centering
  \includegraphics[width=0.80\textwidth]{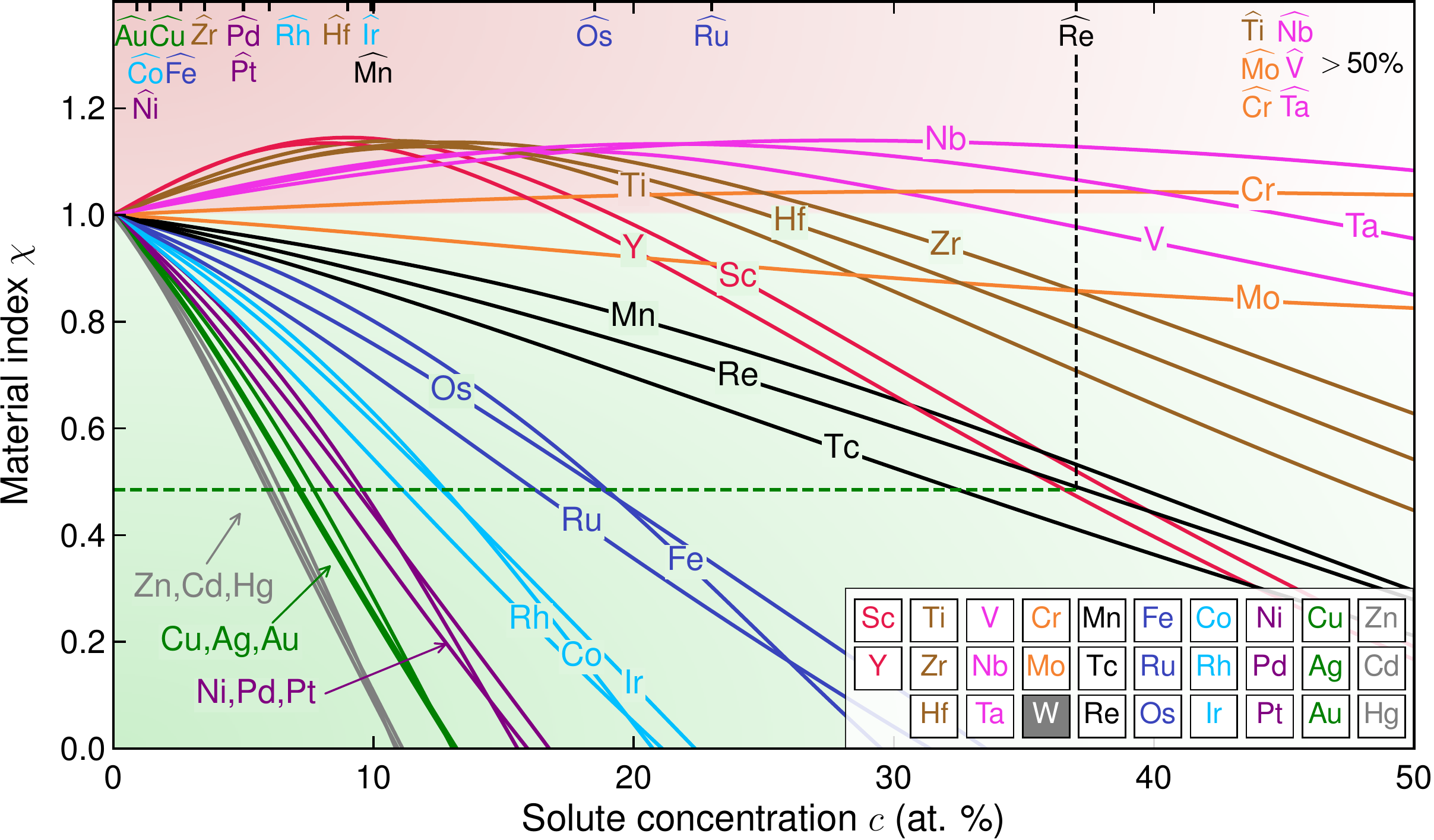}
  \caption{\label{fig:vca_w_qe}\(\vert\) \textbf{Prediction of \(\chi\) as a function of solute concentration for binary W-TM alloys using VCA as implemented in QE.}}
  \end{figure}

\begin{figure}[!htbp]
  \centering
  \includegraphics[width=0.80\textwidth]{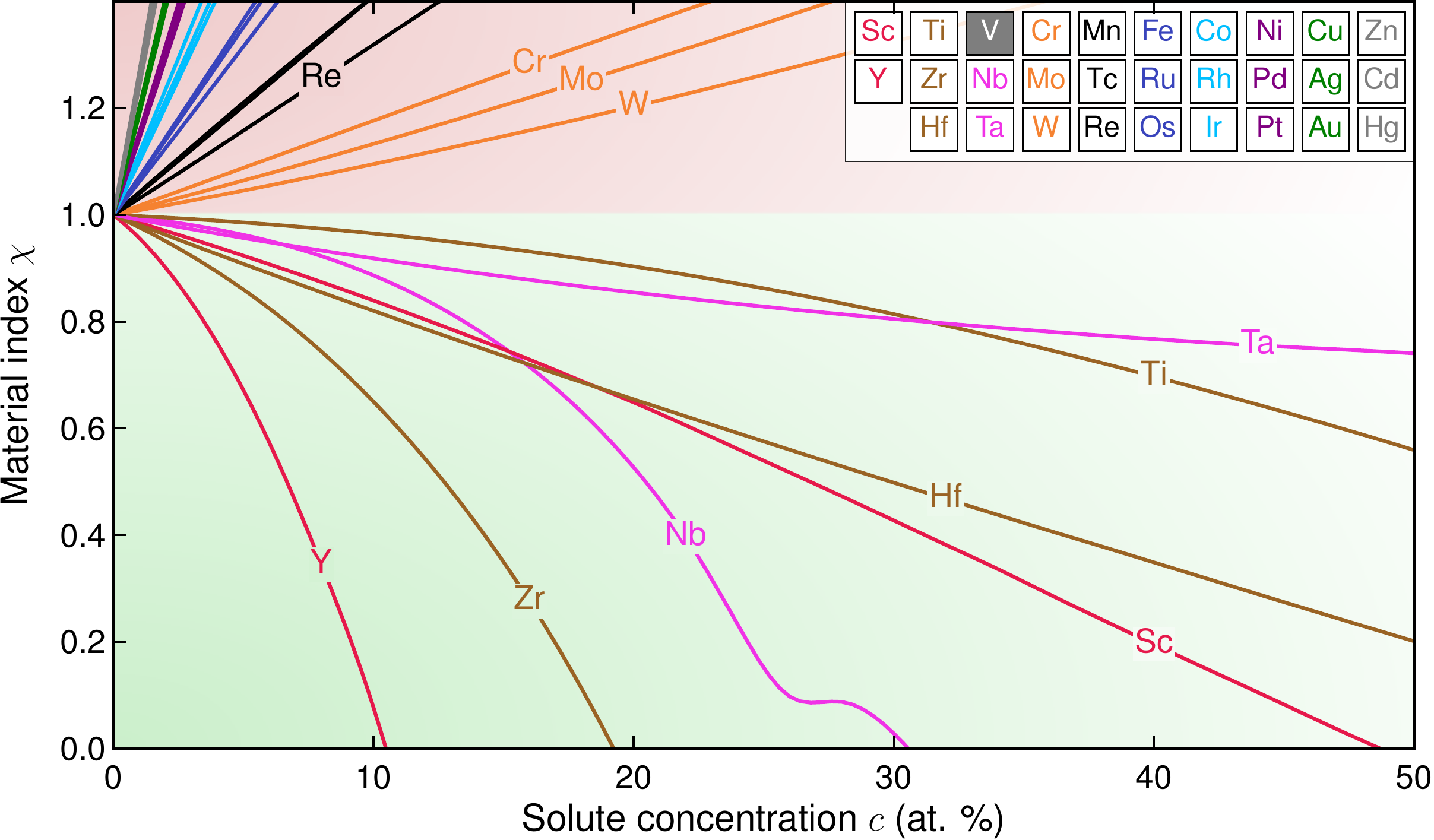}
  \caption{\label{fig:vca_v_vasp}\(\vert\) \textbf{Prediction of \(\chi\) as a function of solute concentration for binary V-TM alloys using VCA as implemented in VASP.}}
\end{figure}

\begin{figure}[!htbp]
  \centering
  \includegraphics[width=0.5\textwidth]{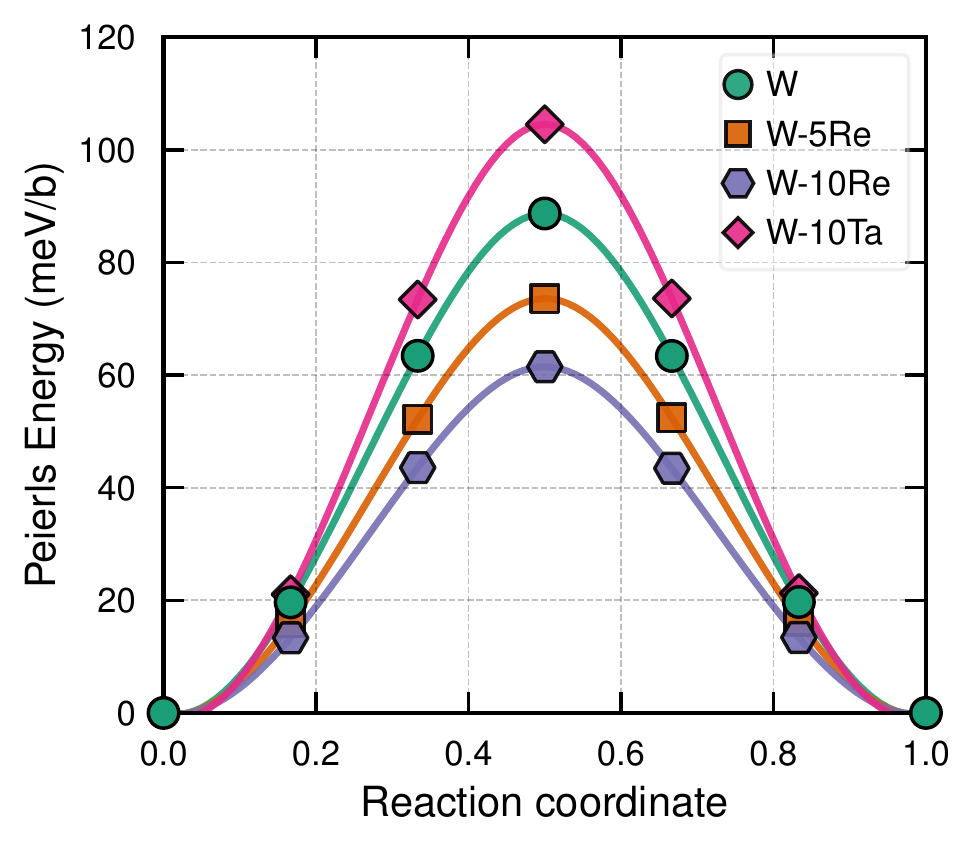}
  \caption{\label{fig:vca_wtms_pe}\(\vert\) \textbf{Peierls barrier in W-Re and W-Ta alloys calculated using NEB with VCA as implemented in VASP.}  }
\end{figure}

\begin{figure}[!htbp]
  \centering
  \includegraphics[width=0.8\textwidth]{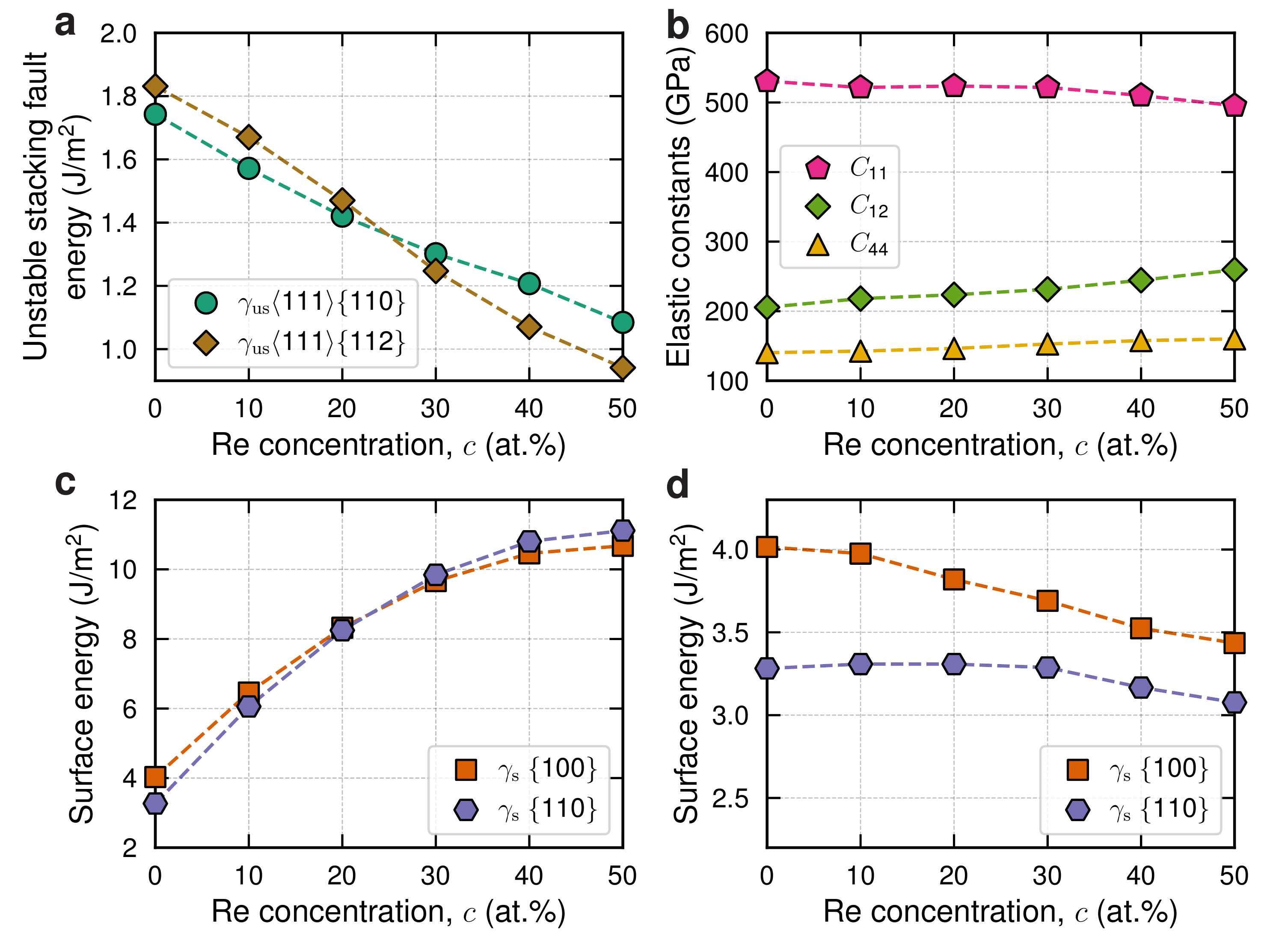}
  \caption{\label{fig:usf_sf_wre_calc}\(\vert\) \textbf{Unstable stacking fault energy, elastic constants and surface energy calculated in BCC W-Re alloys by DFT VCA in VASP and QE.} \textbf{a} Unstable stacking fault energy \(\gamma_\text{us}\) along the \(\langle  111 \rangle\) direction on the \(\{110\}\) and \(\{112\}\) planes. \textbf{b} Elastic constants of the BCC lattice. \textbf{c}-\textbf{d} Surface energy \(\gamma_{\text{s}}\) of \(\{100\}\) and \(\{110\}\) planes as a function of Re concentration. The calculations of (\textbf{a}-\textbf{c}) are performed in VASP, while (\textbf{d}) in QE.}
\end{figure}

\begin{figure}[!htbp]
  \centering
  \includegraphics[width=1.0\textwidth]{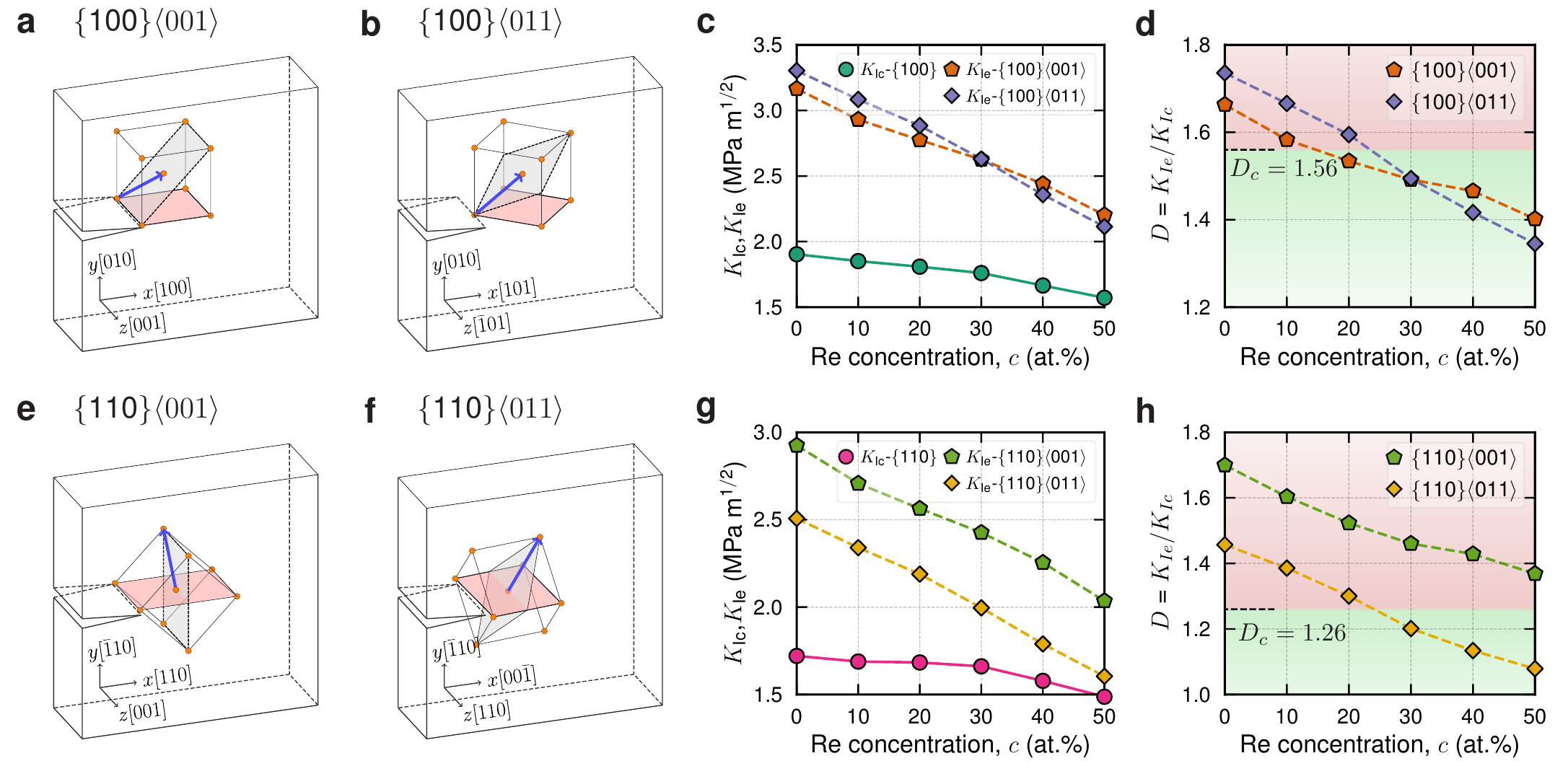}
  \caption{\label{fig:kie_kic_wre_calc}\(\vert\) \textbf{ Crack systems and critical stress intensity factors (\(K_\textnormal{Ic}\) and \(K_\textnormal{Ie}\) ) calculated in BCC W-Re alloys.} \textbf{a},\textbf{b},\textbf{e},\textbf{f}, Schematics of four crack systems susceptible to cleavage fracture in BCC structures. \textbf{c}-\textbf{g} Critical stress intensity factor for cleavage \(K_\textnormal{Ic}\) and dislocation emission \(K_\textnormal{Ie}\). \textbf{d}-\textbf{h} Ductility criterion \(D = K_{\textnormal{Ie}}/K_{\textnormal{Ic}}\) as a function of Re concentration. For all cases, \(K_\textnormal{Ie}\) and \(D\) all have opposite trend with Re concentration \(c\), while \(K_{\textnormal{Ic}}\) are insensitive to Re additions. \(D\) satisfies the \(D_{c}\) threshold starting from 20at.\% Re. The intrinsically brittle cracks become ductile at room temperature for \(\{100\}\) and \(\{110\}\) planes with 20at.\%Re or more.}
\end{figure}

\begin{figure}[!htbp]
  \input{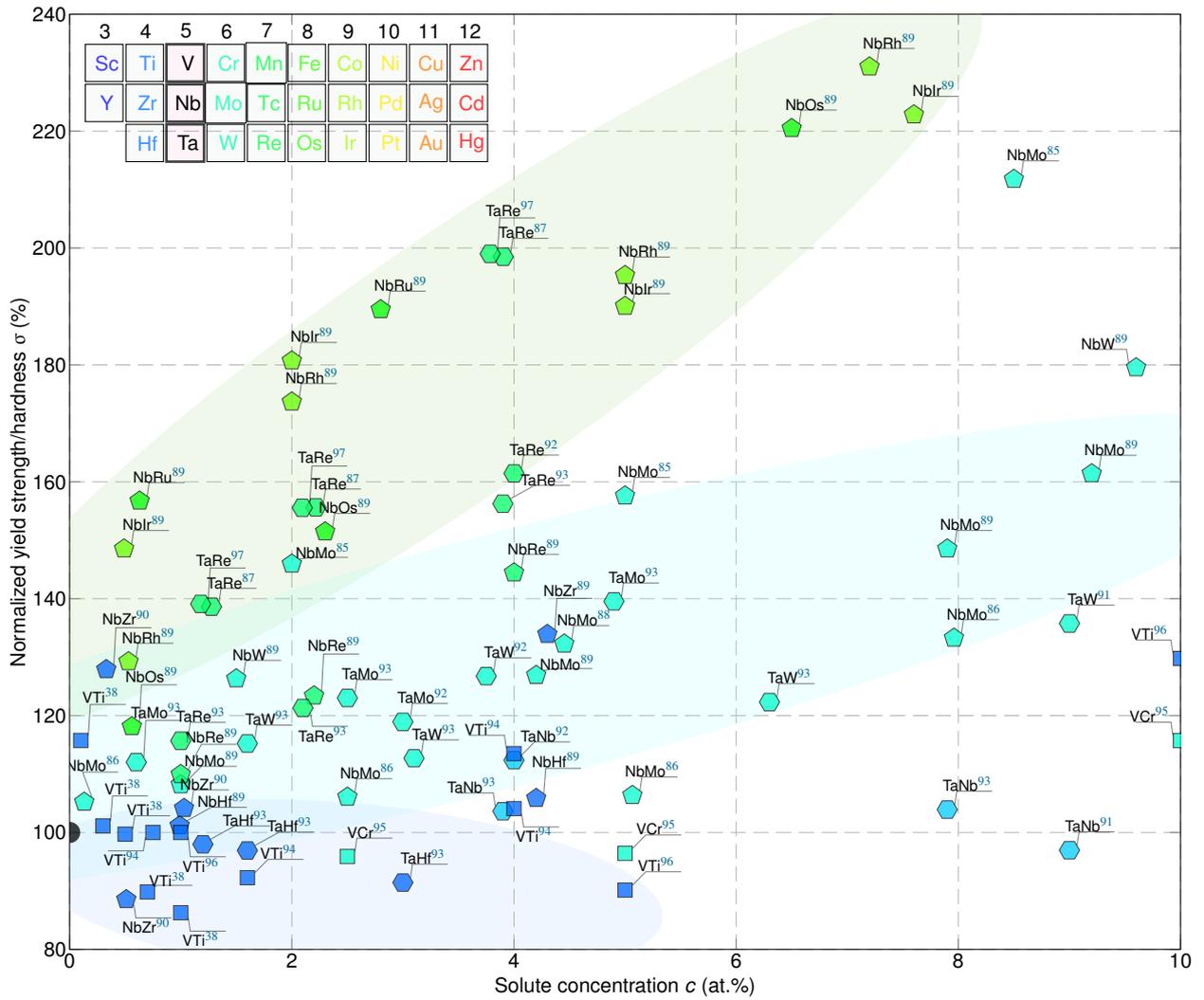}
  \caption{\label{fig:hs_gp_v_exp}\(\vert\) \textbf{Normalized yield strength/hardness of Group V (V, Nb, Ta) binary BCC transition metal alloys at 77/78 K.} In general and at low concentrations, solutes on the left and right of Group V metals have softening and hardening effects, respectively. }
\end{figure}

\begin{figure}[!htbp]
  \input{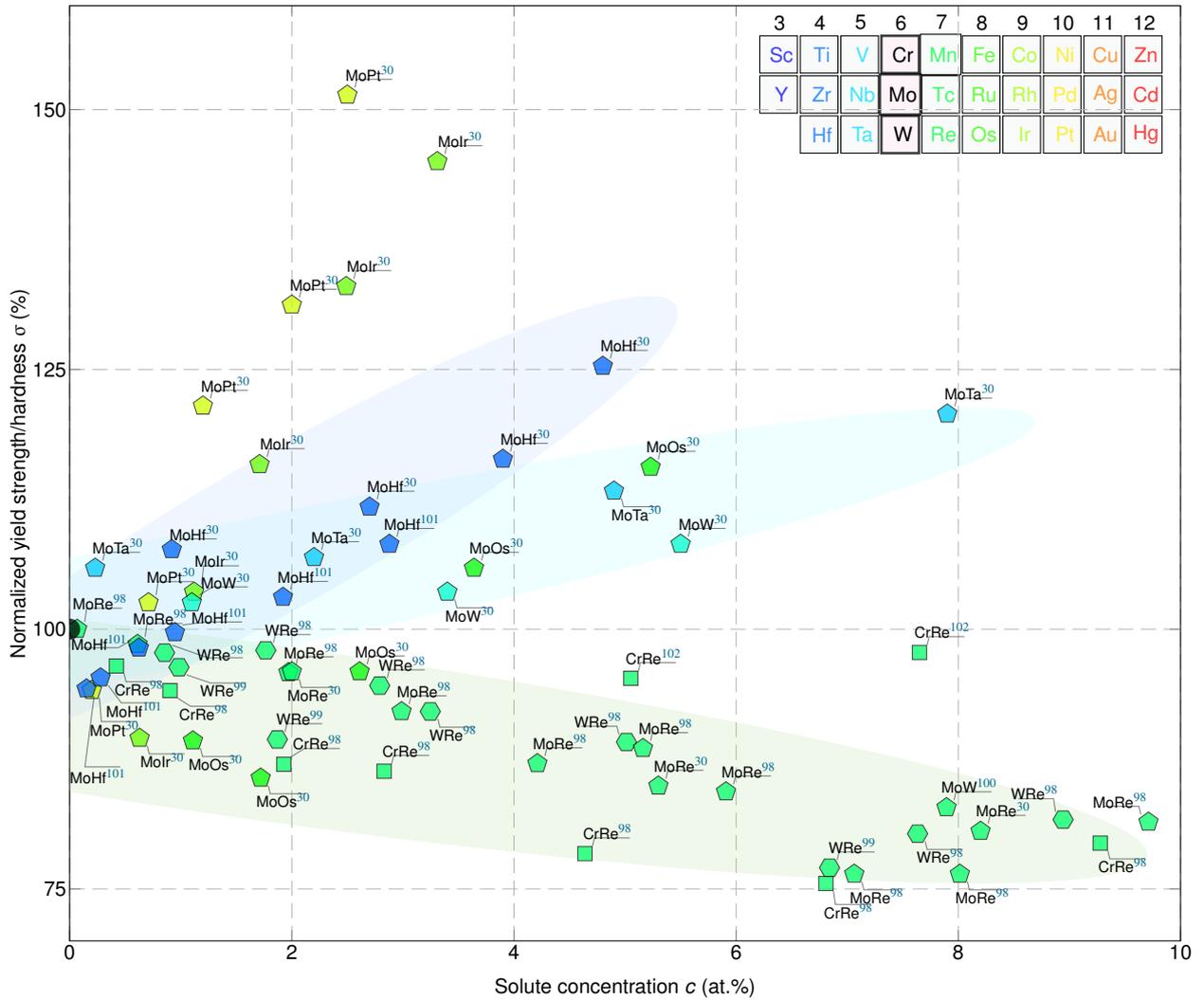}
  \caption{\label{fig:hs_gp_vi_exp}\(\vert\) \textbf{Normalized yield strength/hardness of Group VI (Cr, Mo, W) binary BCC transition metal alloys at 77/78 K.} In general and at low concentrations, solutes on the left and right of Group VI metals have hardening and softening effects, respectively. Pt and Ir exhibit solid solution softening at dilute concentrations (\textasciitilde{}0.5 \%) and rapid hardening above \textasciitilde{}0.5\%. }
\end{figure}


\clearpage

\begin{table}[!htbp]
\centering
\begin{tabular} {x{2.5cm} x{1.25cm} x{2.8cm} x{3cm} x{3cm} x{1.2cm}}
\textbf{Calculations} & \textbf{Element} & \textbf{Valence states} & \textbf{Cutoff energy} (eV) & \(k\)\textbf{-mesh} (unitcell) & \textbf{Sigma} \\
   \hline
&   V  & \(3s^{2}3p^{6}3d^{4}4s^{1}\) & 550                & 21 \texttimes{} 21 \texttimes{} 21   & 0.2 \\
&   Nb & \(4s^{2}4p^{6}4d^{4}5s^{1}\) & 550                & 19 \texttimes{} 19 \texttimes{} 19   & 0.1 \\
&   Ta & \(5p^{6}5d^{4}6s^{1}\)       & 550                & 19 \texttimes{} 19 \texttimes{} 19   & 0.1 \\
&   Cr & \(3s^{2}3p^{6}3d^{5}4s^{1}\) & 600                & 22 \texttimes{} 22 \texttimes{} 22   & 0.1 \\
\multirow{1}{1.8cm}{\textbf{Energy difference \(\Delta E\)}} &   Mo & \(4s^{2}4p^{6}4d^{5}5s^{1}\) & 600 & 20 \texttimes{} 20 \texttimes{} 20 & 0.1 \\
&   W  & \(5s^{2}5p^{6}5d^{5}6s^{1}\) & 550                & 20 \texttimes{} 20 \texttimes{} 20   & 0.05 \\
&   Fe & \(3s^{2}3p^{6}3d^{7}4s^{1}\) & 700                & 23 \texttimes{} 23 \texttimes{} 23   & 0.05 \\
&   Li & \(2s^{1}\)                   & 400                & 19 \texttimes{} 19 \texttimes{} 19   & 0.2 \\
&   Na & \(2p^{6}3s^{1}\)             & 550                & 31 \texttimes{} 31 \texttimes{} 31   & 0.2 \\
&   K  & \(3s^{2}3p^{6}4s^{1}\)       & 550                & 24 \texttimes{} 24 \texttimes{} 24   & 0.2 \\
&      &  &  &  & \\
  \multirow{2}{1.8cm}{\textbf{Dislocation core, \(\gamma\)-surface}} & Li & \(2s^{1}\) & 400 & 19 \texttimes{} 19 \texttimes{} 19 & 0.2 \\
& Ta & \(5d^{4}6s^{1}\) & 550 & 24 \texttimes{} 24 \texttimes{} 24 & 0.2 \\
&   W  & \(5d^{5}6s^{1}\)             & 550                & 31 \texttimes{} 31 \texttimes{} 31 & 0.2 \\
\end{tabular}
\caption{\label{tab:dft_details}\(\vert\) \textbf{Valence states, plane wave energy cutoffs, k-mesh densities and smearing parameters used in standard DFT-VASP calculations.}}
\end{table}

\begin{table}[!htbp]
  \centering
  \begin{tabular} {x{2.5cm} x{4.0cm} x{4.0cm} }
    \textbf{Element} & \multicolumn{2}{c}{\textbf{DFT VCA valence state}} \\
     & VASP  & QE \\
    \hline
    Sc &  \(3d^1 4s^2\)  & \(3s^2 3p^6 3d^1 4s^2\) \\
    Ti &  \(3d^3 4s^1\)  & \(3s^2 3p^6 3d^2 4s^2\) \\
    V  &  \(3d^4 4s^1\)  & \(3s^2 3p^6 3d^3 4s^2\) \\
    Cr &  \(3d^5 4s^1\)  & \(3s^2 3p^6 3d^5 4s^1\) \\
    Mn &  \(3d^6 4s^1\)  & \(3s^2 3p^6 3d^5 4s^2\) \\
    Fe &  \(3d^7 4s^1\)  & \(3s^2 3p^6 3d^6 4s^2\) \\
    Co &  \(3d^8 4s^1\)  & \(3s^2 3p^6 3d^7 4s^2\) \\
    Ni &  \(3d^9 4s^1\)  & \(3s^2 3p^6 3d^8 4s^2\) \\
    Cu &  \(3d^{10} 4s^1\) & \(3s^2 3p^6 3d^{10} 4s^1\) \\
    Zn &  \(3d^{10} 4s^2\) & \(3s^2 3p^6 3d^{10} 4s^2\) \\
    Y  &  \(4s^2 4p^6 4d^1 5s^2\)  & \( 4s^2 4p^6 4d^1    5s^2 \) \\
    Zr &  \(4s^2 4p^6 4d^2 5s^2\)  & \( 4s^2 4p^6 4d^2    5s^2 \) \\
    Nb &  \(4s^2 4p^6 4d^4 5s^1\)  & \( 4s^2 4p^6 4d^3    5s^2 \) \\
    Mo &  \(4d^5 5s^1\)  & \( 4s^2 4p^6 4d^5    5s^1 \) \\
    Tc &  \(4d^6 5s^1\)  & \( 4s^2 4p^6 4d^5    5s^2 \) \\
    Ru &  \(4d^7 5s^1\)  & \( 4s^2 4p^6 4d^7    5s^1 \) \\
    Rh &  \(4d^8 5s^1\)  & \( 4s^2 4p^6 4d^7    5s^2 \) \\
    Pd &  \(4d^{9} 5s^1\)  & \( 4s^2 4p^6 4d^{10} 5s^2 \) \\
    Ag &  \(4d^{10} 5s^1\) & \( 4s^2 4p^6 4d^{10} 5s^1 \) \\
    Cd &  \(4d^{10} 5s^2\) & \( 4s^2 4p^6 4d^{10} 5s^2 \) \\
    Hf &  \(5d^3 6s^1\)  & \( 5s^2 5p^6 5d^2    6s^2 \) \\
    Ta &  \(5d^4 6s^1\)  & \( 5s^2 5p^6 5d^3    6s^2 \) \\
    W  &  \(5d^5 6s^1\)  & \( 5s^2 5p^6 5d^4    6s^2 \) \\
    Re &  \(5d^6 6s^1\)  & \( 5s^2 5p^6 5d^5    6s^2 \) \\
    Os &  \(5d^7 6s^1\)  & \( 5s^2 5p^6 5d^6    6s^2 \) \\
    Ir &  \(5d^8 6s^1\)  & \( 5s^2 5p^6 5d^7    6s^2 \) \\
    Pt &  \(5d^9 6s^1\)  & \( 5s^2 5p^6 5d^8    6s^2 \) \\
    Au &  \(5d^{10} 6s^1\) & \( 5s^2 5p^6 5d^{10} 6s^1\) \\
    Hg &  \(5d^{10} 6s^2\) & \( 5s^2 5p^6 5d^{10} 6s^2\)
  \end{tabular}
  \caption{\label{tab:vs_state_vca_vasp_qe}\(\vert\) \textbf{Valence states used in the DFT VCA calculations in VASP and QE.} Other parameters are described in the Methods section.}
\end{table}

\begin{table}[!htbp]
\centering
\begin{tabular}{x{2.3cm} x{1.5cm} x{8cm} x{2.5cm}}
\hline
& \textbf{Element} & \textbf{Potential formulation and reference} & \textbf{No. of Potentials} \\
\hline
& V & EAM~\cite{mendelev_2007_prb}, MEAM~\cite{lee_2001_prb,maisel_2017_prm}, BOP~\cite{vitek_2014_msmse}, GAP~\cite{byggmastar_2020_prm} V1, V2 & 7 \\
& Nb & EAM~\cite{zhang_2016_jcp,fellinger_2010_prb}, MEAM~\cite{lee_2001_prb,yang_2019_cms}, BOP~\cite{vitek_2014_msmse}, ADP~\cite{starikov_2017_cms}, SNAP~\cite{zheng_2020_npjcm}, GAP~\cite{byggmastar_2020_prm} & 8 \\
& Ta & EAM~\cite{ravelo_2013_prb,li_2003_prb,zhou_2004_prb,chen_2019_cms}, MEAM~\cite{lee_2001_prb}, BOP~\cite{vitek_2014_msmse}, ADP~\cite{pun_2015_am}, SNAP~\cite{zheng_2020_npjcm,thompson_2015_jcp}, GAP~\cite{byggmastar_2020_prm} & 10 \\
Transition metals & Cr & EAM~\cite{mendelev_2020_openkim}, MEAM~\cite{choi_2017_cms}, BOP~\cite{vitek_2014_msmse}, ADP~\cite{mishin_2018_msmse} & 4 \\
& Mo & EAM~\cite{ackland_1987_pma,zhou_2004_prb}, MEAM~\cite{lee_2001_prb,park_2012_prb,kim_2017_calphad}, BOP~\cite{vitek_2014_msmse}, ADP~\cite{starikov_2017_jnm}, SNAP~\cite{zuo_2020_jpca,zheng_2020_npjcm,chen_2017_prm}, GAP~\cite{byggmastar_2020_prm} & 11 \\
& W & EAM~\cite{setyawan_2018_jap,zhou_2004_prb,olsson_2009_cms,marinica_2013_jpcm,bonny_2017_jap,chen_2019_cms}, MEAM~\cite{lee_2001_prb,lenosky_2017_openkim,park_2012_openkim}, BOP~\cite{vitek_2014_msmse}, SNAP~\cite{zheng_2020_npjcm}, GAP~\cite{byggmastar_2020_prm} & 12 \\
& Fe & EAM~\cite{morris_2008_prb,proville_2012_nm,bonny_2009_pm,chiesa_2011_jpcm,chamati_2006_ss,meyer_1998_prb,mendelev_2003_pm,ackland_1997_pma,malerba_2010_jnm,zhou_2004_prb,olsson_2009_cms,starikov_2021_prm}, MEAM~\cite{lee_2001_prb,kim_2009_am,liyanage_2014_prb,asadi_2015_prb}, BOP~\cite{byggmastar_2020_jnm}, ReaxFF~\cite{aryanpour_2010_jpca}, GAP~\cite{dragoni_2018_prm}, ANN~\cite{mori_2020_prm} & 20 \\
& & & \\
& Li & EIM~\cite{zhou_2010_lmp}, MEAM~\cite{alam_2015_jpcs,groh_2015_msmse}, SNAP~\cite{zuo_2020_jpca} & 4 \\
Alkaline metals & Na & EIM~\cite{zhou_2010_lmp} & 1 \\
& K & EIM~\cite{zhou_2010_lmp} & 1 \\
\hline
\end{tabular}
\caption{\label{tab:potential}\(\vert\) \textbf{Interatomic potentials for seven transition and three alkaline metals used in this work.}}
\end{table}

\begin{table}[!htbp]
\centering
\begin{tabular}{x{4.5cm} x{1.5cm} x{1.5cm} x{1.5cm} x{1.5cm} x{1.5cm}}
  & \textbf{Element} &  \multicolumn{4}{c}{\textbf{Structure energy difference}}  \\
 & &                      \(\Delta E\) (meV/atom) & & &  \\
  \textbf{Transition metals} & & this work & & & \\
  \hline
              & V  & 243 & 243~\cite{lin_2014_msmse} & 246~\cite{shang_2010_cms} & \\
  Group V     & Nb & 322 & 324~\cite{lin_2014_msmse} & 324~\cite{shang_2010_cms} & \\
              & Ta & 238 & 247~\cite{lin_2014_msmse} & 244~\cite{shang_2010_cms} & \\
  \hline
              & Cr & 412 & 389~\cite{lin_2014_msmse} & 386~\cite{shang_2010_cms} & \\
  Group VI    & Mo & 424 & 425~\cite{lin_2014_msmse} & 409~\cite{shang_2010_cms} & \\
              & W  & 483 & 495~\cite{lin_2014_msmse} & 474~\cite{shang_2010_cms} & \\
  \hline
  Group VIII  & Fe & 138 & & 141~\cite{shang_2010_cms} & \\
  & & \\
  \textbf{Alkaline metals} & & & & & \\
  \hline
  & Li & -1.44 & & -1.5~\cite{shang_2010_cms} & -1.37~\cite{hutcheon_2019_prb} \\
  & Na & -0.61 & & -0.3~\cite{shang_2010_cms} & -0.69~\cite{xie_2008_njp} \\
  & K  & -0.33 & & -0.3~\cite{shang_2010_cms} & -0.32~\cite{xie_2008_njp} \\
  \end{tabular}
  \caption{\label{tab:delta_E_bcc_fcc}\(\vert\) \textbf{Energy differences of the BCC and FCC structures of 10 elements calculated by DFT in VASP.} The FCC structure is highly unfavourable (\(\Delta E > 138\) meV/atom) in all transition metals, while alkaline metals have negligibly small energy differences (\(\Delta E\) \(\sim\)1 meV/atom) between the BCC and FCC structures.}
\end{table}

\begin{table}[!htbp]
  \centering
  \begin{tabular} {x{4.5cm} x{2.5cm} x{3.0cm} x{3.0cm} x{3.0cm} }
    \multicolumn{2}{c}{\textbf{Calculation}} &\multicolumn{3}{c}{\textbf{Supercell vectors}} \\
                         & &\(\mathbf{c}_1\) & \(\mathbf{c}_2\) & \(\mathbf{c}_3\) \\
    \hline
   Surface energy &  \{100\} plane & [1,0,0] & [0,1,0] & [0,0,1] \\
                  &  \{110\} plane & 1/2[\(\overline{1}\),1,1] & 1/2[1,\(\overline{1}\),1] & [1,1,0] \\
    \hline
   Unstable stacking  & 1/2$\langle 111 \rangle$\{110\} & 1/2[\(\overline{1}\),1,1] & 1/2[1,\(\overline{1}\),1] & [1,1,0] \\
   fault energy       & 1/2$\langle 111 \rangle$\{112\} & [\(\overline{1}\),1,0] & 1/2[\(\overline{1},\overline{1}\),1] & [1,1,2] \\
    \hline
    \(\gamma\)-surface & \{110\} plane & 1/2[\(\overline{1}\),1,1] & 1/2[1,\(\overline{1}\),1] & [1,1,0]
  \end{tabular}
  \caption{\label{tab:supercell_dft}\(\vert\) \textbf{Supercells used in the calculations of surface energies, unstable stacking fault energies and \(\gamma\)-surface by DFT. } Other parameters are described in the Methods section. }
\end{table}

\begin{table}[!htbp]
\centering
\begin{tabular}{x{2cm} x{1.2cm} x{1.2cm} x{1.2cm} x{1.2cm} x{1.2cm} x{1.2cm} x{1.2cm} x{1.2cm} x{1.2cm} x{1.2cm}}
  \hline
  \textbf{Element} & \(E_\text{c}\) & \(r_\text{e}\) & \(\alpha\) & \(\beta^{0}\) & \(\beta^1\)  & \(\beta^2\) & \(\beta^3\) & \(t^1\) & \(t^2\) & \(t^3\) \\
  \hline
            Li     & 1.6700  & 3.0233  & 3.1636     & 1.7854        & 1.2538       & 4.5937      & 2.9397      & 4.1275  & 3.3688  & 0.2672 \\
            V1     & 5.3100  & 2.6154  & 4.7719     & 4.9475        & 2.7601       & 4.4149      & 3.2709      & 4.7765  & -3.9531 & 2.5272 \\
            V2     & 5.3100  & 2.6154  & 4.7283     & 5.6477        & 5.5320       & 2.3578      & 2.3319      & 5.9285  & -7.2050 & 4.9702 \\
  \hline
  \textbf{Element} & \(A\)  & \(C_{\text{min}}\) & \(C_{\text{max}}\) & \(d_{\text{repuls}}\) & \(d_{\text{attrac}}\)  & \(r_c\) & \(\Delta r\)  \\
  \hline
            Li     & 1.0109 & 0.0006             & 2.7284             & 0.0795                & 0.0047                 & 6.6521  & 4.0468        \\
            V1     & 0.5562 & 0.1838             & 3.0049             & 0.0134                & -0.0259                & 7.5334  & 5.2633        \\
            V2     & 0.4915 & 0.2286             & 3.0239             & 0.0208                & 0.0554                 & 7.8342  & 5.5395        \\
  \end{tabular}
  \caption{\label{tab:meam_pot_param}\(\vert\) \textbf{Parameters of new MEAM interatomic potentials for Li and V (V1, V2).} The units of the cohesive energy \(E_\text{c}\) and equilibrium distance \(r_\text{e}\) are eV and \AA{}, respectively.}
\end{table}

\begin{table}[!htbp]
\centering
\begin{tabular}{x{2.5cm} x{2.5cm} x{2.5cm}}
  \hline
  \textbf{Models} & \multicolumn{2}{c}{\textbf{Peierls barrier (meV/$\mathbf{b}$)}} \\
  & \textbf{Drag method}& \textbf{NEB method} \\
  \hline
  Pure W      & 88.5  & 88.7 \\
  W-5\%Re  & 73.7 & 73.6 \\
  W-10\%Re & 60.7 & 61.5 \\
  W-10\%Ta & 104.5 & 104.5 \\
  \end{tabular}
  \caption{\label{tab:drag_vs_neb}\(\vert\) \textbf{Peierls barrier of W and W-Re/Ta alloys calculated by the drag and NEB methods in VASP.}}
\end{table}

\clearpage

\end{document}